\documentclass[reprint,amsmath,amssymb,aps,superscriptaddress,floatfix]{revtex4-1}

\usepackage{graphicx}
\graphicspath{ {../plots/} }
\usepackage{dcolumn}
\usepackage{bm}
\usepackage{natbib} \usepackage{hyperref}
\usepackage{indentfirst} 
\usepackage{float} 
\usepackage{blindtext} 
\usepackage[font=scriptsize,labelfont=bf,format=hang,
justification=centering]{caption,subcaption} 
\usepackage{array}

\newcommand\grad{\nabla}  
\newcommand\be{\begin{equation}} \newcommand\en{\end{equation}}
 \newcommand\inv[1]{\frac{1}{#1}}
\newcommand\half{\frac{1}{2}}

\newcommand\diff[2]{\frac{\textrm{d}{#1}}{\textrm{d}{#2}}}

\newcommand\diffp[2]{\frac{\partial{#1}}{\partial{#2}}}
\newcommand\diffpa[3]{\left.\diffp{#1}{#2}\right|_{#3}}

\begin{document}

\title{
Near-field simulations of pellet ablation for  disruptions mitigation in tokamaks}

\author{Nicolas Bosviel} \affiliation{
  Department of Applied Mathematics and Statistics, Stony Brook University, Stony Brook, NY\\
} \author{Paul Parks} \affiliation{
  General Atomics, San Diego, CA\\
} \author{Roman Samulyak} \affiliation{
  Department of Applied Mathematics and Statistics, Stony Brook University, Stony Brook, NY\\
} \affiliation{
  Computational Science Initiative, Brookhaven National Laboratory, Uptown, NY\\
}

\begin{abstract}
  Detailed numerical studies of the ablation of a single neon pellet in the plasma disruption mitigation parameter space have been
  performed. Simulations were carried out using FronTier, a hydrodynamic and  low magnetic Reynolds number MHD code with 
  explicit tracking of material interfaces.  FronTier's physics models resolve the pellet surface ablation and the formation of a dense, cold cloud of ablated material, the
  deposition of energy from hot plasma electrons passing through the
  ablation cloud,  expansion of the ablation cloud along magnetic field lines and the radiation losses. 
 A local thermodynamic equilibrium model based on Saha equations has been used to resolve atomic processes in the cloud and
 Redlich-Kwong corrections to the ideal gas equation of state for cold and dense gases have been used near the pellet surface.
 The FronTier pellet code is the next generation of the code described in [R. Samulyak, T. Lu, P. Parks, Nuclear Fusion, (47) 2007, 103--118]. 
 It has been validated against the semi-analytic improved Neutral Gas Shielding model in the 1D spherically symmetric
  approximation. Main results include quantification of the influence of atomic processes and Redlich-Kwong corrections on the pellet ablation in spherically  symmetric approximation and verification of analytic scaling laws in a broad range of pellet and plasma parameters. 
  Using axially symmetric MHD simulations, properties of ablation channels and the reduction of pellet ablation rates in magnetic fields of increasing strength have been studied. While the main emphasis has been given to neon pellets for the plasma disruption mitigation, selected results on deuterium  fueling pellets have also been presented.
 \end{abstract}

\maketitle

\section{Introduction}
\label{sec:introduction}
Effective and reliable disruption mitigation is critical for the International Thermonuclear Experimental Reactor (ITER) and
future burning plasma devices. The most likely candidate for the ITER plasma disruption mitigation system (DMS) is the
Shatterd Pellet Injection (SPI). In SPI, a large pellet, composed of a
frozen mixture of neon and deuterium, is injected inside a breaker tube
causing fragmentation. The plume of pellet fragment ablates in the plasma, radiates, and  induces a thermal
quench. A prototype DMS using SPI, tested on DIII-D, has been largely
successful~\cite{Commaux16, Shiraki16, Shiraki_Commaux16} and it is
now critical to develop methods to extrapolate those results to the
much higher temperatures and magnetic fields of ITER.

The ablation of cryogenic pellets and SPI fragments in tokamaks is an intrinsic multiscale and multiphysics phenomenon. 
The spatial scales of the problem range from sub-millimeter-size (the scale of high density gradients in the ablation cloud near the pellet surface, affecting the
ablation rates) to multi-meter-size (the distances traveled by the ablated material in tokamak plasma). Tokamak codes such as M3D-C1 and NIMROD have performed simulations
of the tokamak plasma dynamics in the presence of ablating impurities (jets), pellets and SPI fragments \cite{Izzo06,Kim19,Lyons19}.  Such global scale numerical simulations cannot 
resolve physics processes determining pellet ablation rates and employ semi-analytic models for impurities. In contract, pellet ablation models that resolve all relevant physics processes 
occurring on the pellet surface and in the ablation cloud and compute pellet ablation rates, constrain themselves to domains of the order of 10  - 50 cm around the pellet, and
use tokamak plasma parameters as input. Borrowing the terminology adopted in electromagnetism, we call such models near-field pellet models.  

Near-field pellet ablation models have been explored in the past in theoretical and simulation works. 
Previous and on going analytical and semi-analytical studies
developed an elegant and robust theory of the mechanisms of single
pellet ablation \cite{Parks_transonic,Kuteev_1995, Parks_Ishizaki04} and when supplemented with additional physics ($\mathbf{J}\times \mathbf{B}$
force) provide valuable insight for penetration studies \cite{Samulyak07}. See \cite{Pegourie_2007} and references therein for the summary and other details.

During the pellet ablation, hot plasma electrons impact on the pellet surface causing a rapid ablation and the fiormaiton
of a dense, cold cloud of neutral atoms expanding
isotropically. This cloud partially shields the pellet from the
incoming plasma energy flux and is the main factor controlling the
ablation rate. 
The pellet material heats up, undergoes atomic processes (multiple ionization / recombination for neon and dissociation and ionization for deuterium pellets), and expands along 
magnetic fields lines by $\mathbf{J}\times \mathbf{B}$ forces.
From~\cite{Parks_1992,Samulyak07} the magnetic
Reynold's number $R_m = \mu_0Lv\sigma \ll 1$, where $L$ is the length scale, $v$ is the field transverse velocity and $\sigma$ is the ablation cloud transverse conductivity in the vicinity of the pellet, moreover the magnetic
$\beta = \mu_0P/(B^2/2) \ll 1$ in the far field, under these circumstances the magnetic
field is taken to be constant and externally given, supporting the
electrostatic MHD approximation. The process of grad-B drift of the ablated material across magnetic field lines \cite{Rozhanskij_1994,Parks_2005} establishes 
a finite length of the cloud providing shielding to the pellet (the shielding length).

In our work, numerical methods and algorithms for the near-field pellet ablation model are implemented as
magnetohydrodynamic extensions of the FronTier-Lite~\cite{Fix06}
project and can be considered as the next generation of the pellet ablation code described in \cite{Samulyak07}.
FronTier-Lite provides an application programming interface (API) for front tracking \cite{Glimm1998ThreeDimensionalFT}, a hybrid Lagrangian-Eulerian method
in which a dynamically evolving interface (a Lagrangian mesh) is computed on a fixed volume-filling Eulerian mesh. 
Front tracking preserves and maintains sharp discontinuities by prohibiting taking derivatives across interfaces of steep gradients. The pellet surface
separating the pellet interior from the surrounding ablated material and the interface between the ablated material and the ambient plasma are explicitly 
tracked in our model. Tracking of the interface between the ablated material and the ambient plasma is a significant improvement compared to
the previous model~\cite{Samulyak07}, in which only the pellet surface was explicitly resolved, as it greatly improves the distribution of 
hydrodynamic states across the ablation cloud and eliminates the numerical diffusion across the ablated cloud - background plasma interface.
The use of high order hyperbolic WENO solvers  is another improvement compared to \cite{Samulyak07}.
Contrary to the MUSCL based scheme used in \cite{Samulyak07} WENO solver does not require solving an exact Riemann
problem at each cell interface. This facilitates the implementation of more elaborate equation of states. The code is validated with
comparison to the 1D theoretical predictions and scaling laws for an ideal neon cloud. The present work improves also several physics models of
 \cite{Samulyak07} such as the kinetic electron heating model and the EOS and adds new models such as radiation that was not applicable for 
 hydrogen pellets. 

Due to the axially-symmetric approximation, the 
grad-B drift of the ablated material across magnetic field lines \cite{Parks_Sessions_Baylor} cannot be explicitly resolved. We instead impose a fixed shielding length of the ablation cloud
obtained by theoretical estimates. This restriction will be removed in our forthcoming paper which will report full 3D simulations of the ablation of single 
pellets and SPI fragments using an adaptive Lagrangian particle code \cite{EPS-2018} developed as an extension of 
the Lagrangian particle hydrodynamic method \cite{LP}.  

The paper is organized as follows. In Sec.\ref{sec:equations} introduces equations
governing the evolution of the ablation cloud in the low magnetic Reynolds number MHD
approximation as well as the equation of state electronic heat flux, conductivity and radiation models. In
Sec.\ref{sec:implementation}, we describe in more details the numerical implementation of each component of the model. In Sec.\ref{sec:results} we present
verification of our model with semi-analytical results as well as MHD studies in a broad range of tokamak magnetic fields. While the main emphasis has been given to neon pellets in the plasma disruption regime, selected results on deuterium fueling pellets have also been presented.
Finally, we conclude the paper in Sec.\ref{sec:conclusion} with an overview of the present work and future studies using full 3D SPI numerical models.

\section{Model approximations and governing equations}
\label{sec:equations}
    \subsection{Equations for compressible hydrodynamics with electromagnetic terms\\}
    Following \cite{Samulyak07}, we assume low magnetic Reynold's number approximation for MHD processes in the ablated pellet cloud,
    \[
      \frac{\delta B}{B} \sim R_m \ll 1
     \]
where $\delta B$ is the current induced magnetic field. A detailed analysis of the field distortion near an ablating hydrogen pellet can be found in~\cite{Parks_distorsion}. Downstream from the pellet, the cloud pressure is such that $\beta = P/(B^2/2\mu_0) \ll 1$ and the external magnetic field is unperturbed by the ablation flow. The  magnetic field is therefore assumed constant in time and uniform $\mathbf{B}(\mathbf{x},t) \equiv \mathbf{B}$. The governing system of equation simplifies as follows
    \begin{subequations}    
      \label{eq:Euler}
      \begin{align}
        \diffp{\rho}{t} &= -\grad(\rho\mathbf{u}), \label{euler:mass}\\
        \rho\left(\diffp{}{t} + \mathbf{u} \cdot \grad\right)\mathbf{u} &= -\grad P + \mathbf{J}\times\mathbf{B}, \label{euler:momentum}\\
        \rho\left(\diffp{}{t} + \mathbf{u} \cdot \grad\right)e &= -P\grad\cdot\mathbf{u} + \inv{\sigma}\mathbf{J}^{2} - \grad\cdot\mathbf{q}, \label{euler:energy}\\
        P &= P(\rho,e), \label{euler:EOS}
      \end{align}
    \end{subequations}
where $\mathbf{u}$, $\rho$ and $e$ are the velocity, density and
    specific internal energy of the fluid, respectively, $P$ is the
    pressure, $\mathbf{J}$ is the current density and $\sigma$ is the fluid
    conductivity. The electron heat flux is represented by an external
    heat source -$\grad.\mathbf{q}$ described in detail in Section \ref{subsec:heat_flux}.  In the system eq.\ref{eq:Euler} we
    neglect the effects of heat conduction and viscosity. Finally, the
    system is closed by an equation of state (EOS)
    \eqref{euler:EOS}. Several EOS models will be discussed in Section  \ref{subsec:EOS}.

We ignore pellet motion effects, and carry out the simulations in the system of cylindrical coordinates ($r$, $z$, $\theta$). Since the coordinate $\theta$ is eliminated ($\diffp{\cdot}{\theta}=0$), the flow is constrained to axial symmetry with respect to the $z$-axis passing through the pellet center. The current density is obtained from the Ohm's law 
\be
        \mathbf{J} = \sigma (-\grad\theta + \mathbf{u}\times\mathbf{B}).
\en
Assuming a uniform distribution of the electric potential due to the pellet charging ($\grad\theta = 0$) and the axial symmetry of the ablation flow along magnetic field lines, this equation reduces to
\be
        J_\theta = \sigma u_rB \label{eq:J}.
\en

    \subsection{Equation of state\\}
    \label{subsec:EOS}
    Details of transonic regimes of the ablation flow strongly
    depend on the equation of state used to close the system of
    equations (\ref{eq:Euler}). Most of our verification results in spherically symmetric approximation (\ref{par:ideal}) were
    obtained using the ideal gas EOS model. For more realistic MHD simulations, it
    is critical that the EOS be able to capture accurately atomic and other complex processes in the flow. In the next subsection, we briefly
    describe our model based on the Redlich-Kwong
    equation of state (RK EOS), which we use to probe deviations of gas states in the cold dense region near
    the pellet surface from the ideal gas state. The EOS model that resolves atomic processes  will be
    discussed in~\ref{subsubsec:Saha}.


     \subsubsection{Redlich-Kwong equation of state}
     \label{subsubsec:RK}
     The region of the cloud closest to the pellet is formed of
        dense and cold ablated material. We implemented the
        Redlich-Kwong~\cite{RK} equation of state (RK EOS) to better
        capture this thin layer where the vapor might deviate from the ideal gas assumption.
        RK EOS improves the accuracy of the Van der Waals model
        by proposing a temperature dependence for the attractive term
        \begin{equation}
          P = \frac{RT}{V_m - b} - \frac a{T^{1/2}V_m(V_m + b)},
          \label{eq:RK}
        \end{equation}
        where $V_m$ is the molar volume and
        \begin{subequations}
          \begin{align}
          a&=\frac{0.42748R^2T^{5/2}_{\scriptsize \mathtt{crit}}}{P_{\scriptsize \mathtt{crit}}},
          \\
          b&=\frac{0.08664RT_{\scriptsize \mathtt{crit}}}{P_{\scriptsize \mathtt{crit}}},
          \end{align}
          \end{subequations}
        where $T_{\scriptsize \mathtt{crit}} = 44.5$K and $P_{\scriptsize \mathtt{crit}} = 27.6 $ bars.

        We need to derive explicit expressions for all
        thermodynamic functions for the RK EOS as they are necessary
        for the numerical implementation of our EOS
        library. From the second law of thermodynamic,
        \begin{equation}
          TdS = de + PdV_m,
        \end{equation}
        we have,
        \begin{equation}
          \diffpa e{V_m}T = -P + T\diffpa S{V_m}T.
        \end{equation}
        Using the Maxwell relation,
        \begin{equation}
          \diffpa s{V_m}T = \diffpa P T{V_m},
        \end{equation}
        and \eqref{eq:RK}, we obtain,
        \begin{subequations}
          \begin{align}
          \diffpa e{V_m}T &= -P + T\diffpa P T {V_m} \\
          &= -P +\frac{RT}{V_m-b} +
            \frac a{2T^{1/2}V_m(V_m+b)} \\
            &= \frac32\frac a{T^{1/2}V_m(V_m+b)}.
          \end{align}
          \end{subequations}
        Performing integration, 
        \begin{equation}
          e = -\frac32\frac a{b T^{1/2}}\ln{\frac {V_m+b}{V_m}} + A(T).
        \end{equation}
        Applying the ideal EOS limit, we find the unknown function
        $A(T)$:
        \begin{equation}
          A(T) = \frac{RT}{\gamma-1}.
        \end{equation}
        Therefore, the molar internal energy is
        \begin{equation}
          e = \frac{RT}{\gamma-1} -\frac32\frac a{b T^{1/2}}\ln{\frac {V_m+b}{V_m}}.
        \end{equation}
        Substituting (8) and (9) into the second law of thermodynamics
        and performing integration, we find the expression for the entropy,
        \begin{equation}
          \begin{split}
            S =& \frac R{\gamma-1}\ln{T} + R\ln{(V_m-b)} \\
            &- \frac12\frac a{bT^{3/2}}\ln{\frac{V_m+b}{V_m}}.
          \end{split}
          \end{equation}
        The Gr\"uneisen gamma can be found by taking differential of
        the entropy equation:
        \begin{equation}
        \begin{aligned}          
            \Gamma &= -\frac VT\diffpa TVS \\ 
                   & =  \cfrac{\left[\cfrac{RV_m}{V_m-b} + \cfrac{1}{2}\cfrac a{T^{3/2}(V_m+b)}\right]}{
                     \left[\cfrac R{\gamma-1} + \cfrac a{bT^{3/2}}\textrm{ln}{\cfrac{V_m+b}{V_m}}\right]}.                  
        \end{aligned}
        \end{equation}

     \subsubsection{EOS resolving atomic processes}
     \label{subsubsec:Saha}
    The fast electrons coming from the ambient plasma slow down
         in the ablated material cloud and deposit energy into the gas, causing ionization and heating. This chain of events
         is responsible for thermal ionization, which usually
         dominates fast electron impact ionization. The degree of
         ionization is very low in proximity to the surface of the
         pellet (mostly neutral atoms) and it becomes progressively
         higher further downstream as the temperature, and thus
         thermal ionization, steadily increases. The distribution of
         the multiply ionized states are found by solving a
         coupled set of Saha equations, which depends only on local
         values of the temperature and mass density.  For high-Z pellets, multiple ionization levels
         introduce energy sinks that strongly influence the
         temperature and conductivity of the cloud.

         Using the expression for specific internal energy from  \cite{Zeldovich} and neglecting electronic excitations, we obtain
           \begin{equation}
           e = \frac{3}{2}(1 + \alpha_e)\frac{kT}{m_i} +
           \inv{m_i}\sum_{m}Q_m\alpha_m,
         \end{equation}
         where $m_i$ is the mass of
         an ion (atom), $\alpha_e = n_e/n_t$ is the degree of ionization of the gas,
         and $\alpha_m = n_m/n_t$ is the concentration of m-ions. $\alpha_m$ and
         $\alpha_e$ are subject to the conservation conditions,

         \[ \sum_{m}n_m = n_t, \\ \sum_{m}\alpha_m = 1, \\ \sum_{m}mn_m =
           n_e, \\ \sum_{m}m\alpha_m = \alpha_e,
         \]
         where $n_t$, $n_m$, $n_e$ are the number density of respectively,
         all the nuclei, the $m^{th}$ ions and electrons.  The pressure is
         given by,
         \begin{equation}
           P = \rho(1+\alpha_e)\frac{kT}{m_i}.
         \end{equation}
         The system of Saha equations relating particle concentrations is as follows
         \begin{equation}
           \frac{\alpha_{m+1}\alpha_e}{\alpha_m} = \frac{2}{n_t}\frac{u_{m+1}}{u_m}\left(\frac{2\pi m_e kT}{h^2}\right)^{3/2}\textrm{exp}\left(-\frac{I_{m+1}}{kT}\right),
         \end{equation}
         where $h$ is the Planck constant and $u_m$ (for $m = 1, ..., Z$)
         are known electron partition functions. This system together
         with the conservation equations of mass and charge suffice to
         completely determine the particle concentrations $\alpha_m$
         and $\alpha_e$. Substituting parameters of neon,  the Saha system becomes [cgs-eV units]
         \begin{equation}
           \label{eq:Saha_neon}
           \frac{\alpha_{m+1} \alpha_e}{\alpha_m} = 6.035\times10^{21}\frac{T^{3/2}}{n_t}\textrm{exp}\left(\frac{-I_{m+1}}{T}\right).
         \end{equation}

         Since  it is prohibitively expensive to
         solve the coupled Saha system  of equations in a time-dependent hydrodynamic
         simulation,  a tabular EOS was used.  2D arrays were constructed using independent
         variables ($\rho$, $e$) and ($\rho$, $P$) incremented evenly
         in $\mathrm{log}_{10}$ space on a fine mesh over the range of
         interest for $\rho$, $e$, $P$. The ionization fractions were
         obtained by solving (\ref{eq:Saha_neon}) on these meshes and the
         pertinent thermodynamic quantities computed and saved. During runtime, we
         access these quantities using standard libraries for table look-up and cubic spline interpolation algorithms.

    \subsection{Electron heat flux model\\}
    \label{subsec:heat_flux}

   The electron heat flux model is similar to the one described
in~\cite{Samulyak07,Parks_Ishizaki04} but has been updated for high-Z
elements~\cite{Parks_2020} where the normalized heat flux is given by,
    \begin{align} \mathbf{\tilde q_{\pm}} = \tilde q_+ \: \hat z -
\tilde q_- \: \hat z \Longrightarrow -\mathbf{\grad}.\mathbf{q} =
-\left(\diff{\tilde q_+}{z} - \diff{\tilde q_-}{z}\right)
    \end{align}
    where "+" ("-") refers to right (left) going plasma
electrons, and $-\mathbf{\grad}.\mathbf{q}$ is the corresponding heat
deposition rate per unit volume. The 3-D linearized Fokker-Planck
kinetic equation is solved for the electron distribution function
$f(E,\mu,z)$ where $E$, $\mu$, $z$ are the energy, velocity space
variables and cosine pitch-angle with respect to the magnetic field,
respectively. Although slowing down and pitch-angle scattering on a
neutral target is weaker than it is on partially ionized target, there
is not a great difference between scattering by a neutral atom with
atomic number $Z$ and the atom in a partially ionized state with ionic
charge $Z_*$, provided that it is only weakly ionized, $Z_* \sim
1-2$. The expression for $Z_*$ depends weakly on the electron energy
$E \approx 2\:T_{e_\infty}$ and is provided as tabulated data for neon
and hydrogen species in the code. The numerical solution for parallel
electron heat flux can be accurately fit to a Bessel function,
\begin{subequations}
  \begin{align}
    q_{\parallel} &= q_\infty\:u\:K_2(u^{1/2}),\\ \notag \\
\tilde q_{\pm} &= \half \frac{q_{\parallel}}{q_\infty} = \half
\:u_{\pm}\:K_2(u_{\pm}^{1/2}),\\ \notag \\ q_\infty &=
\left(\frac{2}{\pi\:m_e}\right)^{1/2}\:n_{eff}\:T_{e_\infty}^{3/2}, \\
\notag \\ n_{eff} &= (1-0.001\:A)\:e^{-\Phi}\:n_{e_\infty}, \\ \notag
\\ A &= 23.92\:\ln\left(1+0.2014(1+Z_*)\right).\label{eq:A}
    \end{align}
  \end{subequations}
  Here $q_\infty$ is the half-space incident
Maxwellian heat flux, $A$ is the surface albedo (reflectivity) due to
collisional backscattering, $e^{-\Phi}$ is the decrease due to
electrostatic shielding, $K_n$ is the modified Bessel function of
order $n$ and,
\begin{subequations}
 \begin{align} u_\pm &= \frac{\tau_\pm}{\tau_{eff}}, \\ \tau_+ &=
\int\limits_{-\infty}^{z}n_e(z',r)dz', \quad \tau_- =
\int\limits_{z}^{\infty}n_e(z',r)dz', \\ \tau_{eff} &=
\frac{\tau_\infty}{0.625+0.55\sqrt{1+Z_*}},
     \end{align}
   \end{subequations}
   where $n_e = Z\:n_t = \rho\:Z\:/\:m$, with
$n_t$ and $\rho$ the total number density of nuclei and mass density
respectively of the fluid and $m$ the mass of the atom. Here $\tau_{\pm}$
are the density integrals of the (bound) ablation electrons and
$\tau_{eff}$ is an effective energy flux attenuation thickness of the
hot, streaming electrons due to slowing down and pitch-angle
scattering.

The Bessel function form is the analytical solution when the pitch-angle scattering term is neglected (an explicit derivation of this claim is provided in the appendix of~\cite{Parks_Ishizaki04}), in which case,
\begin{align}
 \tau_{eff} \rightarrow \tau_\infty = \frac{T_{e_\infty}^2}{8\:\pi\:e^4\:\ln\Lambda} \quad \mathrm{and} \quad \ln\Lambda = \ln\left[\sqrt{\frac{e}{2}}\:\frac{E}{I}\right]
  \end{align}
is the Bethe stopping power logarithm pertaining to inelastic scattering of fast electrons of atomic (bound) electron in the neutral gas layer, where $e$ is the Napier constant, $I$ is the mean excitation energy for neutral atoms required for computing the energy loss of fast electrons passing through matter. The analysis is restricted to the standard Bethe regime $E>I$, which corresponds to $T_{e_\infty} > I/2$ and the Coulomb logarithm is evaluated at $E \approx T_{e_\infty}$ since that is the average energy per particle transported into the cloud/pellet system from a distribution of semi-isotropic incident Maxwellian electrons. Using values recommended by~\cite{ICRU}, $I_{D_2} = 19.2$ eV, $Z=1$, $I_{Ne} = 135.5$ eV, $ Z=10$, $I_{Ar} = 188$ eV, $Z=18$. For partially ionized atoms, the corresponding I-values are larger and increase monotonically with ionic charge state. However, the dense, cold region near the pellet, which provides most of the shielding, is typically composed of a mix of neutral atoms and predominately singly charged ions, and the latter have only modestly ($\sim30\%$) larger $I$-values than neutral-atom ones~\cite{Stopping_power}. In the current work the small difference is discarded since atomic processes are of minor importance to the ablation rate. An analogous “Coulomb logarithm” $\ln\Lambda'$ arises from multiple small-angle elastic scattering collisions between fast electrons and the shielded atomic nucleus. For keV electrons, the use of the Thomas-Fermi atomic field model indicates that $\ln\Lambda' < \ln\Lambda$ for all atomic number $Z$. For convenience, the model discounts the difference between the two Coulomb logarithms by using the same formula in both energy loss and elastic scattering processes.

Finally the heat source $-\grad.\textbf{q}$ from
    the energy deposition by hot, long mean-free path electrons
    streaming into the ablation cloud along the magnetic field lines is given by,
    \begin{equation}
      -\grad.\textbf{q} = \frac{q_{\infty}Z n_t(r,z) \textrm{ln}\Lambda}{\tau_{\mathrm{eff}}}\left[g(u_+)+g(u_-)\right],
    \end{equation}
and the heat deposition on the surface is given as,
    \begin{equation}
      q_\pm = q_\infty \half u_\pm K_2(\sqrt{u_\pm}).
    \end{equation}

    \subsection{Conductivity and radiation\\}
    \label{subsec:cond}

    The conductivity model for hydrogenic species has been derived in ~\cite{Samulyak07}. For high-Z impurities the conductivity model was derived in~\cite{Parks_SCIDAC} and we review it here. Observing that LTE conditions must prevail in the ablatant cloud, the Saha system of equations can be used to determine each partial ionization fraction and then the local average ionization level defined by,
\begin{align}
  \overline Z = \frac{\sum\limits_{j=0}^{Z}Z_j\:n_j}{\sum\limits_{j=0}^{Z}n_j} = \frac{n_e}{n_t}              
  \end{align}
where $Z_j$ is the charge state of ion with density $n_j$ and $n_e$ ($n_t$) is the number density of free electrons (nuclei). Evaluating the conductivity for the ablation of high-Z impurities requires a slightly different average charge state sum,
\begin{align}\label{eq:Zeff}
 Z_{eff} = \frac{\sum\limits_jZ_j^2\:n_j}{\sum\limits_{j=0}^{Z}Z_j\:n_j},
  \end{align}
where $Z_{eff} \approx \overline Z$. Using the sum rule the conductivity is,
  \begin{align}
    \sigma = \frac{n_e\:e_i}{m_e\:(\nu_{ei}+\nu_{en})},
    \end{align}
where $\nu_{ei}$ is the electron-ion momentum exchange collision frequency and $\nu_{en}$ is the electron-neutral momentum exchange collision frequency. After evaluating $\nu_{ei}$ and $\nu_{en}$, the conductivity transverse to the magnetic field is given by [eV-cgs]
\begin{align}
  \sigma = \frac{9.7\times10^3 \: T_{e\infty}^{\frac{3}{2}}}{Z_{eff} \: \ln\Lambda \: + \: 0.000443 \: T_{e\infty}^{2.245} \: \frac{n_0}{n_e}}. \label{eq:cond}
    \end{align}
In the absence of neutrals, (\ref{eq:cond}) reduces to the Spitzer transverse conductivity,
\begin{align}    
 \sigma_{\mathrm{Spitzer}} = \frac{9.7\times10^3 \: T_{e\infty}^{3/2}}{Z \: \ln\Lambda},
  \end{align}
where $Z$ is the charge state. 

To account for radiation losses in the neon ablation clouds, we use a non-LTE radiation model in the thin optical limit approximation. The numerical algorithm uses tabulated emissivity table precomputed with the CRETIN code~\cite{CRETIN}.

    \subsection{Surface ablation model\\}
    \label{subsec:surface_ablation}

One detail of our pellet model remains to be explored and it is that of the physics of the phase transition at the pellet surface. After injection in the reactor plasma, the pellet is subjected to a pre-heating period in the sub-microsecond range that ends when its outer layers reach values close to the sublimation energy ($\epsilon_{Ne} = 0.02$ eV, $\epsilon_{H} = 0.005$ eV for neon and hydrogen respectively). From this moment forward the ablation phase begins. Atomic or molecular layers are continuously removed from the pellet providing the source of the expanding cloud. The authors of~\cite{Parks_Turnbull_Foster} give a qualitatively simple picture of the processes at the surface: by the time the energetic plasma electrons reach the surface of the pellet most of their energy has been degraded into thermal energy in the cloud, which drives the cloud expansion and outward flow. As a result they have an effective penetration range in the pellet of the order of the micrometer or less. A thin layer underneath the surface conducts the heat into the pellet and also back to the surface. The heat conducted into the pellet has been shown~\cite{Parks_Turnbull_Foster} to be negligible compared to that conducted to the surface. Because the energy required to vaporize the outer pellet layer is several orders of magnitude smaller compared to the first ionization potential, only a small fraction of the atoms leave the surface ionized ($<0.1$). Therefore, all the energy striking the pellet can be assumed to sustain the sublimation, this process does not raise the temperature of the pellet and the vapor layer around the pellet consists only of neutral atoms and/or molecules.

The system of differential equations~\ref{eq:Euler} governing the ablation requires three boundary conditions at the pellet surface, which are determined by considering,
\begin{itemize}
 \item the transition from solid to vapor,
  \item the energy balance across the solid/vapor transition layer,
   \item the change in flow states along the backward characteristic originating from the Riemann problem at the pellet surface (the flow of vaporized material is subsonic) .
     \end{itemize}

The discussion above justifies using constant temperature for vapor at the pellet surface. The dense and cold vapor near the surface is described by the 
Redlich-Kwong EOS discussed in Section \ref{subsubsec:RK}. We will show in Section \ref{par:RK} that the deviation from the ideal gas EOS causes 
negligibly small changes in the pellet ablation process. This we assume that the vapor temperature near the pellet surface is 
\begin{align}
 T_{\mathrm{vapor}} = 1.75\:T_{\mathrm{criticial}}\:, \label{eq:BC1}
  \end{align}
which for a neon pellet is $T_{\mathrm{vapor}} = 78$ K and $T_{\mathrm{vapor}} = 67$ K for a deuterium pellet, as it assures the ideal gas behavior over
a wide range of pressure ($0-200$ bars) typical of the surface pressure $P_{\mathrm{vapor}}$. 

The energy balance across the solid/gas transition layer is given by,
\begin{align}
  q_{\pm} & = \frac{\rho_{\mathrm{pellet}}u_{\mathrm{pellet}}}{M_a}\left(\epsilon + \int\limits_{T_\mathrm{pellet}}^{T_\mathrm{vapor}}C_p(T)dT \right) \\
         & = \frac{\rho_{\mathrm{vapor}} u_{\mathrm{vapor}}}{M_a}\left(\epsilon - \int\limits_{0}^{T_{\mathrm{pellet}}}C_p^{\mathrm{pellet}}(T)dT + C_p^{\mathrm{vapor}}T_{\mathrm{vapor}}\right),
   \end{align}
where ${\rho_{\mathrm{pellet}}\: u_{\mathrm{pellet}}}$ is the mass flux of the eroded pellet material with  $u_{\mathrm{pellet}}$ the surface recession speed. The second equality comes about because $u_{\mathrm{pellet}} \ll u_{\textrm{vapor}} \Longrightarrow \rho_{\mathrm{pellet}}\: u_{\mathrm{pellet}} = \rho_{\mathrm{vapor}}\: (u_{\mathrm{vapor}} - u_{\mathrm{pellet}}) = \rho_{\mathrm{vapor}}\: u_{\mathrm{vapor}}$. $T_{\mathrm{pellet}}$ is the temperature of the cryogenic solid pellet, $T_{\mathrm{vapor}}$ is the temperature of the vaporized material to satisfy (\ref{eq:BC1}) and $C_p^{\mathrm{vapor}}$ is the heat capacity of ideal gas at constant pressure ($(5/2)R$ for Ne and $(7/2) R$ for D$_2$, with $R = 8.31446$ J/K mol). The cohesive energy of solid $\epsilon$ is the latent heat of sublimation (1902 J/mol for Ne and 1243 J/mol for $\mathrm{D}_2$), the energy that must be supplied to convert the atoms of the solid into well separated neutral atoms at rest. The last undefined term, $C_p^{\mathrm{pellet}}(T)$ is the specific heat at constant pressure of the solid pellet and used to compute the energy of formation of the cryogenic pellet,
\begin{equation}
 \int\limits_{0}^{T_{\mathrm{pellet}}}C_p^{\mathrm{pellet}}(T)\:dT =
  \begin{cases}
   118.7 \; \mathrm{J/mol}, & \text{for Ne}, \\
   15.60  \; \mathrm{J/mol}, & \text{for $\mathrm{D}_2$},
    \end{cases}
     \end{equation}

Finally, to extract the last boundary condition we reuse a derivation proposed in~\cite{Samulyak07}. Since the flow at the pellet is subsonic, the structure of the Riemann problem contains a backward characteristic connecting the vapor to the surface of the solid pellet. Furthermore, noting that the gas is mostly neutrals and very close to ideal at the surface we can write down the characteristic equation,
\begin{align}
 \diff{u_N}{\Lambda_-} - \inv{\rho c} \diff{P}{\Lambda_-} - \alpha\frac{N_0}{r} c u_N = \Gamma \diff{q_\pm}{z} \label{eq:15}
  \end{align}
with the characteristic derivative,
\begin{align}
 \diff{}{\Lambda_-} = \diffp{}{t} + (u_N-c) \diffp{}{\textbf{N}}
  \end{align}
and $c$ is the sound speed, $\Gamma$ is the Gruneisen coeffecient, $\textbf{N}$ is the normal direction to the surface and $\alpha$ is 0, 1, or 2 for rectangular, cylindrical or spherical geometry.These three boundary conditions are sufficient to accurately describe the phase transition process.

\section{Implementation}
\label{sec:implementation}

MHD processes in the pellet ablation cloud are approximated by the low
magnetic Reynolds number MHD equations.  The
boundary  between plasma and cloud can be approximated as
sharp and is explicitly tracked during the simulation. The interface is
represented as a co-dimension one Lagrangian mesh moving on an
Eulerian mesh and we use a front tracking method for the dynamic
motion and topological changes of that interface. Front tracking is
implemented in our code using the FronTier-Lite API
~\cite{Fix06,glimm_axisymmetric,2007JCoPh.226.1532S} which implements
front tracking for inviscid flows and provide robust facilities for
interfacing with the front geometry. In this framework, the front is
represented (in 2D) by a piecewise linear curve moving according to
the contact discontinuity that it represents while the flow fields are
solved on the fixed-in-time spatial grid (Eulerian) representing the
bulk fluid.

Tracking the cloud/plasma interface explicitly keeps the boundary
between the two media sharp and removes interfacial numerical
diffusion by forbidding taking finite differences across thin, steep
gradient regions. Following \cite{Parks_AC}, we neglect
the weak diamagnetic current layer at the cloud boundary and
associated jump in magnetic field pressure, leading us to formulate a
contact discontinuity boundary condition at the interface,
\begin{equation}
  P_{boundary} = P_{\infty}.
\end{equation}
The ablated material is supported by the background plasma $P_{\infty}$.
One of the benefits to adopting front tracking for the cloud boundary is that it enables
us to solve the governing equations only in the region of interest,
that is only for the ablated material. The coupling between the
background plasma and the ablation cloud is done only through the
boundary conditions described above. The advantage of this approach is
twofold: it speeds up our simulations, since only grid cells hit by
the ablated material are updated at each time step, effectively
disregarding grid cells lying currently outside of the cloud; it
removes any unphysical states bound to arise were we trying to solve
the same set of equations for both the cloud and background plasma
over the whole of the computational domain.

The propagation of this interface is done at the beginning of every
time step by computing the time updated states and position of the
front points using a directionnaly split method. A sequence of
generalized Riemann problems are solved for the flow equations
projected onto the directions normal and tangential to the front at
the point being propagated. The implementation of this Riemann solver
and subsequent algorithms for the untangling and redistribution of the interface are
described in details in ~\cite{2007JCoPh.226.1532S} and references therein.

The bulk fluid states are then updated on the rectangular grid
according to the system of MHD equations using operator splitting. The
updated positions and states of the interface points found during the
front propagation routines above establish the region and boundary
conditions in which to solve the system.  The
hyperbolic conservation equations eq.\ref{eq:Euler} are solved using a
fifth order WENO spatial discretization in spherical (1D) and
axysimmetric (2D) coordinates~\cite{Wang_Johnsen,Shu_WENO} combined with
explicit fourth order Runge-Kutta time stepping scheme. Spatial
discretization is based on the integral form of the equations so that
the physical variables are expected to be conserved. In semi-discrete
form the Euler's system becomes,

\begin{subequations}
  \begin{align}
    \diff{\rho_i}{t} &= - \frac{r^\alpha_{i+\half}(\rho u)_{i+\half} - r^\alpha_{i-\half}(\rho u)_{i-\half} }{\Delta V_i}, \\
       \diff{(\rho u)_i}{t} &= - \frac{r^\alpha_{i+\half}(\rho u^2 + P)_{i+\half} - r^\alpha_{i-\half}(\rho u^2 + P)_{i-\half} }{\Delta V_i} + S(r_i), \\
          \diff{E_i}{t} &= - \frac{r^\alpha_{i+\half}((E+P) u)_{i+\half} - r^\alpha_{i-\half}((E+P) u)_{i-\half} }{\Delta V_i},
   \end{align}
\end{subequations}

where  $\alpha=2$ if the flow exhibits spherical symmetry around the origin, $\alpha=1$ for cylindrical symmetry around the $z$-axis, 0 otherwise, and
\begin{equation}
  \Delta V_i = \inv{1+\alpha}(r^{\alpha+1}_{i+\half} - r^{\alpha+1}_{i-\half}),
 \end{equation}
 and
 \begin{equation}
  S(r_i) = \frac{(r^{\alpha} P)_{i+\half} - (r^{\alpha} P)_{i-\half}}{\Delta V_i} - \frac{P_{i+\half} - P_{i-\half}}{\Delta r}.
  \end{equation}

The conserved variables to be evolved are the cell centered values
for the finite difference WENO. We use a local Lax-Friedrich
splitting and characteristic decomposition to approximate the
cell-edge fluxes. The geometric source terms evaluated at the cell
interface are interpolated with the same WENO weights.

The electromagnetic terms are found, in the general case, by solving
the Poisson equation for the electric potential. Since we do not
resolve the electric potential distribution in the ablation channel,
the current density $\textbf{J}$ is readily available from the radial
velocity $u_r$ and the magnetic field $\textbf{B}$. The heat
deposition $-\grad.\textbf{q}$, obtained by integrating the fluid
states along the field lines, is added to the internal energy and
changes the temperature of fluid states and therefore the
conductivity. Finally, the Lorentz force is accounted for by direct
integration of the momentum equation.

Our front tracking API being conceptually physics free allows for the
implementation of somewhat sophisticated equations of states and their
coupling to the spatial discretization schemes. An approximate Riemann
solver following the ideas in~\cite{Glaister} has been designed to
solve the Euler's equations with a general convex equation of state. For an
ideal gas, the pressure is given by
\begin{equation}
  P = (\gamma - 1)\rho e,
\end{equation}
where $\gamma$ is the constant ratio of specific heats. To mimic the ideal case, for a general equation of state of the form $P = P(\rho,e)$ a new dependent variable  $\gamma = \gamma(\rho,e)$ is introduced so that,
\begin{equation}
  P = (\gamma(\rho,e) - 1)\rho e.
 \end{equation}
 Roe averaging is then used to evaluate the left and right eigenvectors. In particular, it should be noted that contrary to the ideal case, $e - \frac{\rho P_\rho}{P_e} \neq 0$ in general and must be adequately approximated in the eigenvector $\left[1, u, v, w, \frac{u^2+v^2+w^2}{2} + e - \frac{\rho P_\rho}{P_e}\right]$.

\section{Results}
\label{sec:results}
The problem of pellet ablating in a hot plasma was studied in the 1D
spherically symmetric and 2D axisymmetric geometries. Following the
terminology of \cite{Samulyak07}, simulations where the
$\mathbf{J}\times \mathbf{B}$ force is ignored/included are termed
\textit{hydrodynamic/MHD}. The 1D hydrodynamic model is primarily used
for benchmark purposes and code verification against an updated transonic flow model \cite{Parks_private_com}
which uses a kinetic solution of the electron distribution function to obtain the heat flux moment for incident 
Maxwellian electrons and for all light element pellets.
The 2D axisymmetric hydrodynamic model studies the effect of anisotropic
heating along the magnetic field lines. It is used for comparison with
the 1D model and serves as a benchmark for the full 2D axisymmetric
MHD model. In the subsequent sections, we present results for the
ablation of a single neon pellet with radius $\mathrm{r}_p$ varying
between 1 mm and 7 mm, plasma electron temperature $T_{e\infty}$
between 1 keV and 8 keV, and plasma number density $n_{e\infty}$ between
$10^{14}$ 1/cc and $4\times10^{14}$ cc$^{-1}$, subject to reductions 
due to the processes described below. Plasma electrons
penetrating through the ablation cloud experience slowing down due to
inelastic scattering and Coulomb scattering. Both processes result in
a small fraction of the incident flux being reflected at the surface
due to back-scattering. It was also remarked in
~\cite{Parks_Ishizaki04} that incident flux of plasma electrons is
partially screened due to electrostatic shielding. Collisional
backscattering and electrostatic effectively reduce the incident
plasma electrons density from $n_{e\infty}$ to
$n_{\mathrm{eff}} =
(1-0.001\mathrm{A})\mathrm{e}^{(-e\phi/T_{e\infty})}n_{e\infty}$ where
\textrm{A} is the reflectivity in percent (surface albedo) of the
incident flux and the exponential term accounts for the electrostatic
shielding where $\phi$ is the potential drop across the cold cloud/hot
background plasma interface.

    \subsection{Ablation studies with hydrodynamic models}
    \label{sec:hydro_models}
      \subsubsection{Spherically symmetric approximation\\}
      \label{sec:spherically_symmetric_approx}
      \paragraph{Ideal gas case\\}
      \label{par:ideal}
In the first numerical experiment, verification studies
            have been performed by benchmarking our code with the        
           updated spherically  symmetric transonic flow model \cite{Parks_private_com} 
            that improves approximations of the Neutral Gas
            Shielding (NGS) model ~\cite{Parks_transonic} and have
            confirmed predictions for the ideal steady-state ablation flow over a
            wide range of pellet sizes and plasma conditions. The results and
            comparisons with the theory are compiled in the following tables and
            were obtained with plasma density (in 1/cc) $n_{e\infty} = 10^{14}
            \Longrightarrow n_{eff} = 1.205\times10^{13}$ and indicated values of plasma
            temperature $T_{e\infty}$ and pellet radius $\mathrm{r_p}$. The
            starred quntities are values at the Mach radius $\mathrm{r^*}$ where the
            Mach number is unity. We provide comparison values for
            the ablation rate G (g/s), the ratio of the pellet radius to Mach
            radius $\mathrm{r_p/r^*}$ and the ratio of surface pressure to the
            pressure at the Mach radius $\mathrm{P_{surface}/P^*}$ in tables~\ref{tab:Te2},~\ref{tab:Te5},~\ref{tab:Te8},~\ref{tab:error}.  
            These simulations have been performed for a neon pellet ($Z=10$, $\gamma = 5/3$), in a
            computational domain extending from 0 to 16 cm with the pellet center at
            the origin $r=0$ and with the grid resolution $\Delta r = 0.005$ cm. The results
            presented in the tables are obtained after verifying that the mesh convergence was reached.
            
            Because the flow is transonic, it is convenient to normalize the flow
            state variables with their values at the sonic radius for comparison
            with the improved NGS. Data in the tables demonstrate a good  agreement  the NGS scaling law. 
            For fixed plasma parameters, the sonic
            radius increases with the pellet radius. Conversely, for fixed pellet
            radius the position of the sonic radius becomes closer to the pellet
            surface as the plasma temperature is increased, similar to figure 2
            of~\cite{Parks_transonic} for hydrogen pellets. However, the
            dependence of $r^*$ on $T_{e_\infty}$ is very weak. The temperature at
            the sonic radius increases with the pellet radius and decreases with
            the plasma electron temperature $T_{e_{\infty}}$. The ratio of the
            pressure at the pellet surface over the sonic pressure is also weakly
            dependent on $T_{e\infty}$, namely $P_{surface}/P^*$ decreases as
            $T_{e\infty}$ increases. This ratio increases with pellet
            radius. In table~\ref{tab:error}, we recorded  errors in \% between
            theory and simulations for varying pellet and plasma parameters.
            We see that for the spherically
            symmetric approximation, the pellet code agrees to less than 1\% with
            the NGS model prediction for almost all cases. We also remark that the
            code consistently underestimates $G$ for all radii $r_p$ and plasma
            temperatures $T_{e_\infty}$ bar $r_p = 1$ mm. The largest error is
            also consistently recorded for $r_p = 2$ mm.
            
            \begin{table*}
              \normalsize \centering
              \begin{subtable}[b]{0.45\linewidth}
                \centering
                \begin{tabular}{|>{\centering\arraybackslash}p{1.5cm}>{\centering\arraybackslash}p{1.5cm}>{\centering\arraybackslash}p{1.5cm}>{\centering\arraybackslash}p{1.5cm}>{\centering\arraybackslash}p{1.7cm}|}
                  \hline
                  $\mathrm{r_p}$ (mm) & G (g/s) & $\mathrm{r_p/r^*}$  & $\mathrm{T^*}$ (eV) & $\mathrm{P_{surface}/P^*}$ \\
                  \hline
                  1      & 25.60    & 0.3343 & 3.920   & 6.531   \\
                  \hline
                  2      & 64.53   & 0.3352 & 6.228   & 6.560   \\
                  \hline
                  5      & 220.9   & 0.3322 & 11.57   & 6.577   \\
                  \hline
                  7      & 347.3   & 0.3305 & 14.74   & 6.618 \\
                  \hline
                \end{tabular}
                \caption{FronTier code}
                \label{tab:FT_Te2}
              \end{subtable}
              \hspace*{2em}
              \begin{subtable}[b]{0.45\linewidth}
                \centering
                \begin{tabular}{|>{\centering\arraybackslash}p{1.5cm}>{\centering\arraybackslash}p{1.5cm}>{\centering\arraybackslash}p{1.5cm}>{\centering\arraybackslash}p{1.5cm}>{\centering\arraybackslash}p{1.7cm}|}
                  \hline
                  $\mathrm{r_p}$ (mm) & G (g/s) & $\mathrm{r_p/r^*}$  & $\mathrm{T^*}$ (eV) & $\mathrm{P_{surface}/P^*}$ \\
                  \hline
                  1      & 25.52    & 0.3366 & 3.897   & 6.534   \\
                  \hline
                  2      & 64.93   & 0.3343 & 6.214   & 6.553  \\
                  \hline
                  5      & 222.2   & 0.3322 & 11.493   & 6.568   \\
                  \hline
                  7      & 348.8   & 0.3317 & 14.40   & 6.571 \\
                  \hline
                \end{tabular}
                \caption{Parks improved NGS model}
                \label{tab:Parks_Te2}
              \end{subtable}
              \caption{Benchmarking the ideal ablated flow characteristics  for
                $n_{e\infty}=10^{14}$ 1/cc, $T_{e\infty} = 2 $keV and varying pellet radius
                $\mathrm{r_p}$ in the spherically symmetric approximation.}
              \label{tab:Te2}
            \end{table*}

            \begin{table*}
              \normalsize \centering
              \begin{subtable}[b]{0.45\linewidth}
                \centering
                \begin{tabular}{|>{\centering\arraybackslash}p{1.5cm}>{\centering\arraybackslash}p{1.5cm}>{\centering\arraybackslash}p{1.5cm}>{\centering\arraybackslash}p{1.5cm}>{\centering\arraybackslash}p{1.7cm}|}
                  \hline
                  $\mathrm{r_p}$ (mm) & G (g/s) & $\mathrm{r_p/r^*}$  & $\mathrm{T^*}$ (eV) & $\mathrm{P_{surface}/P^*}$ \\
                  \hline
                  1      & 117.0    & 0.3350 & 3.374   & 6.521   \\
                  \hline
                  2      & 295.4   & 0.3357 & 5.352   & 6.546   \\
                  \hline
                  5      & 1013   & 0.3327 & 10.00   & 6.574   \\
                  \hline
                  7      & 1591   & 0.3313 & 12.63   & 6.578 \\
                  \hline
                \end{tabular}
                \caption{FronTier code}
                \label{tab:FT_Te5}
              \end{subtable}%
              \hspace*{2em}
              \begin{subtable}[b]{0.45\linewidth}
                \centering
                \begin{tabular}{|>{\centering\arraybackslash}p{1.5cm}>{\centering\arraybackslash}p{1.5cm}>{\centering\arraybackslash}p{1.5cm}>{\centering\arraybackslash}p{1.5cm}>{\centering\arraybackslash}p{1.7cm}|}
                  \hline
                  $\mathrm{r_p}$ (mm) & G (g/s) & $\mathrm{r_p/r^*}$  & $\mathrm{T^*}$ (eV) & $\mathrm{P_{surface}/P^*}$ \\
                  \hline
                  1      & 116.8    & 0.3375 & 3.345   & 6.526   \\
                  \hline
                  2      & 297.5   & 0.3350 & 5.335   & 6.5474  \\
                  \hline
                  5      & 1019   & 0.3326 & 9.872   & 6.565   \\
                  \hline
                  7      & 1600   & 0.3320 & 12.37   & 6.569 \\
                  \hline
                \end{tabular}
                \caption{Parks improved NGS model}
                \label{tab:Parks_Te5}
              \end{subtable}
              \caption{Benchmarking the ideal ablated flow characteristics  for
                $n_{e\infty}=10^{14}$ 1/cc, $T_{e\infty} = 5$ keV and varying pellet radius
                $\mathrm{r_p}$ in the spherically symmetric approximation.}
              \label{tab:Te5}
            \end{table*}

            \begin{table*}
              \normalsize \centering
              \begin{subtable}[b]{0.45\linewidth}
                \centering
                \begin{tabular}{|>{\centering\arraybackslash}p{1.5cm}>{\centering\arraybackslash}p{1.5cm}>{\centering\arraybackslash}p{1.5cm}>{\centering\arraybackslash}p{1.5cm}>{\centering\arraybackslash}p{1.7cm}|}
                  \hline
                  $\mathrm{r_p}$ (mm) & G (g/s) & $\mathrm{r_p/r^*}$  & $\mathrm{T^*}$ (eV) & $\mathrm{P_{surface}/P^*}$ \\
                  \hline
                  1      & 256.7    & 0.3378 & 3.066   & 6.510   \\
                  \hline
                  2      & 649.4   & 0.3372 & 4.885   & 6.550   \\
                  \hline
                  5      & 2232   & 0.3336 & 9.135   & 6.573   \\
                  \hline
                  7      & 3512   & 0.3318 & 11.56   & 6.582 \\
                  \hline
                \end{tabular}
                \caption{FronTier code}
                \label{tab:FT_Te8}
              \end{subtable}%
              \hspace*{2em}
              \begin{subtable}[b]{0.45\linewidth}
                \centering
                \begin{tabular}{|>{\centering\arraybackslash}p{1.5cm}>{\centering\arraybackslash}p{1.5cm}>{\centering\arraybackslash}p{1.5cm}>{\centering\arraybackslash}p{1.5cm}>{\centering\arraybackslash}p{1.7cm}|}
                  \hline
                  $\mathrm{r_p}$ (mm) & G (g/s) & $\mathrm{r_p/r^*}$  & $\mathrm{T^*}$ (eV) & $\mathrm{P_{surface}/P^*}$ \\
                  \hline
                  1      & 257.7    & 0.3381 & 3.055   & 6.520   \\
                  \hline
                  2      & 656.8   & 0.3354 & 4.875   & 6.544  \\
                  \hline
                  5      & 2251   & 0.3329 & 9.024   & 6.563   \\
                  \hline
                  7      & 3535   & 0.3323 & 11.31   & 6.567 \\
                  \hline
                \end{tabular}
                \caption{Parks improved NGS model}
                \label{tab:Parks_Te8}
              \end{subtable}
              \caption{Benchmarking the ideal ablated flow characteristics  for
                $n_{e\infty}=10^{14}$ 1/cc, $T_{e\infty} = 8$ keV and varying pellet radius
                $\mathrm{r_p}$ in the spherically symmetric approximation.}
              \label{tab:Te8}
            \end{table*}

            \begin{table*}
              \normalsize \centering
              \begin{subtable}{0.3\textwidth}
                \centering
                \begin{tabular}{|cccc|}
                  \hline
                  $\mathrm{r_p}$ (mm) & Theory & FronTier & error (\%)\\
                  \hline
                  1      & 25.52 & 25.60 & +0.296 \\
                  \hline
                  2      &  64.93  & 64.53 & -0.615 \\
                  \hline
                  5      & 222.2   & 220.9 & -0.588 \\
                  \hline
                  7      & 348.8   & 347.3 & -0.434 \\
                  \hline
                \end{tabular}
                \caption{$T_{e\infty} = 2 $keV}
                \label{tab:error_Te2}
              \end{subtable}
              \hspace*{1em}
              \begin{subtable}{0.3\textwidth}
                \centering
                \begin{tabular}{|cccc|}
                  \hline
                  $\mathrm{r_p}$ (mm) & Theory & FronTier & error (\%)\\
                  \hline
                  1      & 116.8 & 117 & +0.158 \\
                  \hline
                  2      & 297.5  & 295.4 & -0.695 \\
                  \hline
                  5      & 1019   & 1013 & -0.589 \\
                  \hline
                  7      & 1600   & 1591 & -0.567 \\
                  \hline
                \end{tabular}
                \caption{$T_{e\infty} = 5$ keV}
                \label{tab:error_Te5}
              \end{subtable}
              \hspace*{1em}
              \begin{subtable}{0.3\textwidth}
                \centering
                \begin{tabular}{|cccc|}
                  \hline
                  $\mathrm{r_p}$ (mm) & Theory & FronTier & error (\%)\\
                  \hline
                  1      & 257.7 & 256.7 & -0.402 \\
                  \hline
                  2      & 656.8  & 649.4 & -1.121 \\
                  \hline
                  5      & 2251   & 2232 & -0.854 \\
                  \hline
                  7      & 3535   & 3512 & -0.665 \\
                  \hline
                \end{tabular}
                \caption{$T_{e\infty} = 8$ keV}
                \label{tab:error_Te8}
              \end{subtable}
              \caption{Error in \% in the ablation rate values between theory and our code for
                $n_{e\infty}=10^{14}$, $T_{e\infty} = 2, 5, 8$ keV and varying pellet radius
                $\mathrm{r_p}$ in the spherically symmetric approximation.}
              \label{tab:error}
            \end{table*}

            The error in $G$ grows as the pellet size decreases and
            the plasma temperature increases. The trend to notice is that the code
            overestimates the ablation rate as the intensity of the electron heat
            flux reaching the pellet increases. The heat flux at the pellet can
            increase due to a combination of less shielding from the cold layer
            around the pellet and/or higher unattenuated heat flux $q_\infty$. In
            particular, most of the shielding is provided by a thin layer around
            the pellet and since smaller pellets eject less material as they
            ablate, this region thickness decreases as the pellet size decreases,
            resulting in lower shielding. If this region is not properly resolved,
            the heat flux will not be properly attenuated before hitting the
            pellet. Improving the radial grid resolution allows the steep density
            gradient region near the pellet to be better resolved, and in turn
            provides the predicted shielding sufficient to attenuate the heat
            flux.

             \begin{figure}[h]
             \centering
             \includegraphics[width=\linewidth]{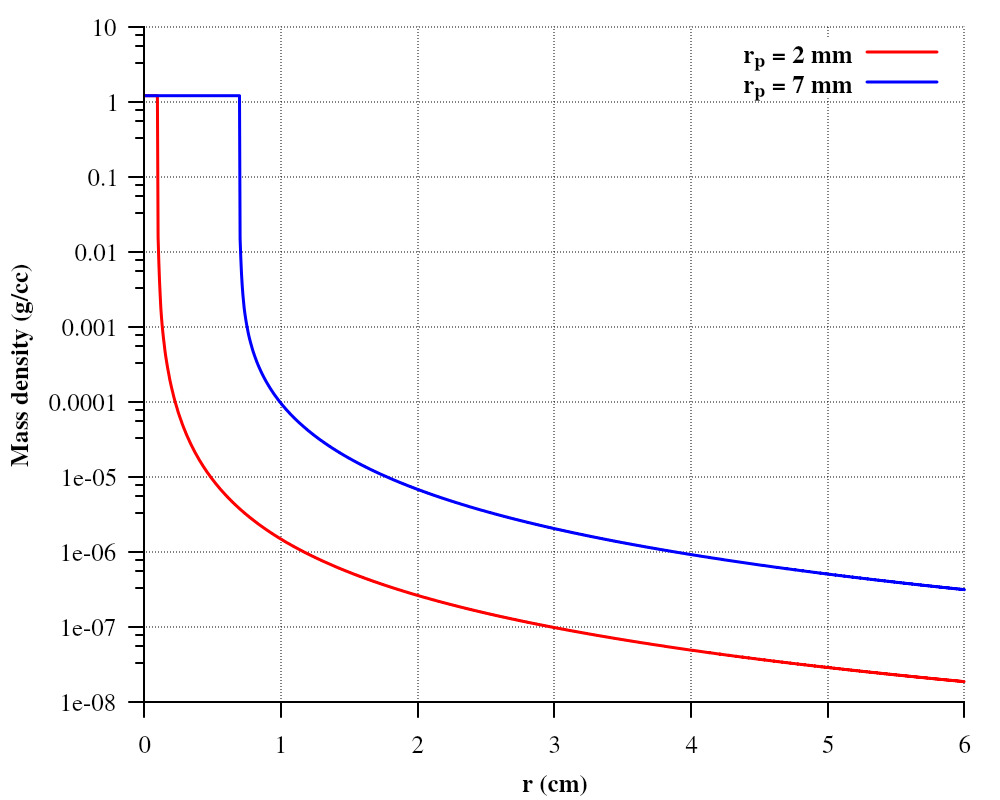}
             \caption[Density in the near pellet layer.]{Mass density profiles for pellet radius $r_p=2$ mm (red) and $r_p=7$ mm and plasma parameters $T_{e_\infty}=2$ keV, $n_{e_\infty} = 10^{14}$ cm$^{-3}$.}
             \label{fig:near_pellet_layer}
             \end{figure}

      \bigskip
       \paragraph{Redlich-Kwong EOS\\}
       \label{par:RK}
     Deviation from the ideal gas model is reflected in the
            equation of state  $P = Z \rho R T$ where $Z$ is a function of
            density and temperature ( $Z=1$ for an ideal
            gas). Typically the vapor state near the pellet has
            compressibility $\mathrm{Z}(\rho,T)<1$ where
            \begin{equation*}
              \begin{split}
                \mathrm{Z}(\rho,T) &= \frac{1}{1-0.08664\mathrm{Z}_c^{-1}\rho / \rho_c} \\
                & - 0.42748\left(\frac{T_c}{T}\right)^{1.5}\frac{1}{\mathrm{Z}_c}\frac{\rho}{\rho_c}\left(\frac{1}{1+0.08664\textrm{Z}_c^{-1}\rho/\rho_c}\right),
              \end{split}
            \end{equation*}
            with $\mathrm{Z}_c = 0.31192$ and $\mathrm{\rho_c} = 0.484$ g/cc.
            Therefore the vapor density exceeds the ideal gas density
            for the same pressure and temperature.

            \begin{table}[H]
              \normalsize \centering
              \begin{tabular}
                {|>{\centering\arraybackslash}p{1.7cm}>{\centering\arraybackslash}p{2.2cm}>{\centering\arraybackslash}p{2.2cm}>{\centering\arraybackslash}p{2.2cm}|}
                \hline
                &$T_{e_\infty} = 2$ keV & $T_{e_\infty} = 5$ keV & $T_{e_\infty} = 8$ keV\\
                \hline
                $r_{p} = 1$ mm & 0.9794    & 0.9991 & 1.057 \\
                \hline
                $r_{p} = 7$ mm & 0.9689 & 0.9768 & 0.9971 \\
                \hline
              \end{tabular}                                  
              \caption[RK compressibility factor.]{Compressibility factor at the pellet surface.}
              \label{tab:Z_RK}
            \end{table}

            In table~\ref{tab:Z_RK}, the compressibility factor at the pellet surface are computed.
            The compressibility factor is usually less than 1. The only exception
            is for $T_{e_\infty} = 8$ keV and $r_p = 1$ mm where the surface
            pressure exceeds 400 bars (417 bars) resulting in $Z > 1$. In
            table~\ref{tab:RK}, the ablation rates obtained using the ideal and real gas
            EOS are recorded and compared. Using a real gas EOS improves the
            shielding and lowers the ablation rate except when $Z>1$. The
            difference in ablation rate is negligible and we conclude that real
            gas effects have no influence on the ablation rate and pellet life
            time and is not used for the reminder of this paper. The flow states in
            the cloud are also negligibly affected by the change in equation of state.q

             \begin{table*}
              \normalsize \centering
              \begin{subtable}[b]{0.48\linewidth}
                \raggedright
                \begin{tabular}{|>{\centering\arraybackslash}p{1.7cm}>{\centering\arraybackslash}p{2cm}>{\centering\arraybackslash}p{2cm}>{\centering\arraybackslash}p{2cm}|}
                  \hline
                  &$T_{e\infty} = 2$ & $T_{e\infty} = 5$ & $T_{e\infty} = 8$\\
                  \hline
                  $\mathrm{r_{p}} = 1$ mm & 25.60    & 117.0 & 256.7\\
                  \hline
                  $\mathrm{r_{p}} = 7$ mm & 347.3 & 1591 & 3512\\
                  \hline
                \end{tabular}
                \caption{Ideal EOS}
                \label{tab:ideal_RK}
              \end{subtable}%
              \begin{subtable}[b]{0.5\linewidth}
                \centering
                \begin{tabular}{|>{\centering\arraybackslash}p{1.7cm}>{\centering\arraybackslash}p{2.35cm}>{\centering\arraybackslash}p{2.35cm}>{\centering\arraybackslash}p{2.35cm}|}
                  \hline
                  &$T_{e\infty} = 2$ & $T_{e\infty} = 5$ & $T_{e\infty} = 8$\\
                  \hline
                  $\mathrm{r_{p}} = 1$ mm & 25.46 \small(-0.55)  & 116.3 \small(-0.6)& 257.8 \small(+0.43)\\
                  \hline
                  $\mathrm{r_{p}} = 7$ mm & 346.2 \small(-0.32)& 1585 \small(-0.38)& 3500 \small(-0.34)\\
                  \hline
                \end{tabular}
                \caption{Non ideal Redlich-Kwong EOS with reduction in \%}                
                \label{tab:RK_RK}
              \end{subtable}
              \caption{Comparison of ablation rates (in g/s) under the ideal and non ideal gas model (with relative change in parenthesis in \%) for varying pellet radii and plasma temperatures $T_{e\infty}$ (in keV). The Redlich-Kwong EOS slightly reduces the ablation rates in most cases.}
              \label{tab:RK}
            \end{table*}

            \begin{figure*}
              \begin{subfigure}[t]{0.495\linewidth}
                \centering \includegraphics[width=8cm,
                height=6cm]{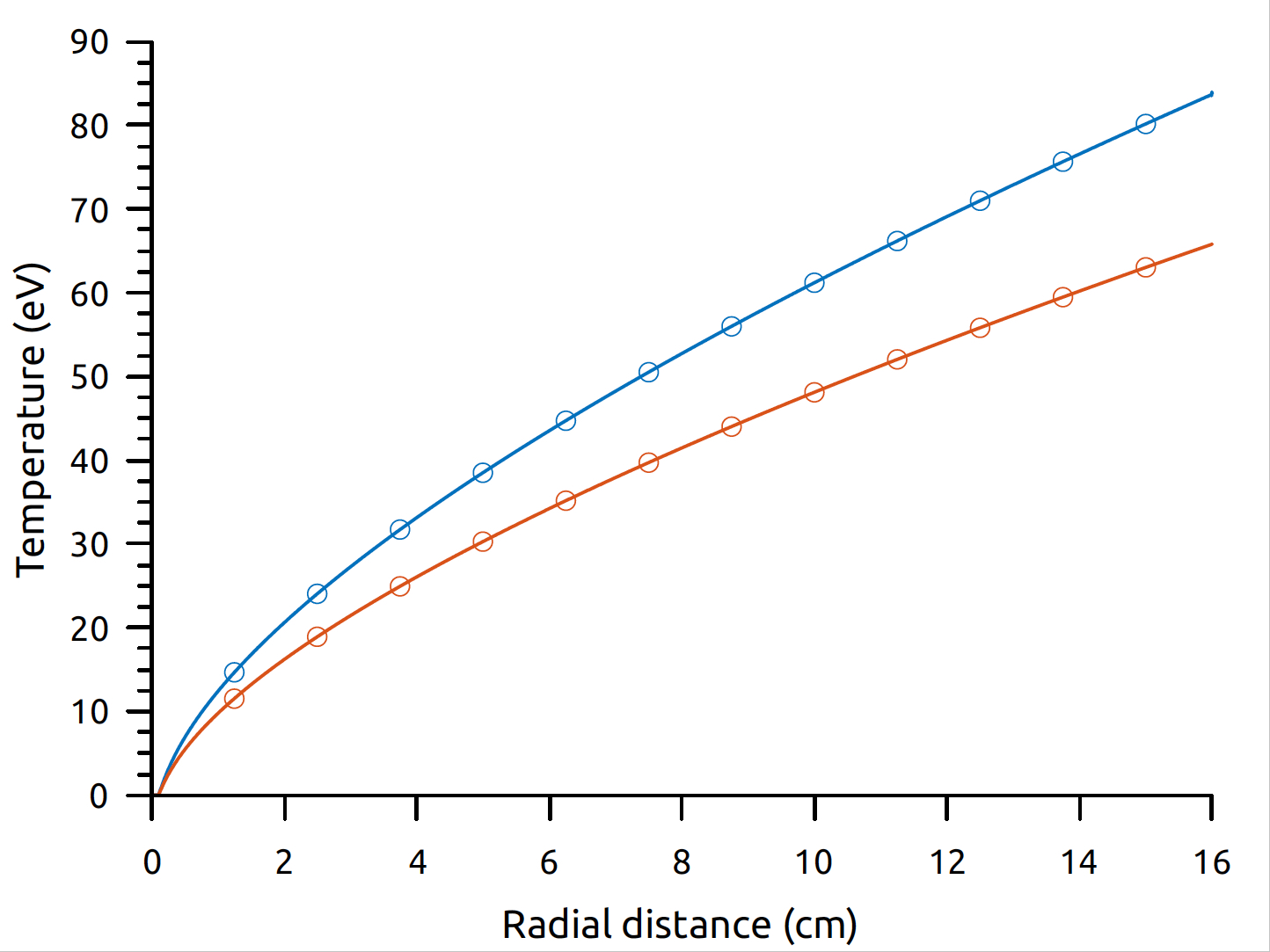}
                \label{fig:RK_ideal_temp}
              \end{subfigure}
              \begin{subfigure}[t]{0.495\linewidth}
                \centering \includegraphics[width=8cm,
                height=6cm]{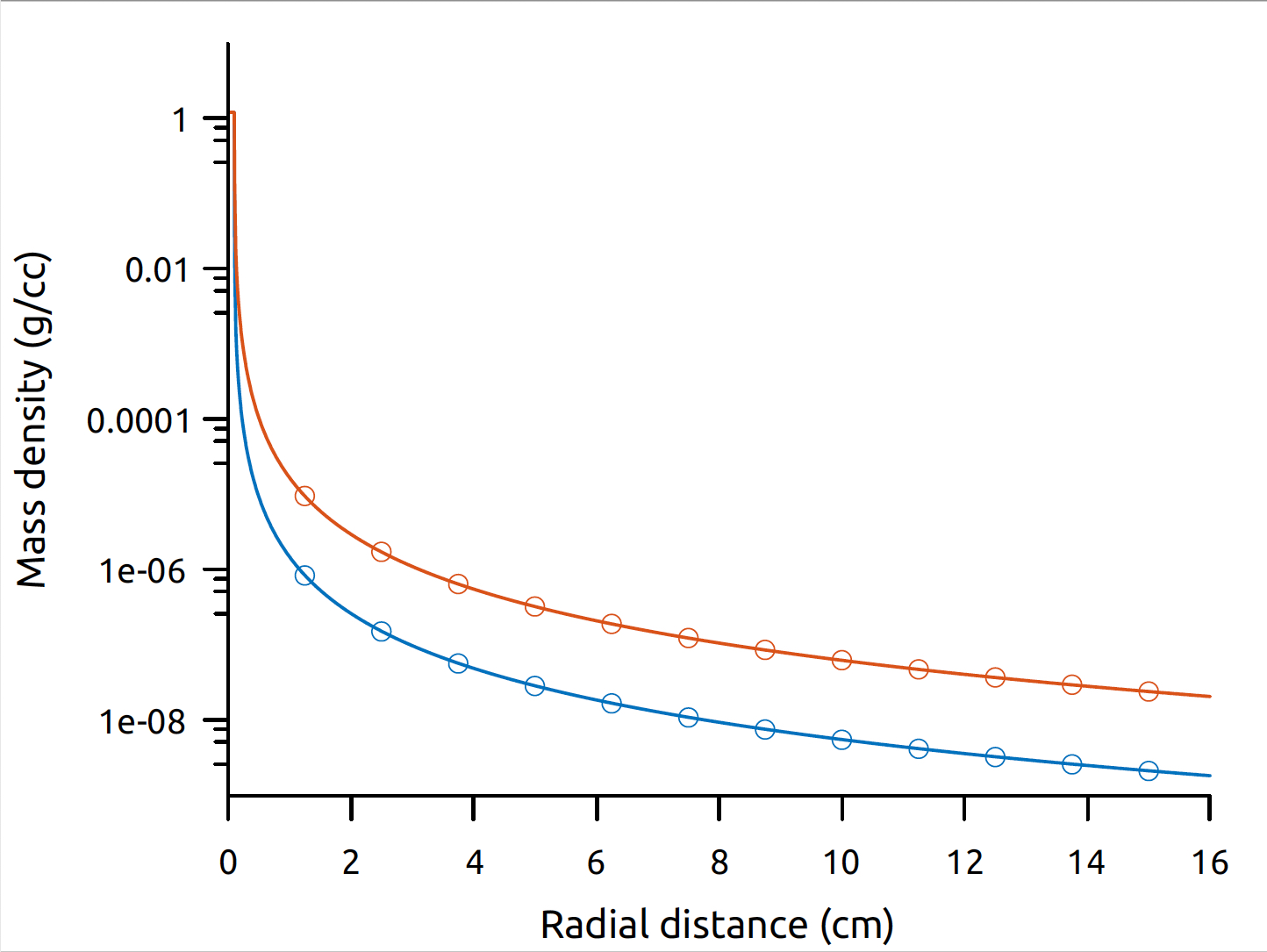}
                \label{fig:RK_ideal_dens}
              \end{subfigure}
              \caption{Temperature and mass density profiles for RK (circles) and ideal EOS (solid line) at
                fixed $\mathrm{r_p}=1$ mm, $n_{e\infty}=10^{14}$ and varying $T_{e\infty} = 2$ keV (blue) and $T_{e\infty} = 5$ keV (orange).}
              \label{fig:normalized_ideal_saha}
            \end{figure*}

       \bigskip

\begin{figure}[H]
              \centering \includegraphics[width=\linewidth]{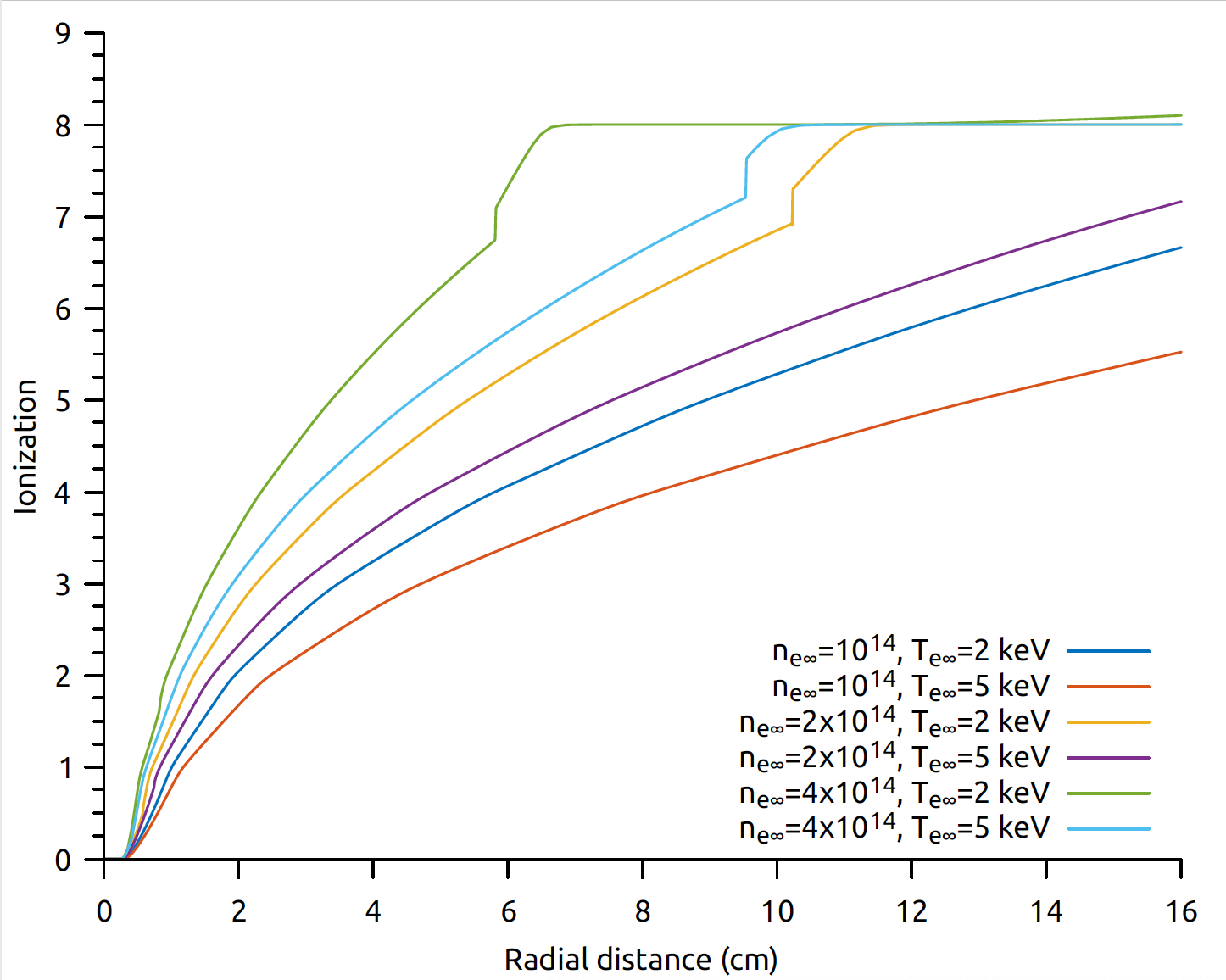}
              \caption[Electron fraction in the spherically symmetric ablation cloud.] {Electron fraction in the cloud for varying plasma particle density $n_{e_\infty}$ (in cm$^{-3}$) and plasma temperature $T_{e_\infty}$ (in keV) and $r_p=2$ mm pellet radius, in the spherically symmetric approximation.}
              \label{fig:ionization_1d}
            \end{figure}

       \paragraph{Saha EOS\\}
       \label{par:saha}
    In this Section, we study the influence of atomic processes such as multiple ionization  of neon and dissociation and ionization of deuterium
     on the pellet ablation.  The degree of ionization is very low in
            proximity to the surface of the pellet (mostly neutral atoms) and it
            becomes progressively higher further downstream as the temperature,
            and thus thermal ionization, steadily increases as is seen
            in figure~\ref{fig:ionization_1d} for neon (Z=10).  We use the term average ionization level interchangeably
            for the electron fraction in the cloud $n_e/n_t$. For high-Z pellets,
            multiple ionization levels introduce energy sinks that influence
            strongly the temperature and in turn the conductivity of the
            cloud.  We report in this subsection
            1D spherically symmetric results when atomic processes are allowed in
            the ablation cloud and compare the ablation rate and flow states with
            the ideal gas case.

            In figure~\ref{fig:ionization_1d} the dependence of the
            ionization level on the plasma parameter is displayed. The ionization
            level grows with plasma density $n_{e\infty}$ but decreases with
            plasma temperature $T_{e\infty}$. From the NGS scaling laws, the mass flow follows the plasma electron
            density trend, scaling with $n_{e_\infty}^{1/3}$, but is more
            sensitive to the temperature, scaling with $T_{e\infty}^{1.64}$. As
            both of these parameters increase, the ablation rate $G$ and cloud
            temperature increase. However, since $G$ is more sensitive to an
            increase in $T_{e\infty}$, for a fixed value of $n_{e\infty}$, the
            increase in temperature in the cloud is not sufficient to overcome the
            increase in $G$ and the cloud becomes denser and more difficult to
            ionize. The reverse is true for a fixed value $T_{e\infty}$ and
            increasing $n_{e\infty}$.

            \begin{table*}
              \normalsize \centering
              \begin{subtable}[b]{0.48\linewidth}
                \raggedright
                \begin{tabular}{|>{\centering\arraybackslash}p{1.8cm}>{\centering\arraybackslash}p{2cm}>{\centering\arraybackslash}p{2cm}>{\centering\arraybackslash}p{2cm}|}
                  \hline
                  &$T_{e\infty} = 2$ & $T_{e\infty} = 5$ & $T_{e\infty} = 8$\\
                  \hline
                  $\mathrm{r_{p}} = 1$ mm & 25.60    & 117.0 & 256.7\\
                  \hline
                  $\mathrm{r_{p}} = 2$ mm & 64.53    & 295.4 & 649.4\\
                  \hline
                  $\mathrm{r_{p}} = 7$ mm & 347.3 & 1591 & 3512\\
                  \hline
                \end{tabular}
                \caption{Ideal EOS ($n_{e\infty}=10^{14}$) \vspace{5mm}}
                \label{tab:ideal_Saha_ne1}
              \end{subtable}%
              \begin{subtable}[b]{0.52\linewidth}
                \raggedleft
                \begin{tabular}{|>{\centering\arraybackslash}p{1.8cm}>{\centering\arraybackslash}p{2.35cm}>{\centering\arraybackslash}p{2.35cm}>{\centering\arraybackslash}p{2.35cm}|}
                  \hline
                  &$T_{e\infty} = 2$ & $T_{e\infty} = 5$ & $T_{e\infty} = 8$\\
                  \hline
                  $\mathrm{r_{p}} = 1$ mm & 25.83 \small(+0.9)  & 118.3 \small(+1.1)& 271.5 \small(+5.8)\\
                   \hline
                  $\mathrm{r_{p}} = 2$ mm & 65.58 \small(+1.6) & 301 \small(+1.9) & 680.9 \small(+4.8)\\
                  \hline
                  $\mathrm{r_{p}} = 7$ mm & 338.3 \small(-2.7)& 1606 \small(+0.94)& 3603 \small(+2.6)\\
                  \hline
                \end{tabular}
                \caption{Saha EOS, difference compared to ideal case in \% ( $n_{e\infty}=10^{14}$) \vspace{5mm}}                
                \label{tab:Saha_Saha_ne1}
              \end{subtable}
              \begin{subtable}[b]{0.48\linewidth}
                \raggedright
                \begin{tabular}{|>{\centering\arraybackslash}p{1.8cm}>{\centering\arraybackslash}p{2cm}>{\centering\arraybackslash}p{2cm}>{\centering\arraybackslash}p{2cm}|}
                  \hline
                  &$T_{e\infty} = 2$ & $T_{e\infty} = 5$ & $T_{e\infty} = 8$\\
                  \hline
                  $\mathrm{r_{p}} = 1$ mm & 32.57    & 148.9 & 328.5\\
                  \hline
                  $\mathrm{r_{p}} = 2$ mm & 81.71    & 374.3 & 826.3\\
                  \hline
                  $\mathrm{r_{p}} = 7$ mm & 438.4 & 2011 & 4445\\
                  \hline
                \end{tabular}
                \caption{Ideal EOS ($n_{e\infty}=2\times10^{14}$)\vspace{5mm}}
                \label{tab:ideal_Saha_ne2}
              \end{subtable}%
              \begin{subtable}[b]{0.52\linewidth}
                \raggedleft
                \begin{tabular}{|>{\centering\arraybackslash}p{1.8cm}>{\centering\arraybackslash}p{2.35cm}>{\centering\arraybackslash}p{2.35cm}>{\centering\arraybackslash}p{2.35cm}|}
                  \hline
                  &$T_{e\infty} = 2$ & $T_{e\infty} = 5$ & $T_{e\infty} = 8$\\
                  \hline
                  $\mathrm{r_{p}} = 1$ mm & 33.19 \small(+1.9)  & 155.4 \small(+4.4)& 350.1 \small(+6.7)\\
                   \hline
                  $\mathrm{r_{p}} = 2$ mm & 82.53 \small(+1) & 387.1 \small(+3.5) & 859.2 \small(+3.8)\\
                  \hline
                  $\mathrm{r_{p}} = 7$ mm & 398 \small(-9.2)& 1932 \small(-3.9)& 4279 \small(-3.7)\\
                  \hline
                \end{tabular}
                \caption{Saha EOS, difference compared to ideal case in \% ($n_{e\infty}=2\times10^{14}$) \vspace{5mm}}                
                \label{tab:Saha_Saha_ne2}
              \end{subtable}
              \begin{subtable}[b]{0.48\linewidth}
                \raggedright
                \begin{tabular}{|>{\centering\arraybackslash}p{1.8cm}>{\centering\arraybackslash}p{2cm}>{\centering\arraybackslash}p{2cm}>{\centering\arraybackslash}p{2cm}|}
                  \hline
                  &$T_{e\infty} = 2$ & $T_{e\infty} = 5$ & $T_{e\infty} = 8$\\
                  \hline
                  $\mathrm{r_{p}} = 1$ mm & 41.52  & 190.2 & 419.6\\
                  \hline
                  $\mathrm{r_{p}} = 2$ mm & 103.7 & 475.2 & 1050\\
                  \hline
                  $\mathrm{r_{p}} = 7$ mm &553.2 & 2536 & 5610\\
                  \hline
                \end{tabular}
                \caption{Ideal EOS ($n_{e\infty}=4\times10^{14}$)\vspace{0mm}}
                \label{tab:ideal_Saha_ne4}
              \end{subtable}%
              \begin{subtable}[b]{0.52\linewidth}
                \raggedleft
                \begin{tabular}{|>{\centering\arraybackslash}p{1.8cm}>{\centering\arraybackslash}p{2.35cm}>{\centering\arraybackslash}p{2.35cm}>{\centering\arraybackslash}p{2.35cm}|}
                  \hline
                  &$T_{e\infty} = 2$ & $T_{e\infty} = 5$ & $T_{e\infty} = 8$\\
                  \hline
                  $\mathrm{r_{p}} = 1$ mm & 42.52 \small(+2.4)  & 202.2 \small(+6.3)& 455.8 \small(+8.6)\\
                   \hline
                  $\mathrm{r_{p}} = 2$ mm & 101.9 \small(+1.7) & 490.0 \small(+3.1) & 1113 \small(+6.0)\\
                  \hline
                  $\mathrm{r_{p}} = 7$ mm & 461.7 \small(-16.5)& 2277 \small(-10.2)& 5267 \small(-6.1)\\
                  \hline
                \end{tabular}
                \caption{Saha EOS, difference compared to ideal case in \% ($n_{e\infty}=4\times10^{14}$) \vspace{0mm}}                
                \label{tab:Saha_Saha_ne4}
              \end{subtable}
              \caption{Comparison of ablation rates (in g/s) under the ideal and ionized gas model for background plasma densities $n_{e\infty}=10^{14}$ and $n_{e\infty}=2\times10^{14}$ 1/cc. The ionized gas model based on the solution of the Saha system reduces the ablation rate only for bigger pellets at more extreme regimes ($T_{e\infty} > 2$keV, $n_{e\infty} > 10^{14}$ 1/cc and $\mathrm{r_p} > 2$ mm) whereas it slightly increases the ablation rate at lower temperature and plasma densities.}
              \label{tab:Saha_1d}
            \end{table*}

            The trend emerging
            from tables~\ref{tab:Saha_Saha_ne1},~\ref{tab:Saha_Saha_ne2},~\ref{tab:Saha_Saha_ne4} is
            that for small pellets ($r_p<7$ mm), the relative change in the ablation
            rate due to atomic processes actually increases as the heat flux
            intensity grows (due to an increase in $n_{_\infty}$ and/or
            $T_{e\infty}$). For bigger pellets, the trend is reversed and we
            observe sizable reduction in the ablation rate at high plasma
            densities. For a fixed plasma electron density, the relative increase
            in ablation rate for small pellet is greater at high plasma
            temperatures, conversely for bigger pellets the reduction in $G$ is
            more important at low plasma temperatures. This effect grows markedly
            with the plasma density $n_{e\infty}$.

\begin{figure*}
              \begin{subfigure}{\linewidth}
                \centering \includegraphics[width=15cm,height=11cm]{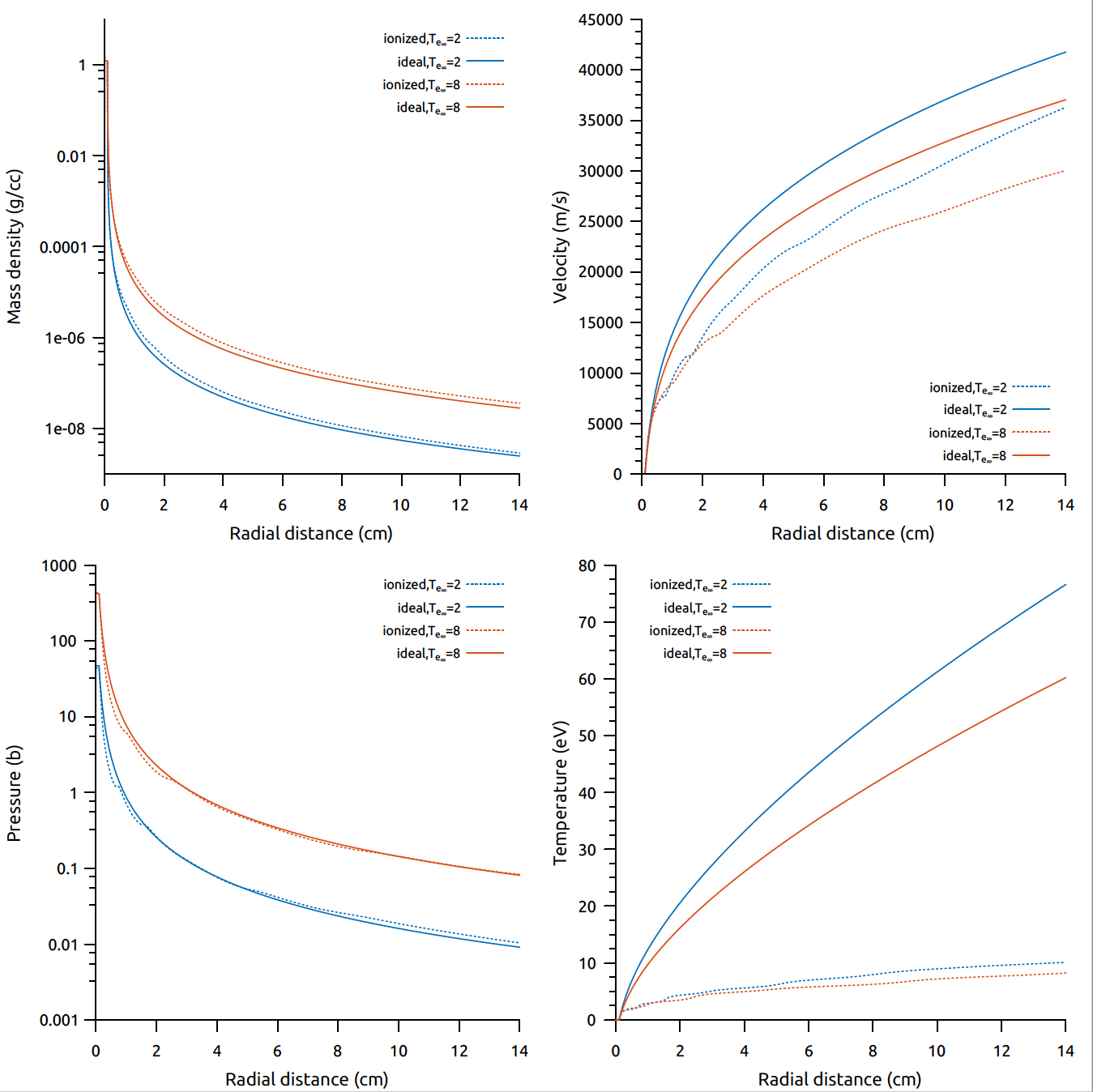}
                \caption{$n_{e\infty}=10^{14}$ 1/cc}
              \label{fig:ideal_saha_ne1_rp1}
            \end{subfigure}
            \vspace{3mm}
              \begin{subfigure}{\linewidth}
                \centering \includegraphics[width=15cm,height=11cm]{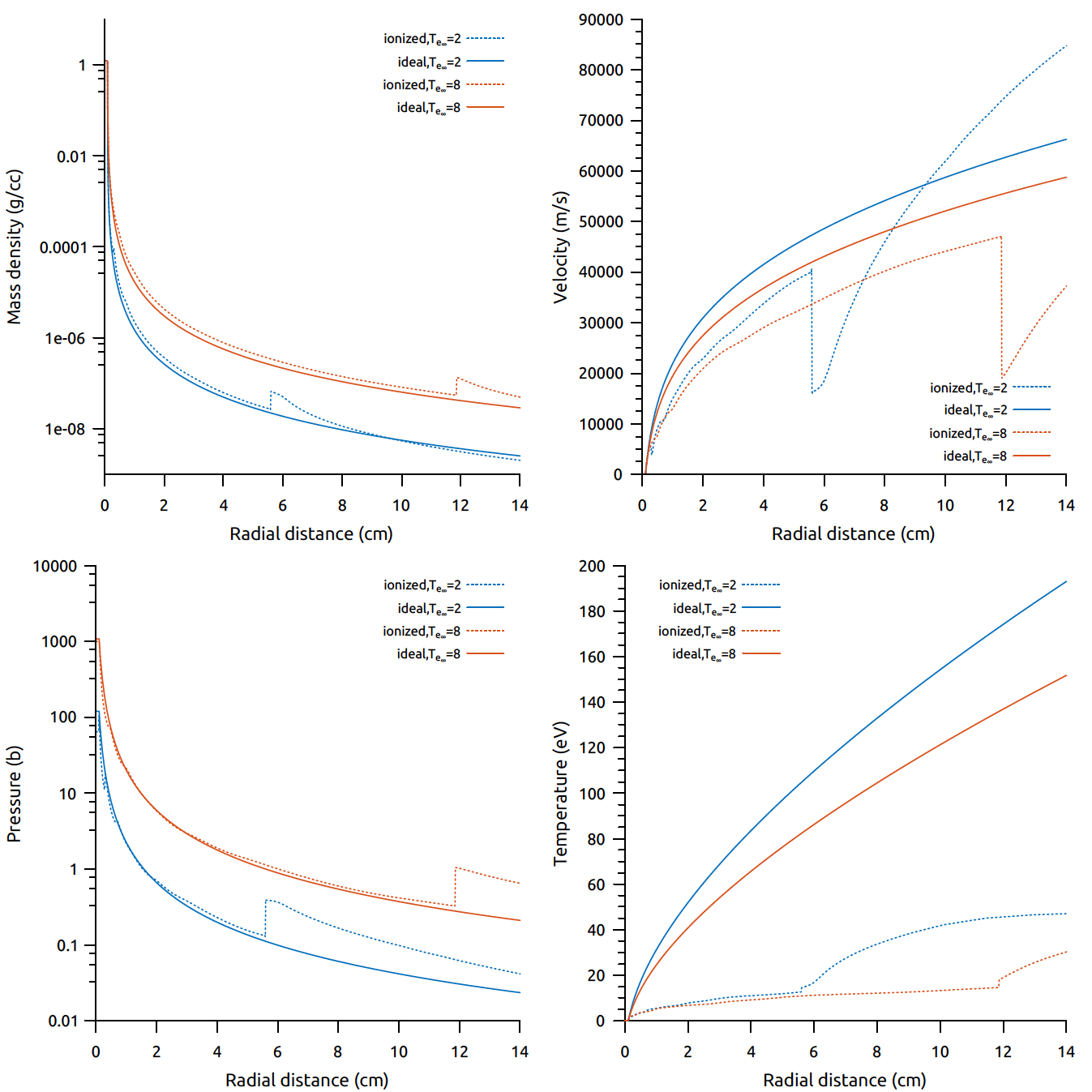}
                \caption{$n_{e\infty}=4\times10^{14}$ 1/cc}
                \label{fig:ideal_saha_ne4_rp1}
            \end{subfigure}
            \caption{Flow field states for ideal (solid line) and ionized gas
              (dashed line) for $\mathrm{r_p}= 1$ mm, varying  $T_{e\infty}=2$ keV (blue),
              $T_{e\infty}=8$ keV (orange) and $n_{e\infty}=10^{14}$ (\ref{fig:ideal_saha_ne1_rp1}), $n_{e\infty}=4\times10^{14}$ (\ref{fig:ideal_saha_ne4_rp1}).}
          \end{figure*}

\begin{figure*}
            \begin{subfigure}{\linewidth}
              \centering \includegraphics[width=15cm,height=11cm]{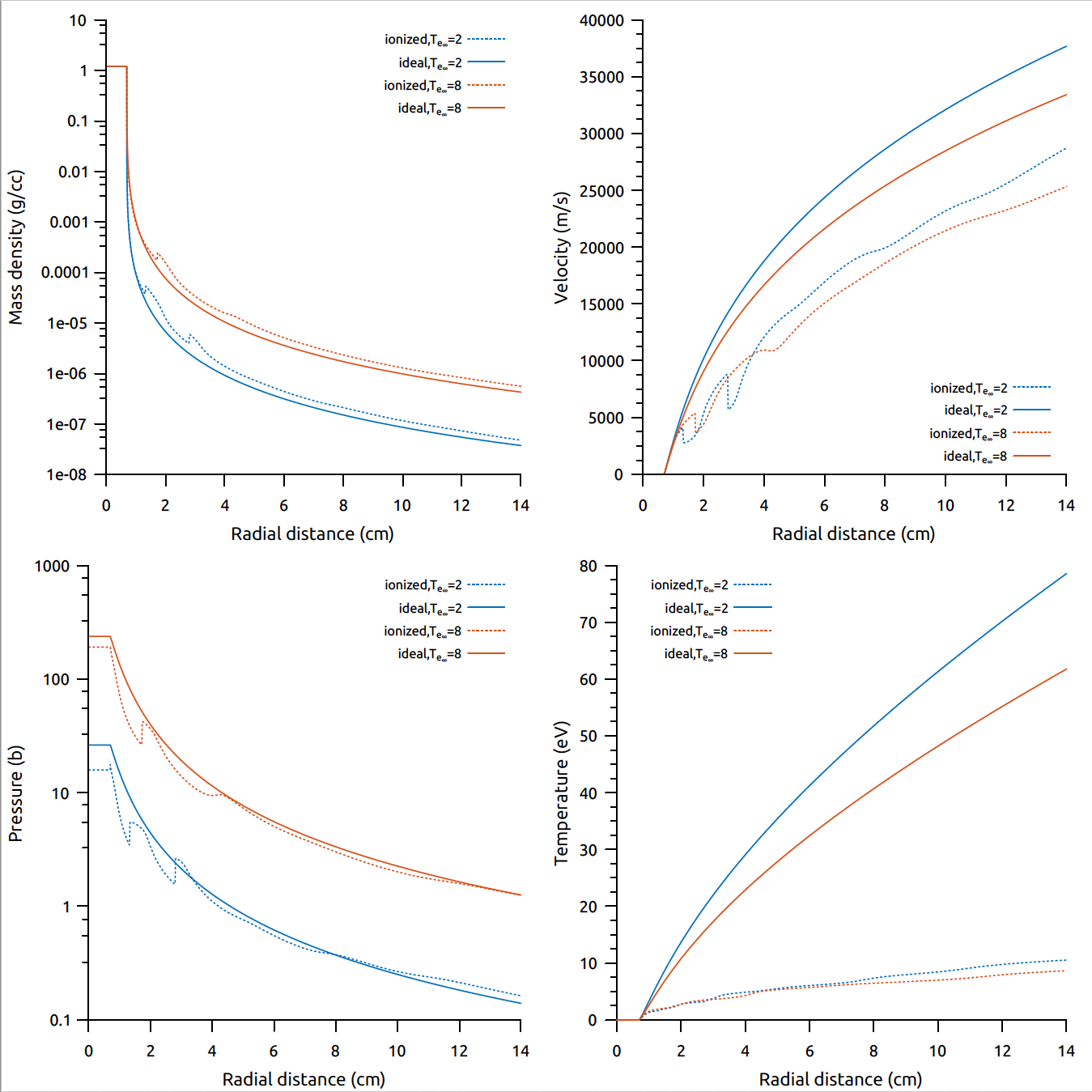}
              \caption{$n_{e\infty}=10^{14}$ 1/cc}
              \label{fig:ideal_saha_ne1_rp7}
            \end{subfigure}
            \vspace{3mm}
            \begin{subfigure}{\linewidth}
              \centering \includegraphics[width=15cm,height=11cm]{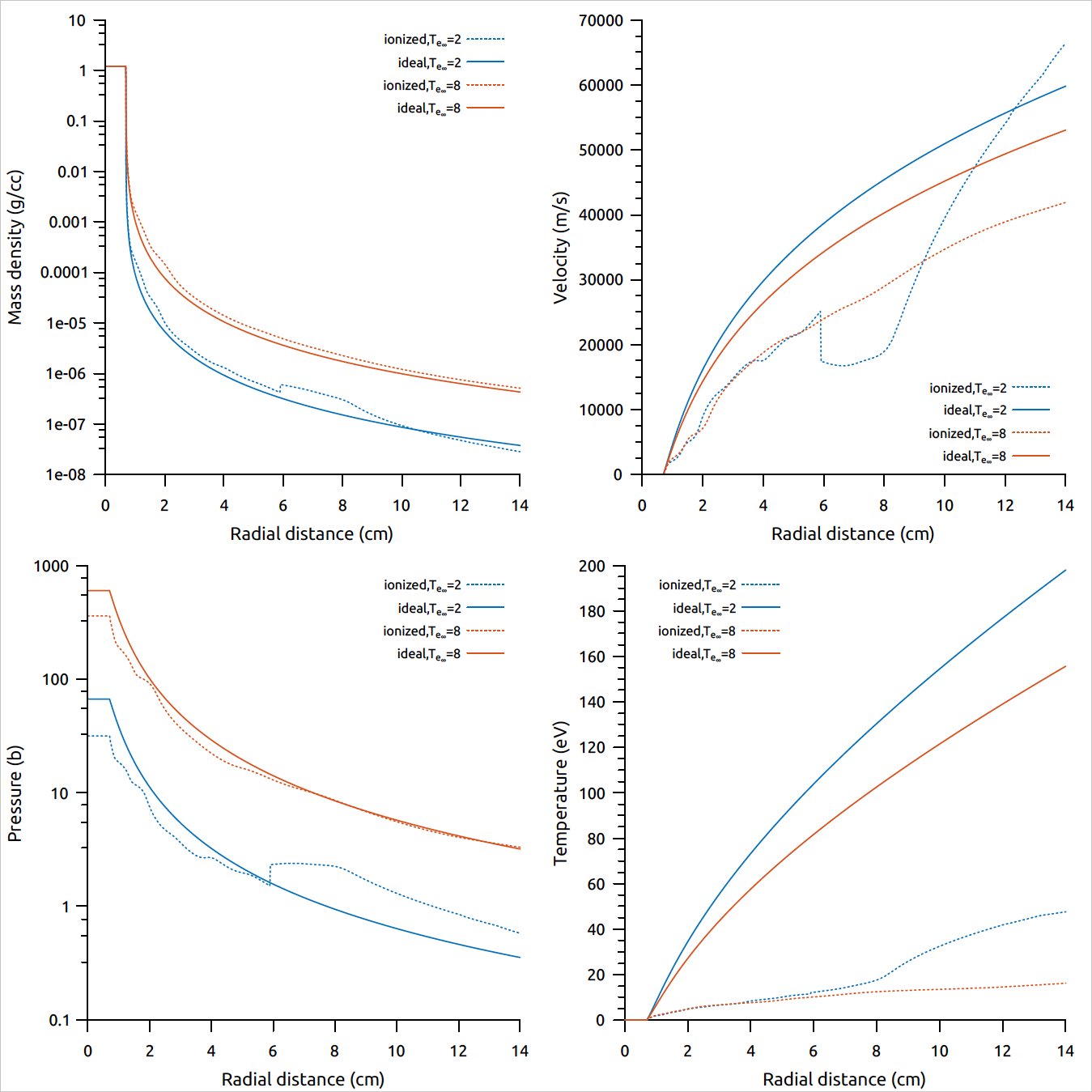}
              \caption{$n_{e\infty}=4\times10^{14}$ 1/cc}
              \label{fig:ideal_saha_ne4_rp7}
            \end{subfigure}
            \caption{Flow field states for ideal (solid line) and ionized gas
              (dashed line) for $\mathrm{r_p}= 7$ mm, varying  $T_{e\infty}=2$ keV (blue),
              $T_{e\infty}=8$ keV (orange) and $n_{e\infty}=10^{14}$ (\ref{fig:ideal_saha_ne1_rp7}), $n_{e\infty}=4\times10^{14}$ (\ref{fig:ideal_saha_ne4_rp7}).}
          \end{figure*}

            The radial flow states profiles (density, pressure, temperature, velocity) are plotted in figures~\ref{fig:ideal_saha_ne1_rp1},~\ref{fig:ideal_saha_ne4_rp1},~\ref{fig:ideal_saha_ne1_rp7},~\ref{fig:ideal_saha_ne4_rp7}
            for varying pellet radii $r_p$, plasma electron density
            $n_{e_\infty}$ and temperature $T_{e_\infty}$. For each flow variable,
            the plotted profiles were obtained from the ideal EOS (no atomic processes, solid line) and LTE EOS with atomic processes (dashed line). 

          For a pellet of radius $r_p = 1$ mm, when
          $n_{e\infty}=10^{14}$ cm$^{-3}$ (figure~\ref{fig:ideal_saha_ne1_rp1}), the
          cloud states varies smoothly along the cloud for both EOS and plasma
          temperatures. The density is consistently higher in the ionized
          cloud. However, the energy sinks introduced by the ionization
          potentials lower the thermal and kinetic energy, the temperature is
          decreased almost eight times in the ionized cloud and the velocity
          reduction is close to 1.5. The differences in these profiles increase
          slightly with $T_{e\infty}$.
          
          In figure~\ref{fig:ideal_saha_ne4_rp1}, the plasma density is
          increased to $n_{e\infty}=4\times10^{14}$ cm$^{-3}$. The degree of
          ionization is higher in this case throughout the cloud compared to
          $n_{e\infty}=10^{14}$ cm$^{-3}$ as can be seen
          in figure~\ref{fig:ionization_1d}. As a result, the energy sinks become
          deeper and lead to the formation of a strong shock in the flow (at
          $r=5.7$ cm for $T_{e\infty}=2$ keV and $r=12$ cm for $T_{e\infty}=8$
          keV). The shock is pushed downstream as $T_{e\infty}$ increases since
          the corresponding ionization level is also reached further downstream. In the region $r<r_{shock}$,
          the flow exhibits the same behavior as
          in figure~\ref{fig:ideal_saha_ne4_rp1}. For a small radius pellet, the
          reduction in velocity is not sufficient to overcome the increase in
          density and to lower the ablation rate.
          
            \begin{figure}
              \begin{subfigure}[t]{1\linewidth}
                \centering
                \includegraphics[width=8cm,height=5cm]{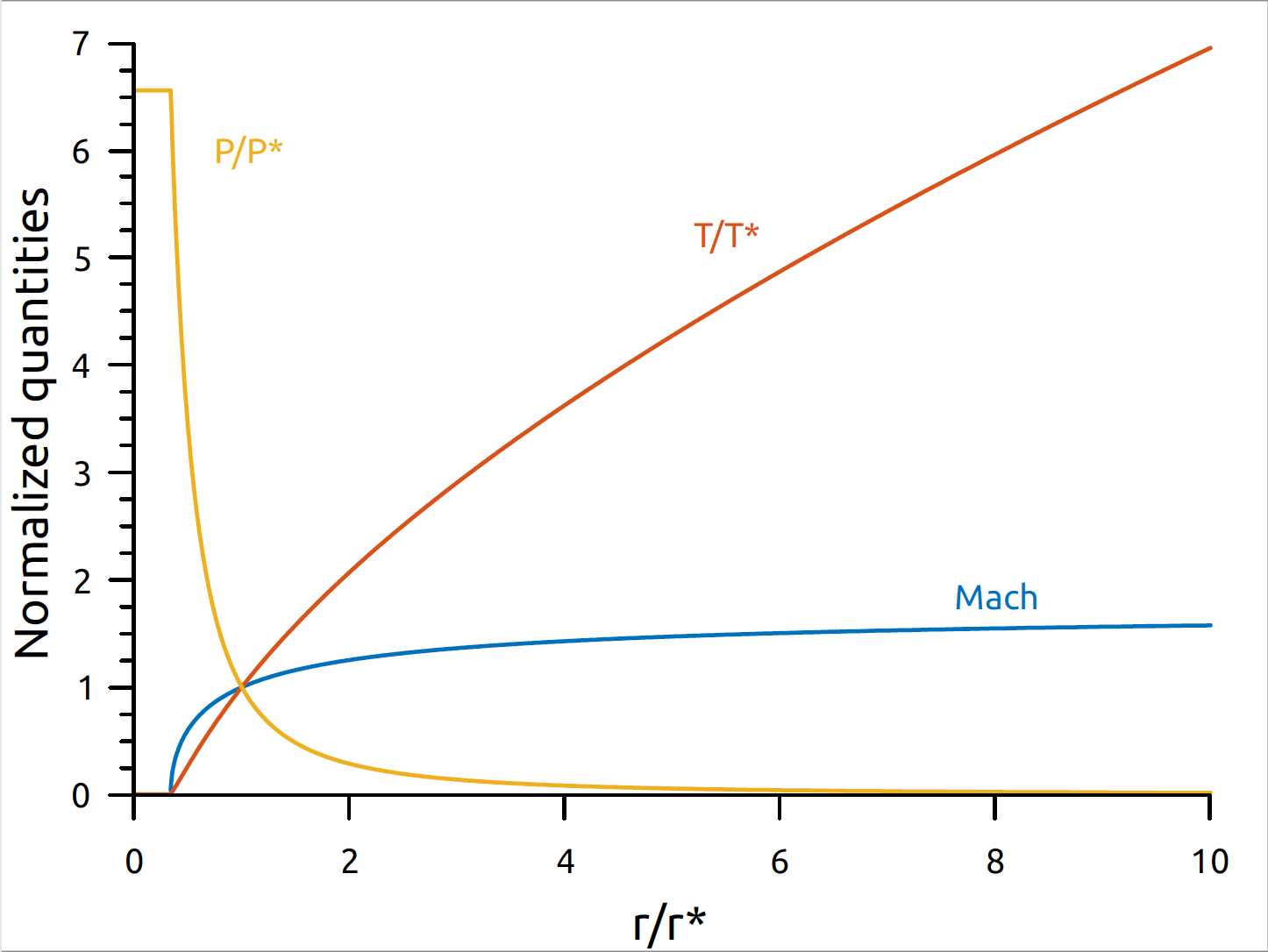}
                \caption{Ideal neon gas: $r^*=5.9565$ mm, $T^*=6.228$ eV, $P^*=6.09$
                  bars}
                \label{fig:normalized_ideal_rp02_Te2_ne1}
              \end{subfigure}
              \begin{subfigure}[t]{1\linewidth}
                \centering \includegraphics[width=8cm,height=6cm]{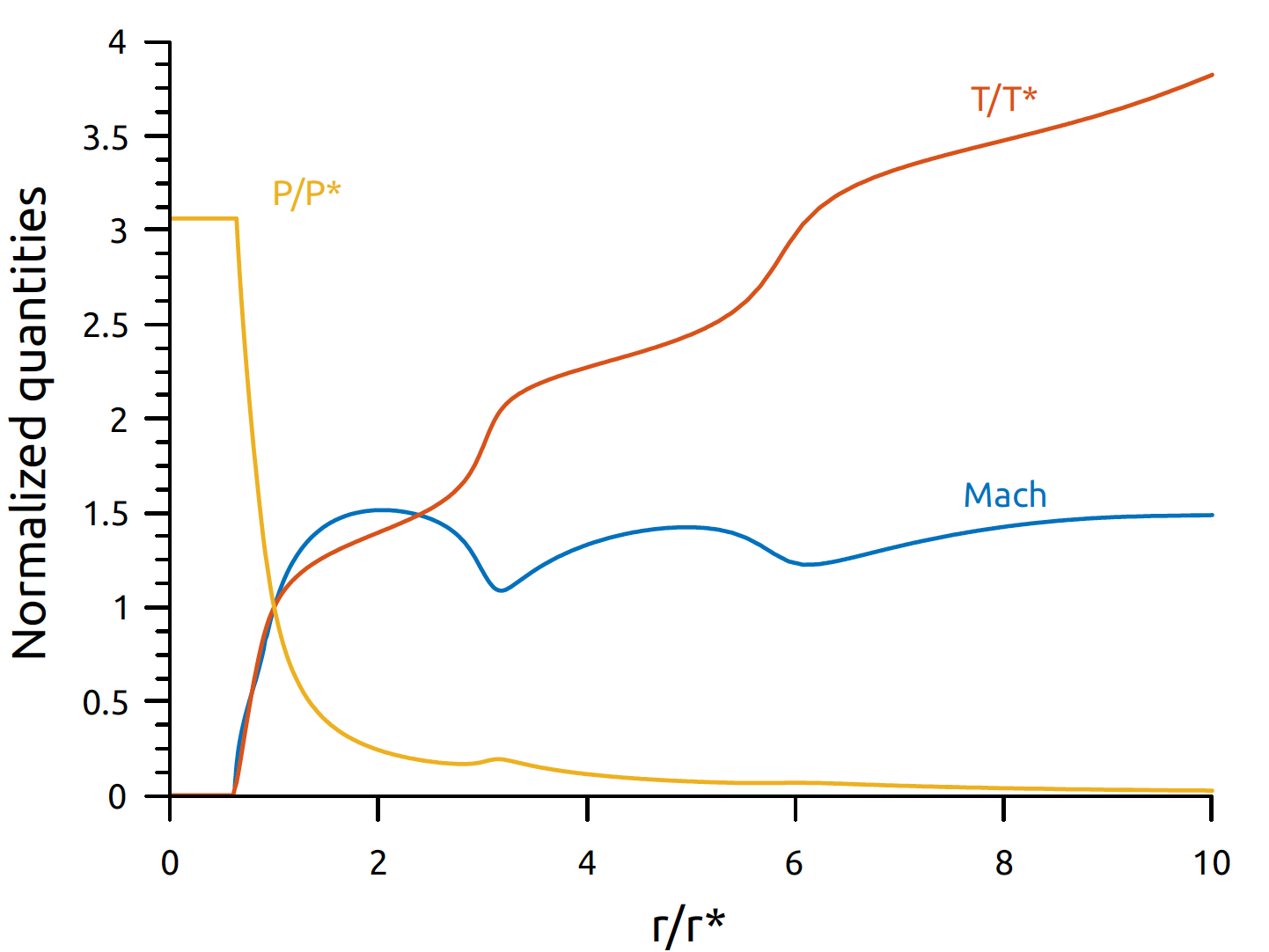}
                \caption{Ionized neon gas: $r^*=3.254$ mm, $T^*=1.349$ eV,
                  $P^*=10.93$ bars}
                \label{fig:normalized_saha_rp02_Te2_ne1}
              \end{subfigure}              
              \caption{Normalized profiles for ideal (\ref{fig:normalized_ideal_rp02_Te2_ne1}) and ionized (\ref{fig:normalized_saha_rp02_Te2_ne1}) neon gas with: $\mathrm{r_p}=2$ mm, $n_{e\infty}=10^{14}$, $n_{eff}=1.205\times10^{13}$,
                $T_{e\infty}=2$ keV.}              
              \label{fig:normalized_ideal_saha}
            \end{figure}

            \begin{figure}
              \centering \includegraphics[width=8cm,height=5cm]{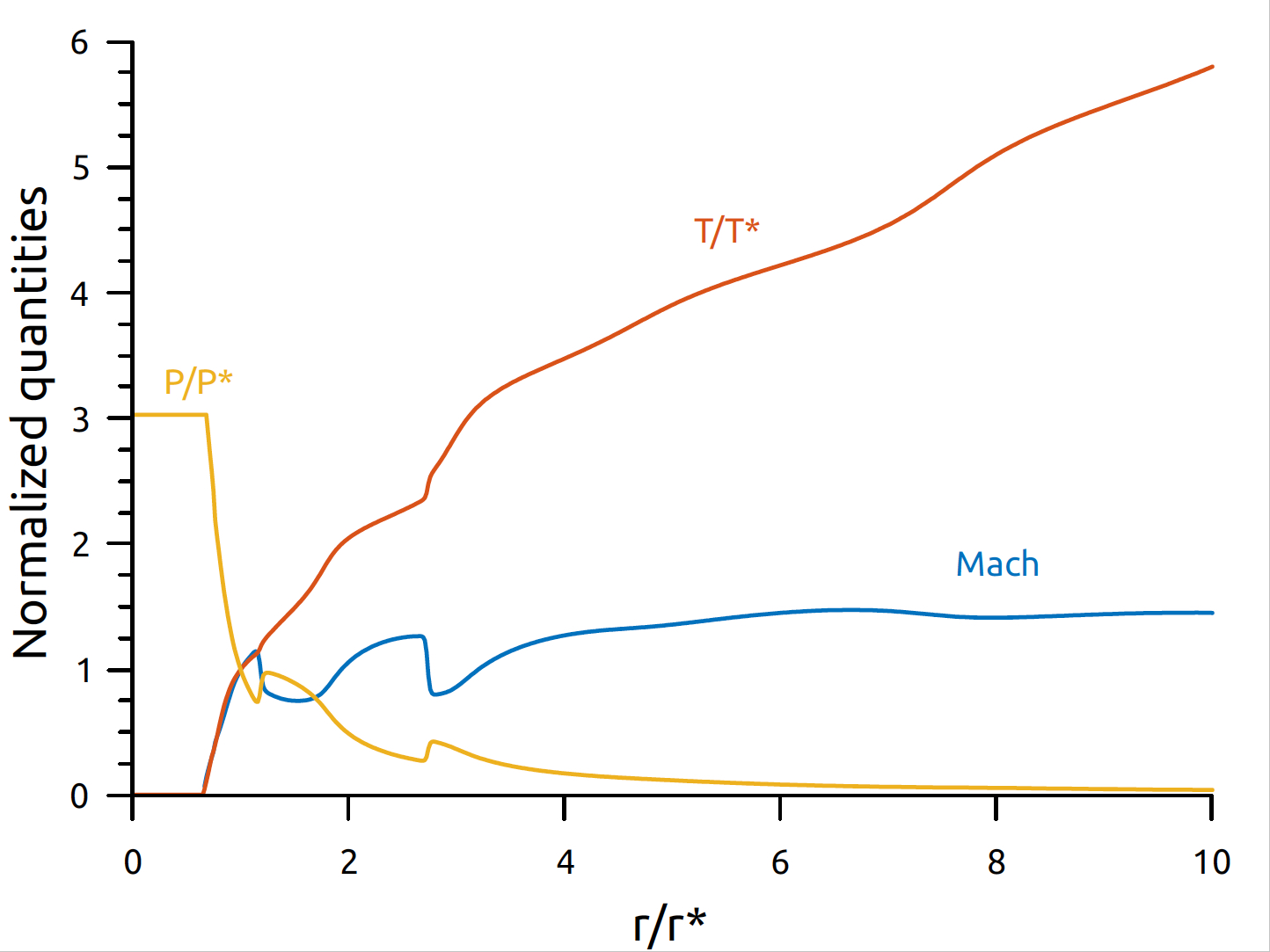}
              \caption{Normalized profiles for ionized neon gas: $r^*=3.03$ mm, $T^*=1.566$ eV,
                  $P^*=21.97$ bars. The plasma paramaters are $\mathrm{r_p}=2$ mm, $n_{e\infty}=4\times10^{14}$,
                $T_{e\infty}=2$ keV.}
              \label{fig:normalized_saha_rp02_Te2_ne4}
            \end{figure}
            
            We now turn our attention to the influence of atomic
            processes on the ablation rate of a molecular deuterium pellet: $D_2$,
            $Z=1$, $\gamma=7/5$. For cryogenic $D_2$ pellets, Ishizaki \textit{et
              al.} in~\cite{Parks_Ishizaki04} found that inclusion of atomic
            processes lowered the ablation rate by a few percents. This
            calculation was confirmed in~\cite{Samulyak07}. We performed similar
            simulations to verify the implementation of our pellet code against
            those results. The numerical EOS for partially dissociated and ionized
            deuterium plasmoid implemented in the pellet code is similar to that
            of neon. The only differences stem from the multiple ionized states of
            neon (Z=10) compared to deuterium (Z=1) and the dissociation events to
            which are subjected $D_2$ molecules. The thermodynamic quantities of
            interest, pressure and energy, are derived
            in~~\cite{Parks_Ishizaki04}. Similarly to the neon case, logarithmic
            tables for the dissociation and ionization fractions were created over
            a relevant range in several sets of thermodynamic quantities. These
            tables were accessed and interpolated during the runs eschewing the
            need to solve two non-linear LTE Saha equations at each location and
            for each time step.

            At this point, we caution the reader that the electronic heat flux used
            in these works were slightly different than what has been described
            in section~\ref{subsec:heat_flux}. Our numerical experiments for this
            verification was performed using the kinetic model developped
            in~\cite{Parks_Ishizaki04} and the same parameters: $r_p=2$ mm,
            $n_{e_\infty} = 10^{14}$ cm$^{-3}$ (no electrostatic shielding),
            $T_{e_\infty} = 2$ keV. For an ideal $D_2$ ablation cloud Ishizaki
            \textit{et al.} reported $G=113$ g/s and Samulyak \textit{et al.}
            reported $G=112$ g/s. For this case we found $G=116.8$ g/s.

            The atomic processes for molecular $D_2$ include dissociation of the
            molecules into atoms upon absorption of the dissociation energy
            $\epsilon_d=4.48$ eV. These deuterium atoms can then be ionized (with
            ionization energy $\epsilon_i = 13.6$ eV). The resulting cloud
            consists of partially dissociated $D_2$ molecules and partially
            ionized $D$ atoms. The dissociation fraction and ionization fraction
            are found locally from the solution of two Saha equations in complete
            analogy with the system of Saha equations for multiply ionized states
            in the case of neon. When both dissociation and
            ionization are accounted for, Ishizaki \textit{et al.} reported a
            reduction in the ablation rate, $G=106$ g/s. Similarly Samulyak
            \textit{et al.} found $G=106.5$ g/s. We report $G=109.2$ g/s, which is
            a reduction of $\sim 6.5\%$, in agreement with Ishizaki and Samulyak
            ($6\%$ and $6.5\%$ respectively). An interesting result that emerged
            from this set of verifications, and which has been theorized by Felber
            \textit{et al.} in~\cite{Felber_1979}, was that dissociation was the
            only contributing factor to the reduction of the ablation
            rate. Indeed, in simulations ignoring ionization of dissociated $D$
            atoms and only allowing dissociation of $D_2$ molecules in the cloud,
            we observed a slightly steeper reduction in the ablation rate,
            $G=106.7$ g/s, confirming the trend observed
            in tables~\ref{tab:Saha_Saha_ne1},~\ref{tab:Saha_Saha_ne2},~\ref{tab:Saha_Saha_ne4} for
            neon that ionization tend to increase the ablation rate for small
            pellets. Because the ionization potential of $D$ is almost three times
            the dissociation energy for $D_2$ the dense gas layer shielding the
            pellet and controlling the ablation rate is essentially completely
            dissociated before any ionization effects come into play. A sufficient
            portion of the incoming electron energy is absorbed and used for
            dissociation in this layer, which in turn reduces the available energy
            for vaporization.

            This observation can be compared to simulations carried out with the
            2019 updated kinetic model of Parks for the heat flux. Using the same
            plasma and pellet parameters as before, the ideal $D_2$ cloud mass
            flow is $G=61.3$ g/s, the mass flow when only dissociation is allowed
            is $G=60.7$ g/s and when dissociation and ionization are allowed
            $G=61.2$ g/s. Here the effects of atomic processes are negligible
            compared to the ideal case. As shown in figure~\ref{fig:diss_heat_04-19} the
            region where most of the dissociation energy is absorbed is much wider
            and further from the pellet with the updated heat flux model than with
            the heat flux model used in~\cite{Parks_Ishizaki04}.

            \begin{figure}[h] \centering\includegraphics[width=\linewidth]{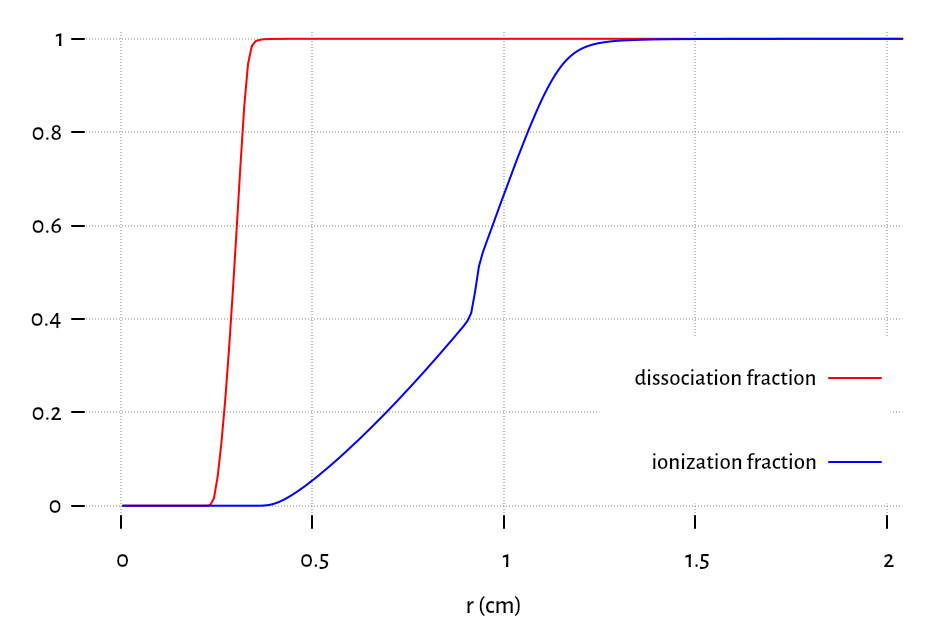}
              \caption[Dissocation and ionization fraction in a $D_2$
              cloud.]{Dissociation and ionization fraction in a $D_2$ ablation cloud
                when the 2004 kinetic heat flux model is employed with parameters:
                $r_p=2$ mm, $n_{e_\infty}=10^{14}$ cm$^{-3}$, $T_{e_\infty} = 2$
                keV. The dissociation and ionization fractions grow to 100\% within a
                few pellet radii.}\label{fig:diss_elec_D2}
            \end{figure}

            \begin{figure}[h] \centering
              \includegraphics[width=\linewidth]{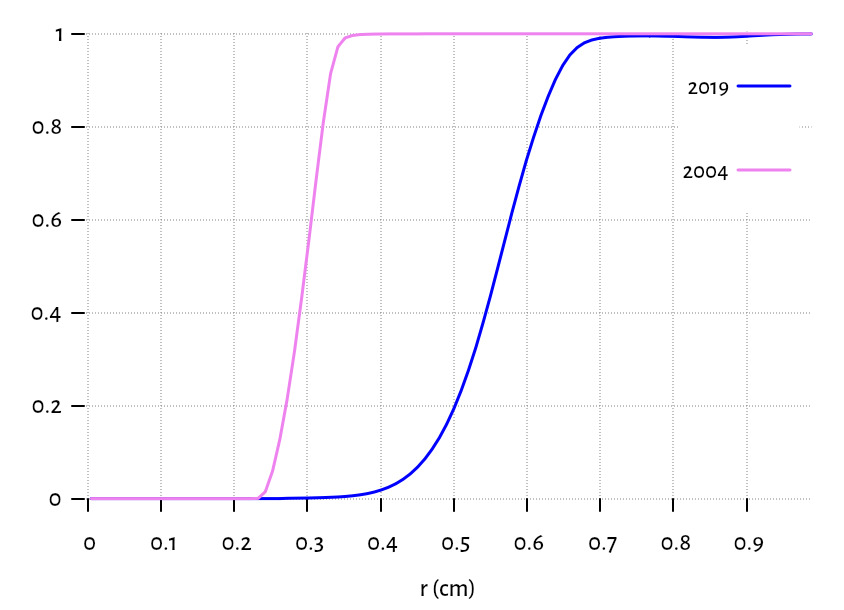}
              \caption[Comparison of dissociation fraction using the 2004 and
              2019 heat flux models for a $D_2$ pellet.]{Comparison of dissociation
                fraction using the 2004 and 2019 heat flux models for canonical plasma
                parameter and $D_2$ pellet with $r_p=2$
                mm.} \label{fig:diss_heat_04-19}
            \end{figure}

            We close this section with an analysis of dimensionless profiles for
            neon pellets when ionization in the cloud is permitted. The flow is subsonic near the
            pellet, the heat flux from the plasma electrons
            accelerates the flow, at the sonic point the expanding cloud balances
            the effect of heating and the flow becomes supersonic. When ionization
            is permitted, a fraction of the energy of the incoming electrons is
            used for ionization rather than for thermal and kinetic energy. The
            presence of energy sinks in the cloud due to ionization losses reduces
            the temperature and decelerates the
            flow (figure~\ref{fig:normalized_saha_rp02_Te2_ne1}). If enough energy is lost for the
            ionization of neon atoms, the flow velocity drops to subsonic values
            (figure~\ref{fig:normalized_saha_rp02_Te2_ne4}). The formation of a shock
            wave constrasts with the smooth flow profiles obtained when using an
            ideal gas EOS with continuous
            heating (figure~\ref{fig:normalized_ideal_rp02_Te2_ne1}). Behind the shock, the pressure and
            temperature jump and the flow is re-accelerated. When the plasma
            number density $n_{e\infty}$ is increased, more energy is deposited in
            the cloud, the degree of ionization corresponding to sufficient energy
            loss for shock formation are reached earlier
            (see figures~\ref{fig:ionization_1d},~\ref{fig:normalized_saha_rp02_Te2_ne4})
            and the first shock front sits closer to the pellet surface. This has
            also been observed for hydrogen
            pellets~\cite{Parks_Ishizaki04,Samulyak07}. High-Z pellets however,
            allow for the possibility of multiply ionized states, which introduce
            multiple energy sinks and results in multiple shock fronts in the
            flow. In figure~\ref{fig:normalized_saha_rp02_Te2_ne4}), a second shock
            further downstream is observed. This secondary shock is located two
            Mach radius away from the first shock. Atomic processes introduce
            marked effects on the flow dynamic, which follow the same trend for a
            wide range of pellet and plasma parameters. This can be contrasted
            with the influence of atomic processes on the ablation rate, which
            trend does depend on the pellet size and the plasma parameters
            (see tables~\ref{tab:Saha_Saha_ne1},~\ref{tab:Saha_Saha_ne2},~\ref{tab:Saha_Saha_ne4}).

       \subsubsection{Axisymmetric approximation\\}
       \label{subsubsec:Axisymmetric approx}
\begin{figure}[!hbtp]
              \centering
              \begin{subfigure}[t]{1\linewidth}
                \includegraphics[width=1.02\textwidth,height=3.5cm]{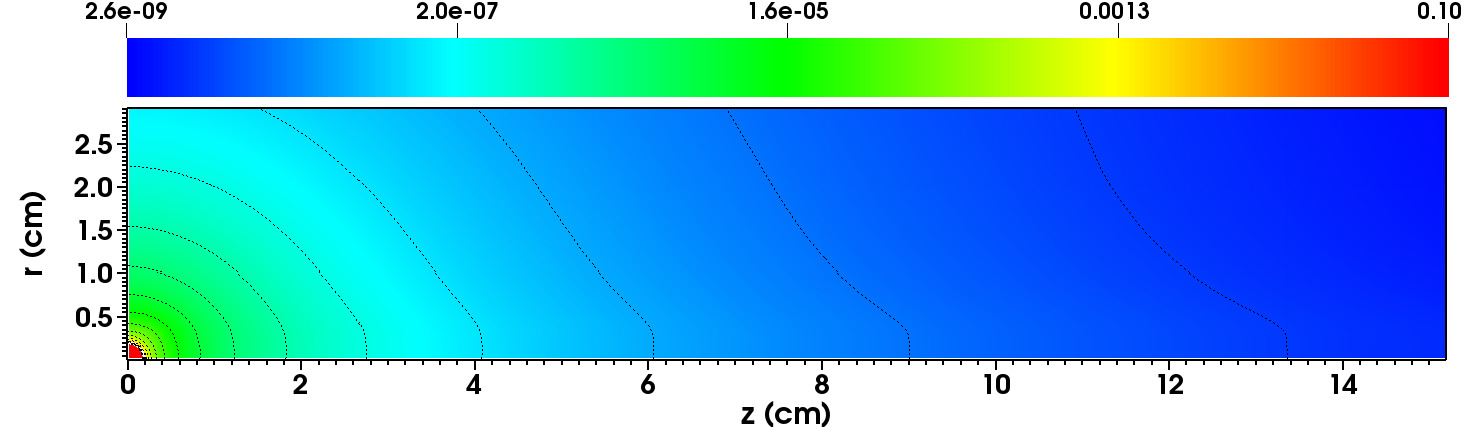}
                \caption{Density (g/cc)}
                \label{fig:dens_ideal_contours}
              \end{subfigure}
              \begin{subfigure}[t]{1\linewidth}
                \includegraphics[width=1.02\textwidth,height=3.5cm]{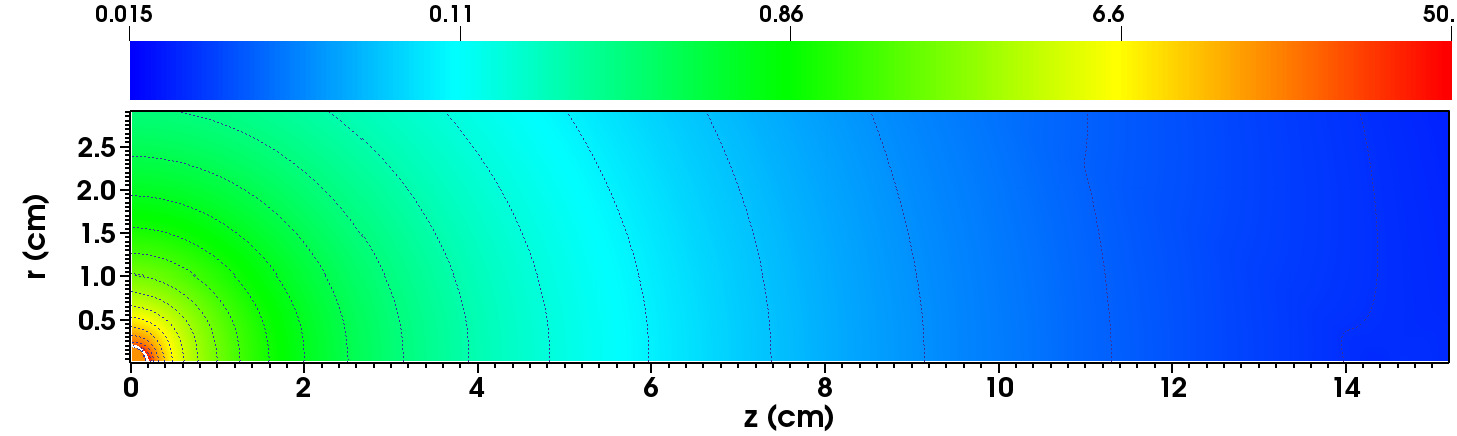}
                \caption{Pressure (bar)}
                \label{fig:pres_ideal_contours}
              \end{subfigure}
              \begin{subfigure}[t]{1\linewidth}
                \includegraphics[width=1.02\textwidth,height=3.5cm]{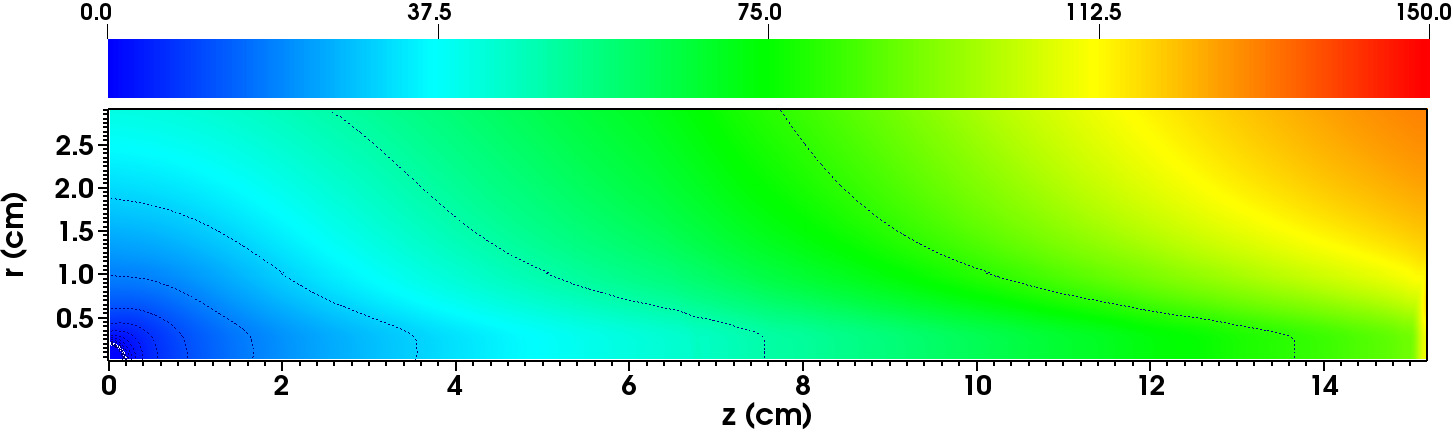}
                \caption{Temperature (eV)}
                \label{fig:temp_ideal_contours}
              \end{subfigure}
              \caption{States distribution for an ideal neon ablatant with
                parameters $T_{e\infty}=2$ keV, $n_{e\infty}=10^{14}$, $\mathrm{r_p}=2$ mm
                in axisymmetric approximation.}
              \label{fig:states_ideal_contours}
            \end{figure}

\begin{figure}
              \centering
              \begin{subfigure}[t]{1\linewidth}
                \includegraphics[width=1.02\textwidth,height=3.5cm]{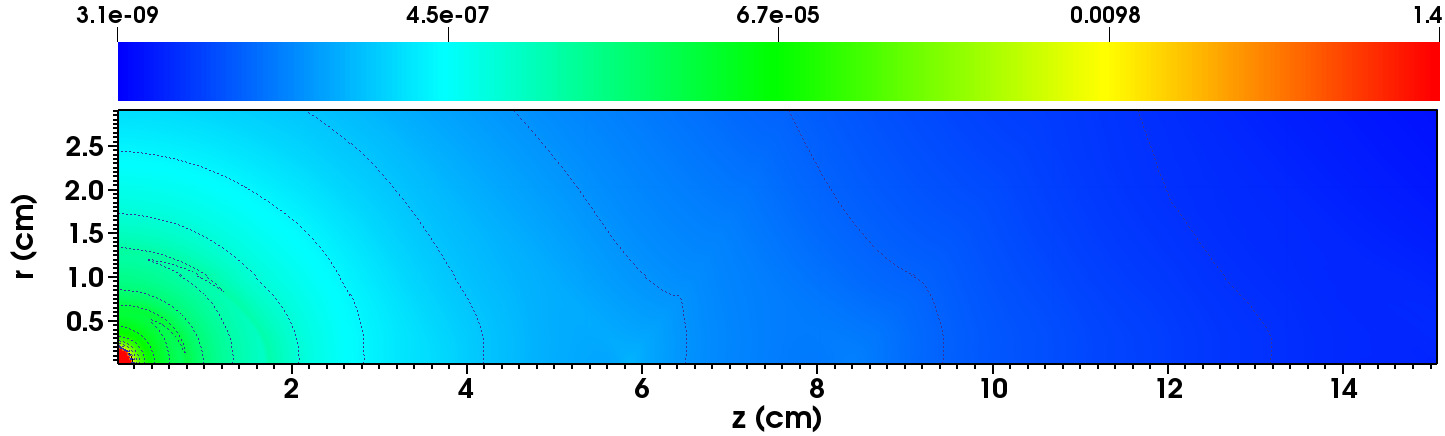}
                \caption{Density (g/cc)}
                \label{fig:dens_0T_contours}
              \end{subfigure}
              \begin{subfigure}[t]{1\linewidth}
                \includegraphics[width=1.02\textwidth,height=3.5cm]{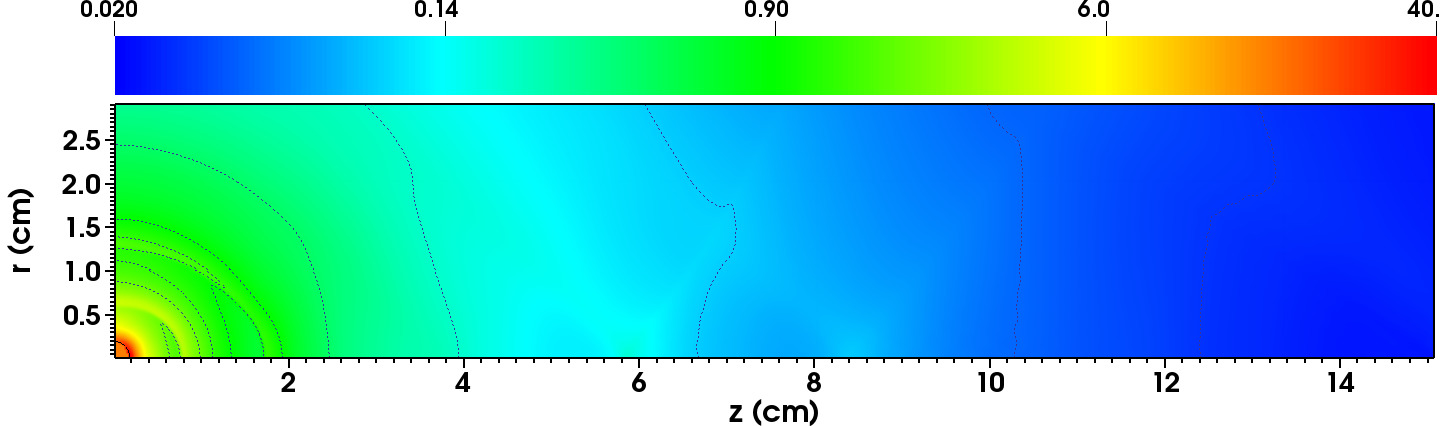}
                \caption{Pressure (bar)}
                \label{fig:pres_0T_contours}
              \end{subfigure}
              \begin{subfigure}[t]{1\linewidth}
                \includegraphics[width=1.02\textwidth,height=3.5cm]{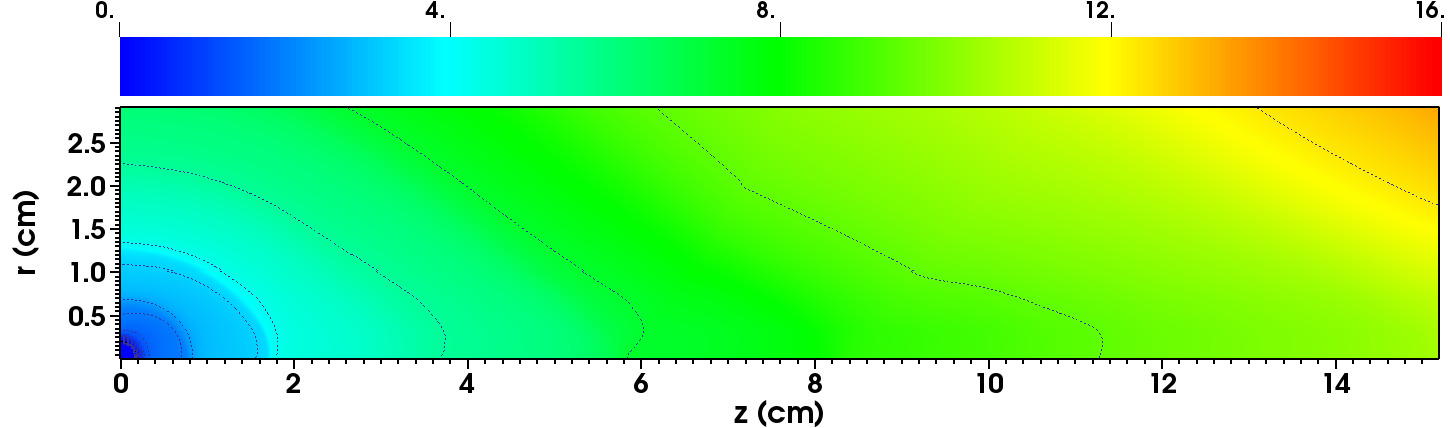}
                \caption{Temperature (eV)}
                \label{fig:temp_0T_contours}
              \end{subfigure}
              \caption{States distribution for an ionized neon gas ablatant with
                parameters $T_{e\infty}=2$ keV, $n_{e\infty}=10^{14}$, $\mathrm{r_p}=2$ mm
                in axisymmetric approximation.}
              \label{fig:states_0T_contours}
            \end{figure}

\begin{figure*}
              \centering \includegraphics[width = 0.85\linewidth,
              height = 0.5\textheight]{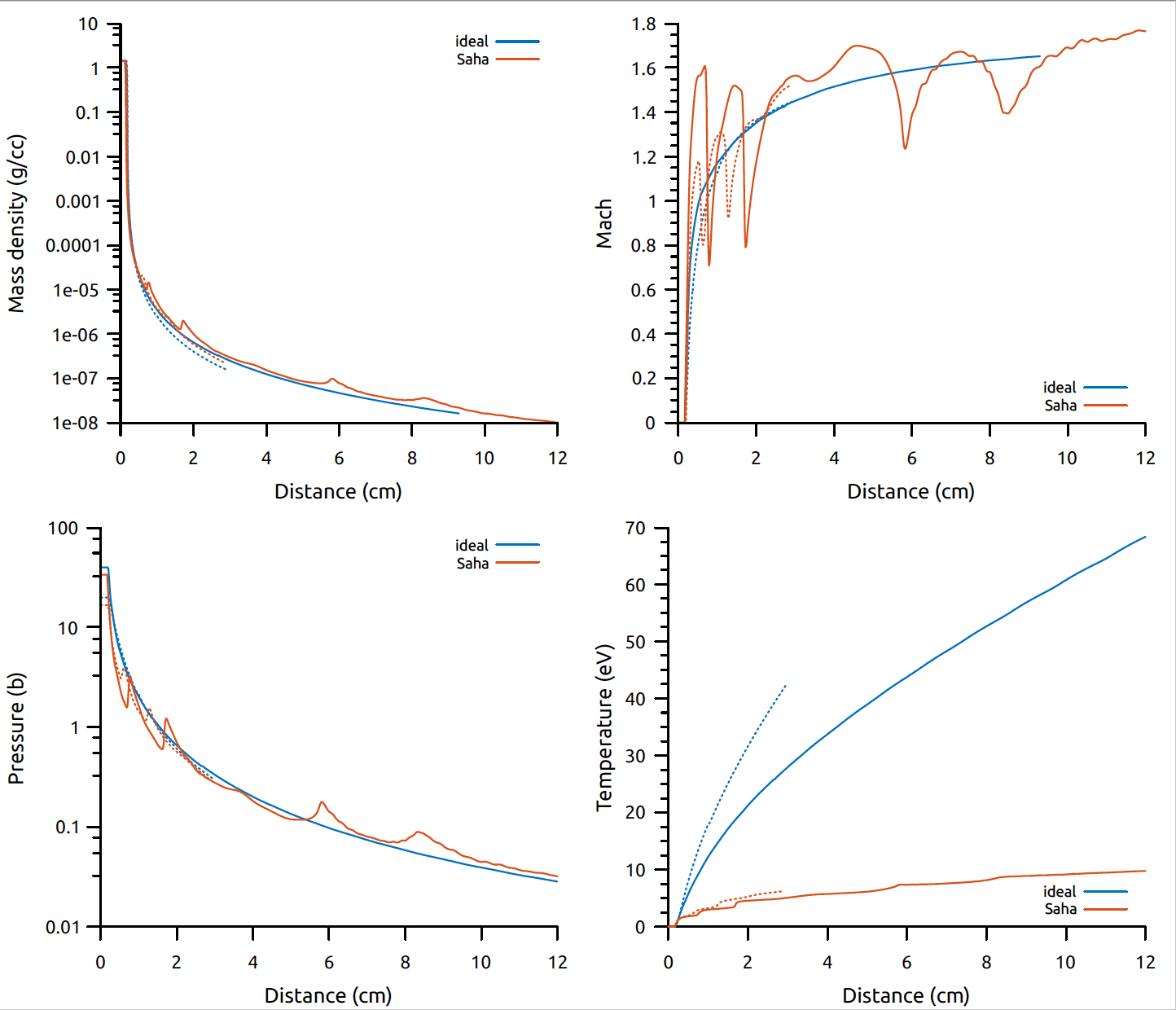}
              \caption{Longitudinal (solid) and radial (dashed) flow
                variables profiles along the r=0 plane and z=0 plane
                respectively, for an ideal and ionizaed cloud.}
              \label{fig:long_rad_all_states_noB_mach}
            \end{figure*}

         We now turn our attention to the 2D cylindrical axisymmetric
         model for the ablation of a single pellet. In what follows, we
         probe the effect of anisotropic heating on the cloud dynamic
         and the ablation rate. To allow comparison with the spherically
         symmetric case, we performed simulations with and without
         ionization in the cloud. The canonical parameters we used for
         the remaining of this work are: $\mathrm{r_p}=2$mm,
         $T_{e\infty}=2$ keV and
         $n_{e\infty}=10^{14} \Longrightarrow
         n_{eff}=1.205\times10^{13}$ 1/cc. The action of the
         $\mathrm{J}\times \mathrm{B}$ force on the ablatant cloud is
         not included in this section but will be investigated next, in Section
         \ref{subsec:ablation_studies_with_axisymmetric_mhd_model}. For
         now, the only effect of the magnetic field is to direct the
         hot electrons along the $z$-axis. We remind the reader that we consider a
         rigid pellet, the interior fluid states inside the pellet
         are constant with zero velocity, and only the surface states are
         updated according to the surface ablation model. The states
         distribution in the 2D region are plotted at steady state in figures
         \ref{fig:states_ideal_contours} and \ref{fig:states_0T_contours}. To
         compare the properties of the flow in the longitudinal and
         radial direction,  the corresponding profiles are plotted in
         figures \ref{fig:long_rad_all_states_noB_mach}. The
         dark curve near the origin in the figures below is the pellet surface. It should also be noted that
         in the absence of the Lorentz force to constrain the cloud
         along the field lines, the cloud/plasma interface is free to
         expand unrestrained and eventually exits the compuational
         domain, which is why it is not visible anymore in figures \ref{fig:states_ideal_contours} and \ref{fig:states_0T_contours}.

            \bigskip

            \paragraph{Hydrodynamic ideal model\\}
The 2D states distribution of the hydrodynamic ideal model
            (neither MHD nor ionization is included) are presented in
            \ref{fig:states_ideal_contours} and the radial and
            longitudinal profiles are plotted in
            \ref{fig:long_rad_all_states_noB_mach}. In this case, even
            if \textbf{B}=0, the expansion is essentially two dimensional due to
            the longitudinal direction being the imposed streaming direction of
            the plasma electrons. However, it was shown in~\cite{Parks_distorsion}
            that the parallel heat flux is approximately independent of the polar
            angle resuting in the flow quantities being approximately spherically
            symmetric.

            The steady state is
            reached after approximately 30 $\mu$s and we record an
            ablation rate of 45.8 g/s.   This corresponds to a $\sim 29.0\%$ decrease in the
            ablation rate compared to the 1D spherical symmetry with
            identical plasma parameters (see table \ref{tab:Saha_1d}).  At steady-state, the
            shape of the cloud is close to spherical but stretches out
            in the longitudinal direction. Similar to the 1D case in figure
            \ref{fig:normalized_ideal_rp02_Te2_ne1}, the flow smoothly
            transitions from subsonic to supersonic across an ellipse-shaped sonic surface with the major axis $r_{r}^* = 7.3$ mm
            and the minor axis $r_{z}^* = 5.35$ mm. Figure
            \ref{fig:long_rad_all_states_noB_mach} shows that the density distribution
            is isotropic over the pellet surface before falling more
            quickly in the radial direction. 
            The pressure distribution is not uniform over the pellet surface and it is
            larger at the poles ($\sim$ 40 bars) compared to the equator
            ($\sim 20$ bars). 
            
            The pressure falls sharply within a small region along the $z$-axis to reach values similar to
            that at the pole; the pressure distribution then becomes
            nearly isotropic in the rest of the cloud
            \ref{fig:pres_ideal_contours}. The surface pressure
            differential was also observed in ~\cite{Parks_Ishizaki04}
            where they reasonned that the build up of ablated material
            at the poles, where the heat flux is fully absorbed by the
            pellet as opposed to the equatorial points where incoming
            electrons stream by unimpeded by the pellet, was
            responsible for the non-uniform pressure
            distribution. The temperature along the z-axis is
            $\sim 34\%$ less everywhere than along the radial
            direction. As explained in \cite{Samulyak07} this is due
            to pellet shadow. The region $r\leq \mathrm{r_p}$ is heated by
            incoming electrons streaming along the positive $z$-axis
            only: electrons streaming from the opposite direction are
            completely absorbed by the pellet when they reached the
            surface and do not participate in the heating of the
            region behind the pellet. In the rest of the domain, where
            $r\geq\mathrm{r_p}$, the cloud is heated from electrons streaming in
            both directions.
            
            \bigskip

            \paragraph{Hydrodynamic model with atomic processes\\}
             The hydrodynamic ideal model did not allow ionization in
            the cloud. We now report simulation results where
            ionization of the neon gas is enabled by shifting to the
            EOS model describe in Section \ref{subsubsec:Saha}. The steady-state is reached in
            50$\mu$s, and the corresponding state distribution are
            plotted in figure \ref{fig:states_0T_contours}. For this case,
            and in the absence of MHD forces, we record an ablation
            rate of $G \sim 47.2$ g/s corresponding to a $28\%$
            reduction compared to the 1D model in table
            \ref{tab:Saha_Saha_ne1}. This is consistent with the
            effect of anisotropic heating on the ablation rate in the
            ideal case described in the previous paragrah. Ionization
            increases the ablation rate by 3\% compared to the 2D
            hydrodynamic ideal model, twice as much as the increase in
            the 1D spherically symmetric approximation (table
            \ref{tab:Saha_1d}). The shape of the cloud remains almost
            spherical with the contour lines of state variables being nearly
            spherical. The flow experiences distinct shocks and
            five transonic regions. The sonic surfaces are
            ellipsoidal except for the first one, closest to the
            pellet, which is practically spherical. The density
            (figure \ref{fig:long_rad_all_states_noB_mach}) is uniform over the
            pellet and follows the ideal gas density distribution. The
            surface pressure continues to exhibit a differential over
            the pellet with values slightly inferior to the ideal gas.
            The temperature is about seven times lower in this
            case compared to the ideal case, and the radial and
            longitudinal temperature profiles reveals that although it
            is present, the effect of the pellet shadow due to
            anisotropic heating is greatly reduced. The lower
            temperature of the cloud is due to energy sinks introduced
            by the multple ionization events and radiation cooling,
            both of which are absent in the ideal case.

            \subsection{Ablation studies with axisymmetric MHD model}
            \label{subsec:ablation_studies_with_axisymmetric_mhd_model}
In this Section, we study the influence of the
            $\mathbf{J}\times \mathbf{B}$ force on the pellet ablation
            flow. The cloud becomes ionized and conductive, the Lorentz force drives the ablation
            cloud along the direction of the magnetic field creating
            an ablation channel parallel to the $z$-axis. As is the
            case for hydrogenic pellets, we expect this phenomenon to be
            dependent on the plasma parameters, magnetic field and
            pellet radius. In ~\cite{Samulyak07} it was found that
            for deuterium pellets the warm up time, i.e a simulation
            parameter used to mimic the ramping up of the magnetic
            field and incoming electron heat flux as seen by the
            traveling pellet, to be a sensitive parameter in the
            formation of the ablation channel. We conducted a series
            of numerical experiments to assess the sensitivity of the
            neon cloud evolution to this parameter. The values that we
            sampled for the warm up time ranged from a few
            microseconds to several hundred microseconds, however we
            found the ablation channel and steady-state ablation rate
            to be independent from this parameter.  
            To reduce the simulation time, we use 10 $\mu$s warm-up time. 
             We impose an effective shielding length of 15 cm; past that mark the
            electron heat flux is cut off and the fluid is allowed to
            escape the computational domain via an outflow boundary
            condition. This effective shielding length is 
            necessary for establishing a non trivial steady-state
            solution to eq.\ref{eq:Euler} as is explained below. 

            Making use of the approximation that the inertial terms in
            the radial force balance are negligibly small in the
            ablation channel part of the flow field where the magnetic
            field inhibits radial expansion, the continuity and
            momentum equations in eq.\ref{eq:Euler} can be reduced to a
            single convective-diffusion equation. Solving this
            equation shows that steady-state is not possible due to
            the fact that the longitudinal integrated density along
            the symmetry axis passing through the pellet, or optical
            depth
            \small$\tau_{axis} = \int\limits_{\infty}^{z_1}\rho_{axis}dz$,
            would be infinite if the flow field were to reach steady
            state. Consequently, the incident plasma electrons would
            be completely shielded and the pellet would cease to
            ablate, contradicting steady-state premise. Thus the upper limit on the
            opacity integral should be finite. However, a finite
            opacity (cloud length) is only possible if we invoke 3D
            effects such as polarization drift of the ablation cloud
            caused either by the pellet motion itself or the grad-B
            drift drive. These effects that shift the elongated
            ablation channel off the pellet axis and lead to an
            effective cloud length will be investigated in the future using
            full 3D simulations.

            We present results
            for the $\mathbf{B}$ field strength of 2, 4, 6 and 9 Tesla
            and fixed canonical plasma parameters: $T_{e\infty}=2$
            keV,
            $n_{e\infty}=10^{14} \Longrightarrow
            n_{eff}=1.205\times10^{13}$ 1/cc and $\mathrm{r_p}=2$ mm.
           The formation of an ablation channel is the main
            difference compared to the nearly spherical expansion of cloud in the
            hydrodynamic case. The flow starts out nearly spherical and remain
            quasi-spherical close to the pellet throughout the computation. As the
            ablated material flows downstream (a few pellet radii) it quickly
            evolves into a cylindric channel symmetric with respect to the longitudinal
            axis. This is due to the combined effect of the streaming hot
            electrons ionizing the ablation cloud and the magnetic field lines
            redirecting the flow in the direction of $\mathbf{B}$ along the
            $z$-axis. The flow redirection is observed on the velocity vector
            field in figure~\ref{fig:2T_velo_2d},~\ref{fig:4T_velo_2d},\ref{fig:6T_velo_2d},\ref{fig:9T_velo_2d}
            for increasing $\mathbf{B}$ field values.            

            \begin{figure}[hbtp]
              \centering \includegraphics[width=0.495\textwidth]{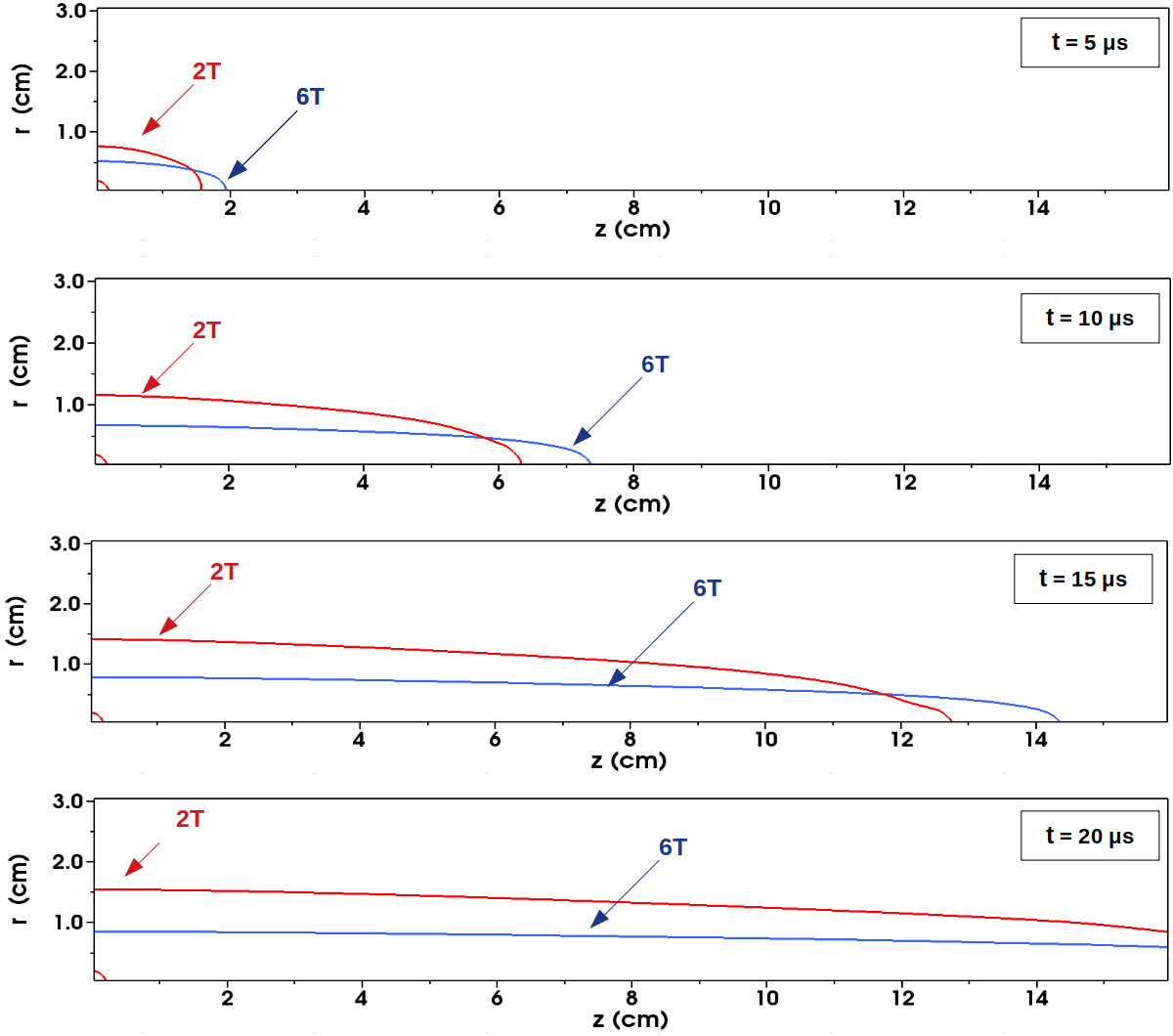}
              \caption{Channel formation}
              \label{fig:channel_formation}
            \end{figure}

            \begin{table}[H]
              \centering
              \begin{tabular}{|>{\centering\arraybackslash}p{1.5cm}|>{\centering\arraybackslash}p{1.5cm}|>{\centering\arraybackslash}p{1.5cm}|>{\centering\arraybackslash}p{1.5cm}|}
                \hline
                2T & 4T & 6T & 9T\\
                \hline
                2.59 cm & 1.75 cm  & 1.07 cm & 1.02 cm\\
                \hline
              \end{tabular}
              \caption{Channel width (cm) for different field strengths.}
              \label{tab:channel_width}
            \end{table}
                                    
\begin{figure}[!h]
\centering \includegraphics[width=0.95\linewidth,height=0.3\textheight]{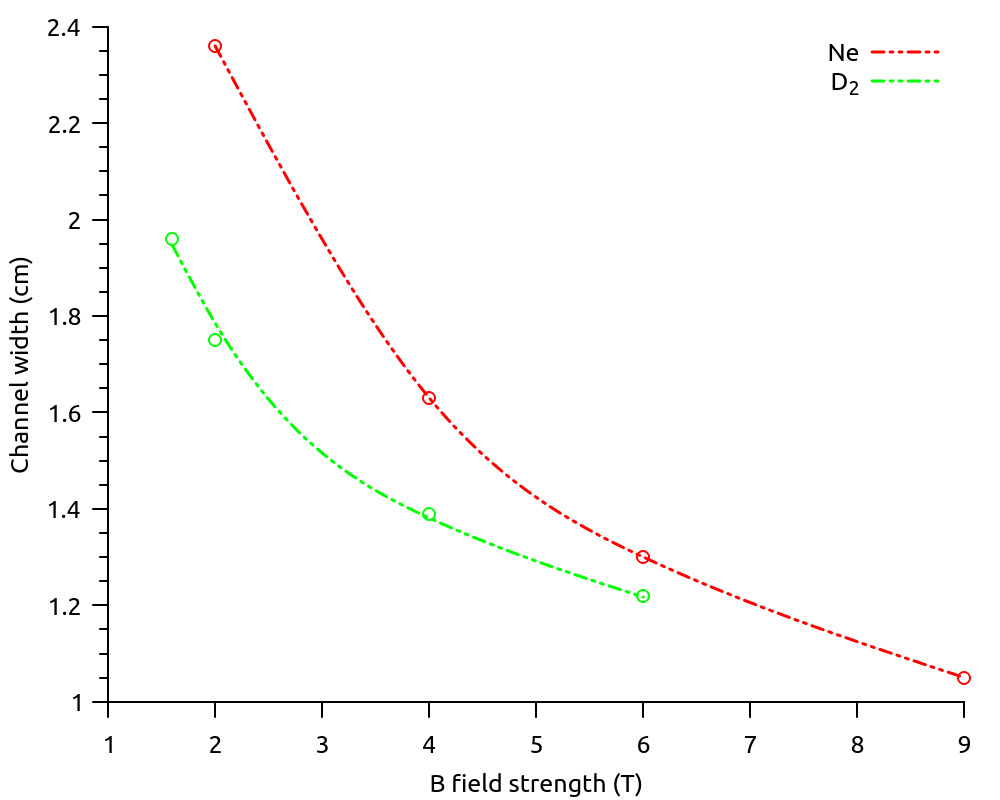}
\caption[Ablation channel width as a function of $B$ for $D_2$ and $Ne$ using FronTier-Lite pellet code.] {Ablation channel width as a function of \textbf{B} for $Ne$ (red) and $D_2$ (green) pellets computed from the FronTier-Lite pellet code.}
\label{fig:rperp_FT}
\end{figure}

\begin{figure}
              \centering
                 \begin{subfigure}{1\linewidth}
                  \includegraphics[width = 0.95\textwidth, height = 0.15\textheight]{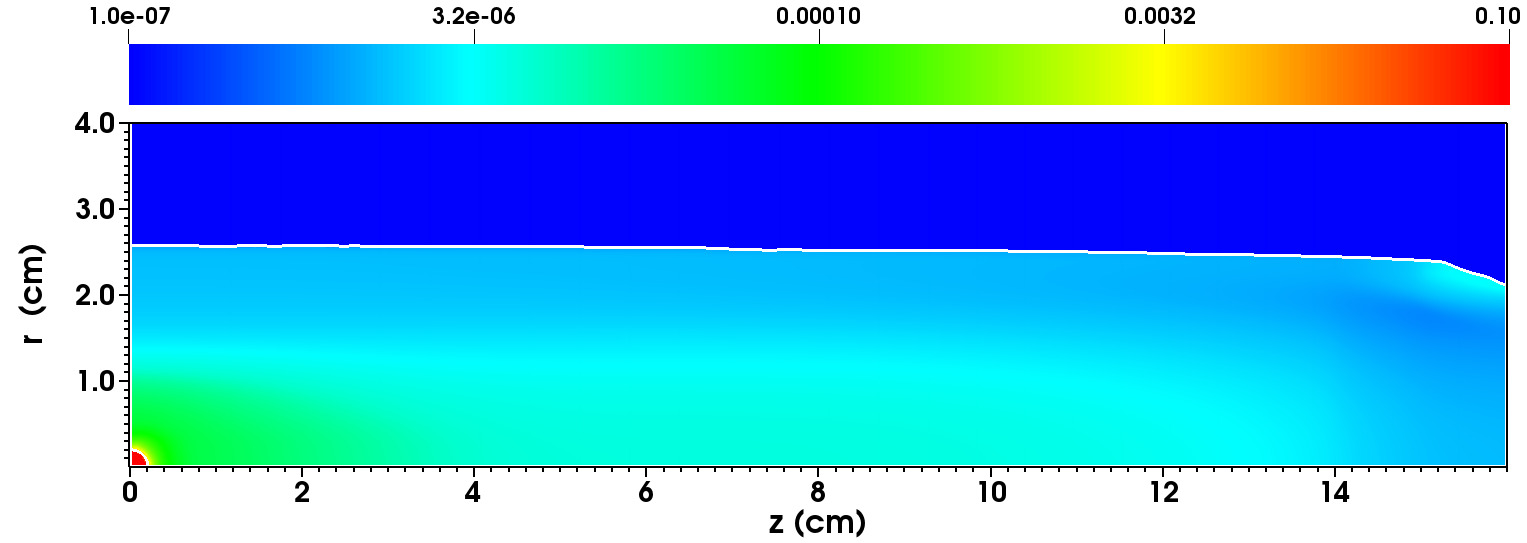}
                  \captionof{figure}{Density (g/cc)}
                  \label{fig:2T_dens_2d}
                \end{subfigure}
                \begin{subfigure}{1\linewidth}
                  \includegraphics[width = 0.95\textwidth, height = 0.15\textheight]{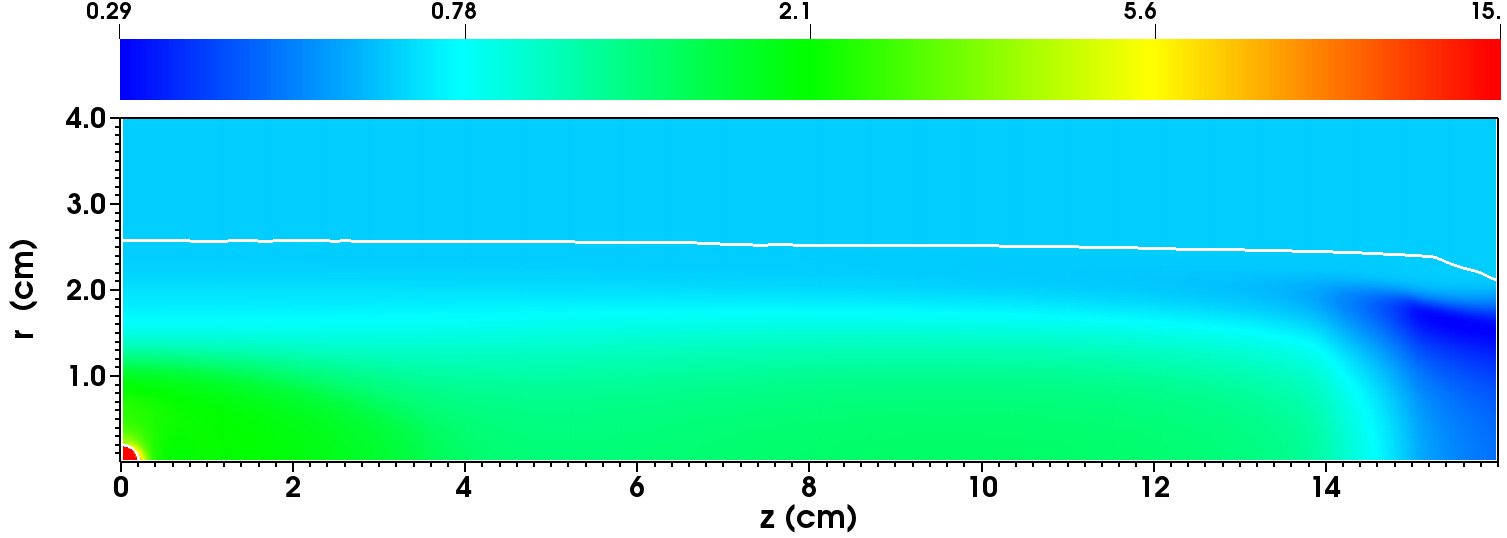}
                  \captionof{figure}{Pressure (b)}
                  \label{fig:2T_pres_2d}
                \end{subfigure}
                \begin{subfigure}{1\linewidth}
                  \includegraphics[width = 0.95\textwidth, height = 0.15\textheight]{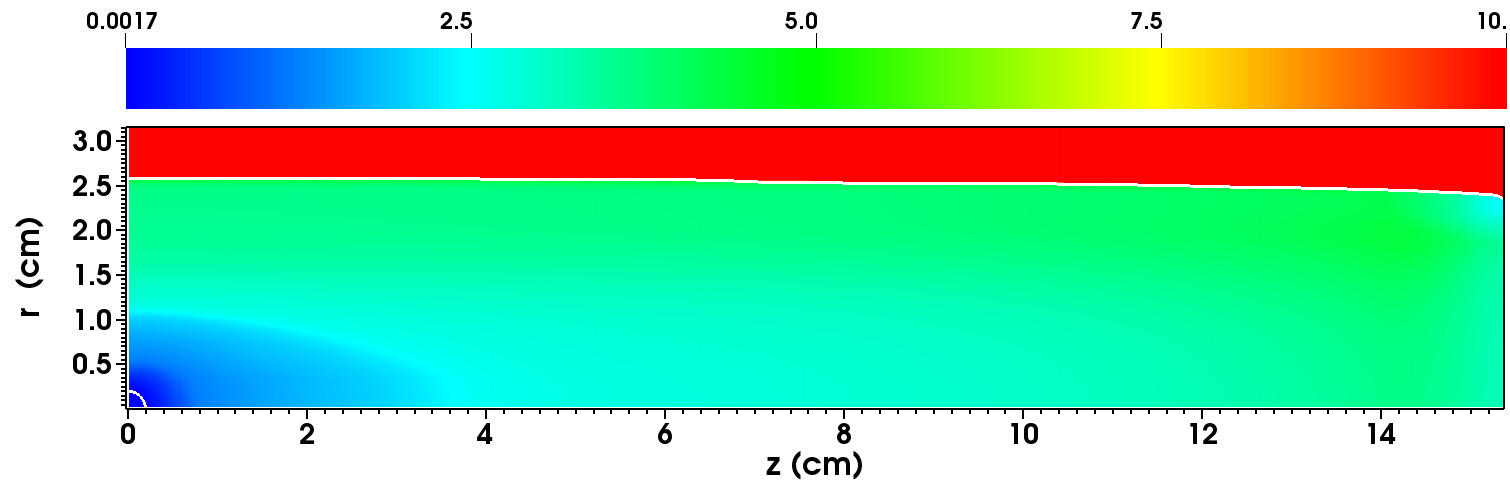}
                  \captionof{figure}{Temperature (eV)}
                  \label{fig:2T_temp_2d}
                \end{subfigure}
                \begin{subfigure}{1\linewidth}
                  \includegraphics[width = 0.95\textwidth, height = 0.15\textheight]{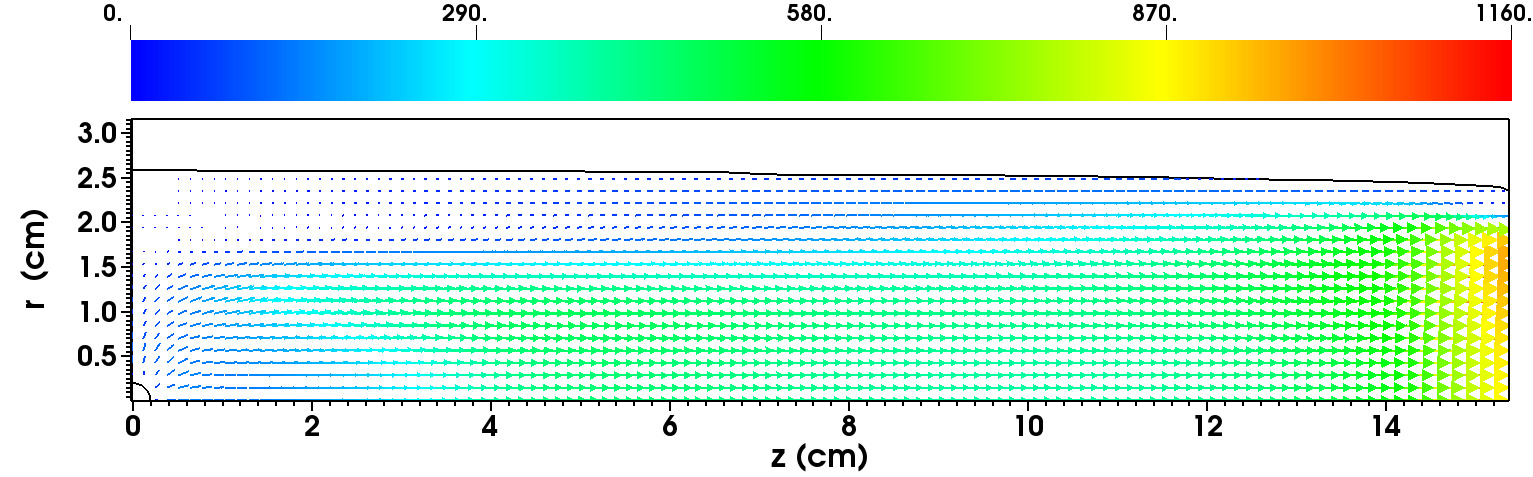}
                  \captionof{figure}{Velocity field (cm/ms)}
                  \label{fig:2T_velo_2d}
                \end{subfigure}
                \caption{States distribution for an ionized neon gas ablatant with
                  parameters $T_{e\infty}=2$ keV, $n_{e\infty}=10^{14}$, $\mathrm{r_p}=2$ mm, $B=2T$ in axisymmetric approximation.}
                \label{fig:2T}
             \end{figure}

\begin{figure}
                \centering
                \begin{subfigure}{1\linewidth}
                  \includegraphics[width = 0.95\textwidth, height = 0.15\textheight]{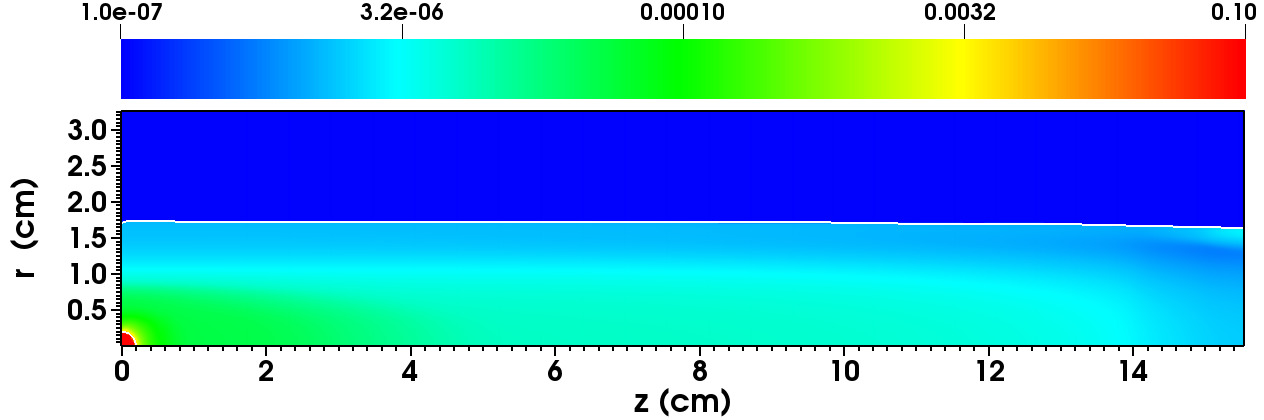}
                  \captionof{figure}{Density (g/cc)}
                  \label{fig:4T_dens_2d}
                \end{subfigure}
                \begin{subfigure}{1\linewidth}
                  \includegraphics[width = 0.95\textwidth, height = 0.15\textheight]{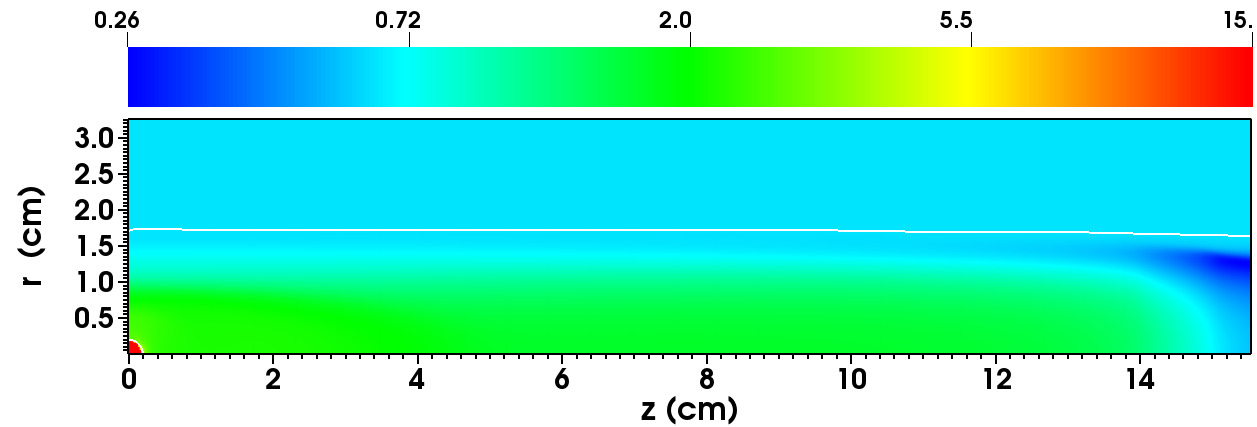}
                  \captionof{figure}{Pressure (b)}
                  \label{fig:4T_pres_2d}
                \end{subfigure}
                \begin{subfigure}{1\linewidth}
                  \includegraphics[width = 0.95\textwidth, height = 0.15\textheight]{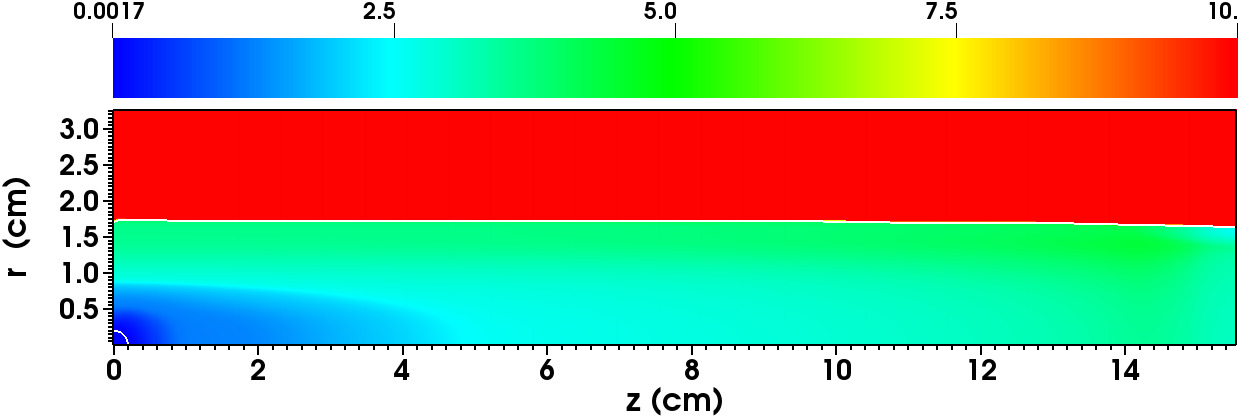}
                  \captionof{figure}{Temperature (eV)}
                  \label{fig:4T_temp_2d}
                \end{subfigure}
                \begin{subfigure}{1\linewidth}
                  \includegraphics[width = 0.95\textwidth, height = 0.15\textheight]{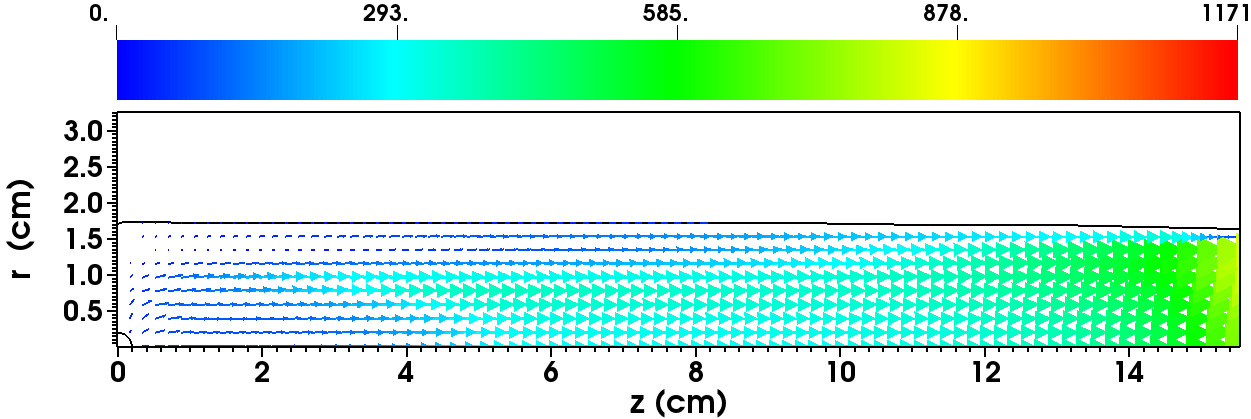}
                  \captionof{figure}{Velocity field (cm/ms)}
                  \label{fig:4T_velo_2d}
                \end{subfigure}
                \caption{States distribution for an ionized neon gas ablatant with
                  parameters $T_{e\infty}=2$ keV, $n_{e\infty}=10^{14}$, $\mathrm{r_p}=2$ mm, $B=4T$ in axisymmetric approximation.}
                \label{fig:4T}
           \end{figure}

\begin{figure}
              \centering
                \begin{subfigure}{1\linewidth}
                  \includegraphics[width = 0.95\textwidth, height = 0.15\textheight]{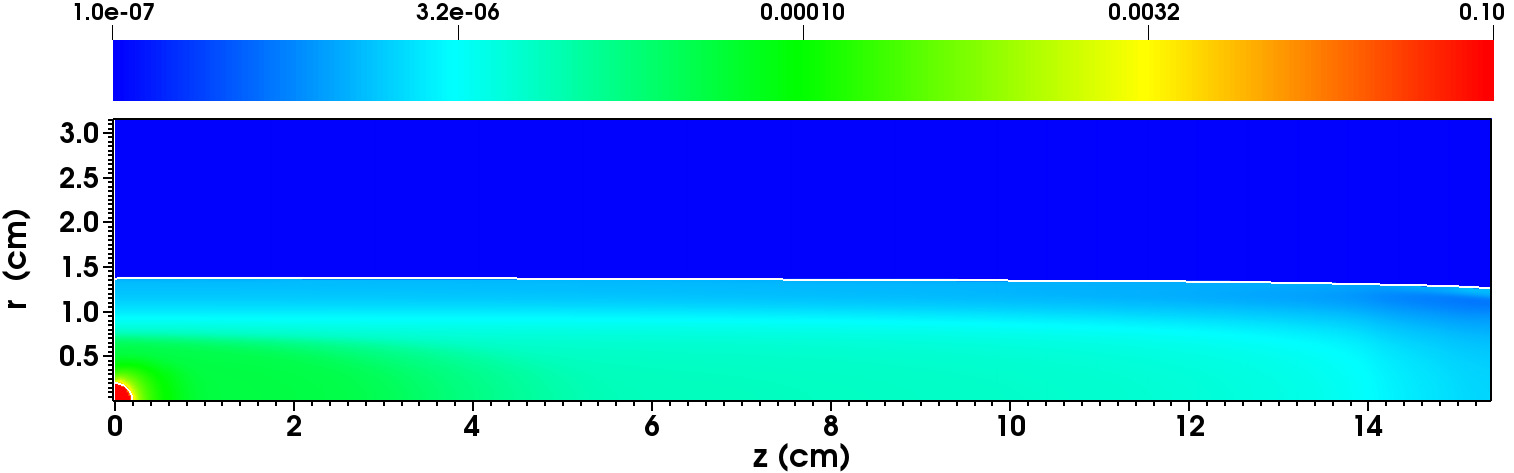}
                  \captionof{figure}{Density (g/cc)}
                  \label{fig:6T_dens_2d}
                \end{subfigure}
                \begin{subfigure}{1\linewidth}
                  \includegraphics[width = 0.95\textwidth, height = 0.15\textheight]{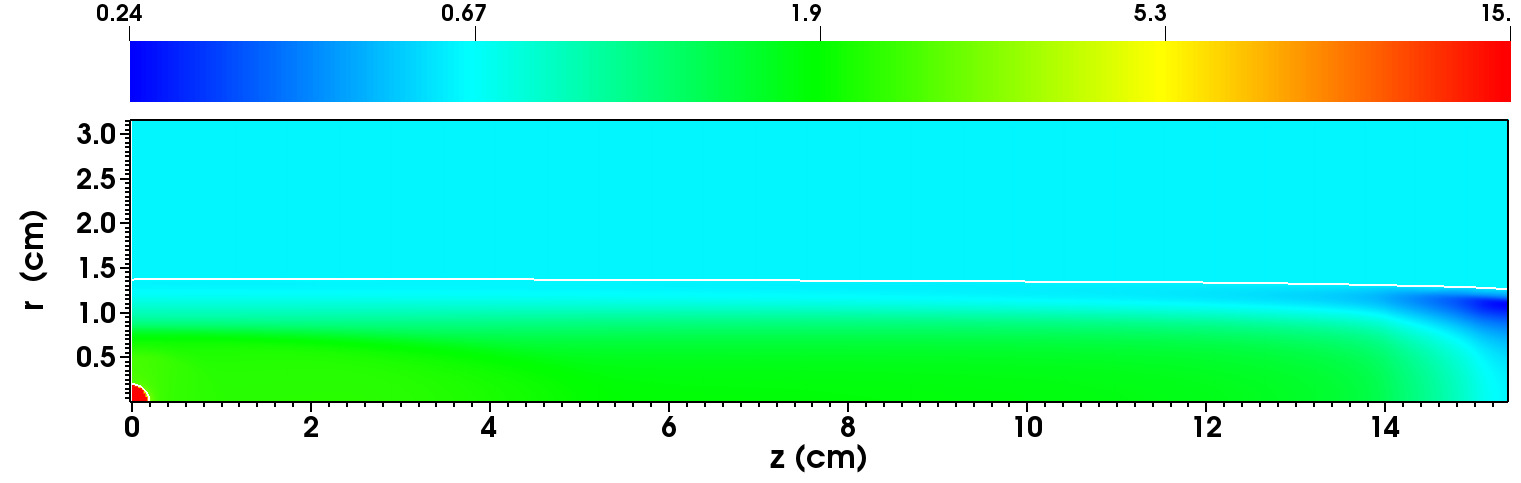}
                  \captionof{figure}{Pressure (b)}
                  \label{fig:6T_pres_2d}
                \end{subfigure}
                \begin{subfigure}{1\linewidth}
                  \includegraphics[width = 0.95\textwidth, height = 0.15\textheight]{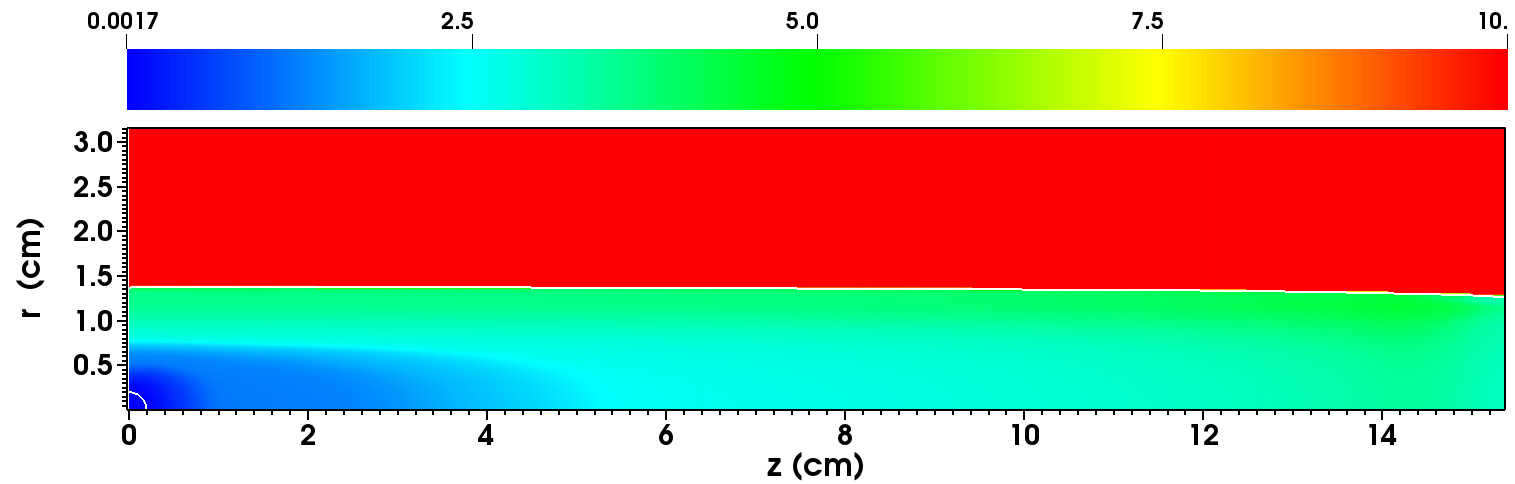}
                  \captionof{figure}{Temperature (eV)}
                  \label{fig:6T_temp_2d}
                \end{subfigure}
                \begin{subfigure}{1\linewidth}
                  \includegraphics[width = 0.95\textwidth, height = 0.15\textheight]{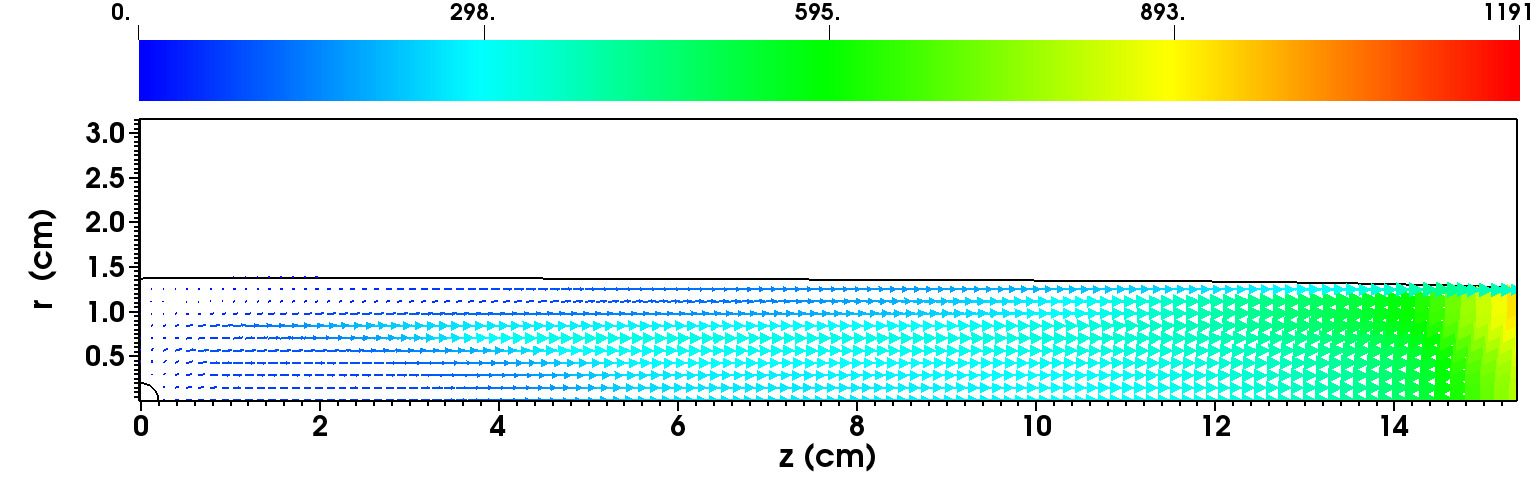}
                  \captionof{figure}{Velocity field (cm/ms)}
                  \label{fig:6T_velo_2d}
                \end{subfigure}
                \caption{States distribution for an ionized neon gas ablatant with
                  parameters $T_{e\infty}=2$ keV, $n_{e\infty}=10^{14}$, $\mathrm{r_p}=2$ mm, $B=6T$ in axisymmetric approximation.}
                \label{fig:6T}
          \end{figure}

\begin{figure}     
                \centering
                \begin{subfigure}{1\linewidth}
                  \includegraphics[width = 0.95\textwidth, height = 0.15\textheight]{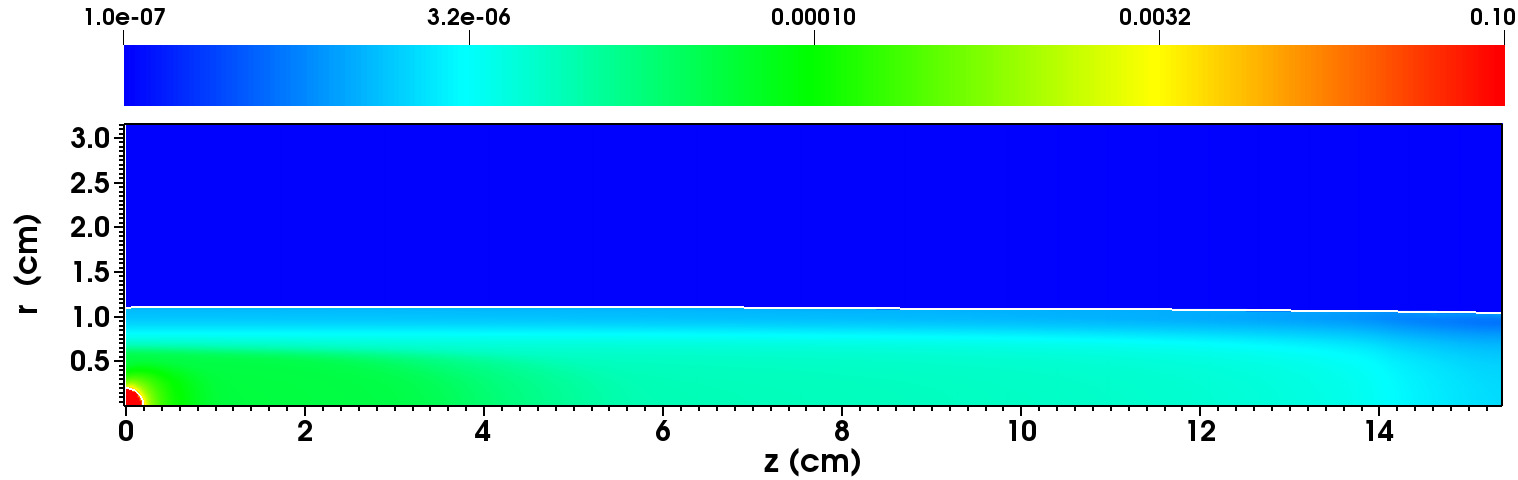}
                  \captionof{figure}{Density (g/cc)}
                  \label{fig:9T_dens_2d}
                \end{subfigure}
                \begin{subfigure}{1\linewidth}
                  \includegraphics[width = 0.95\textwidth, height = 0.15\textheight]{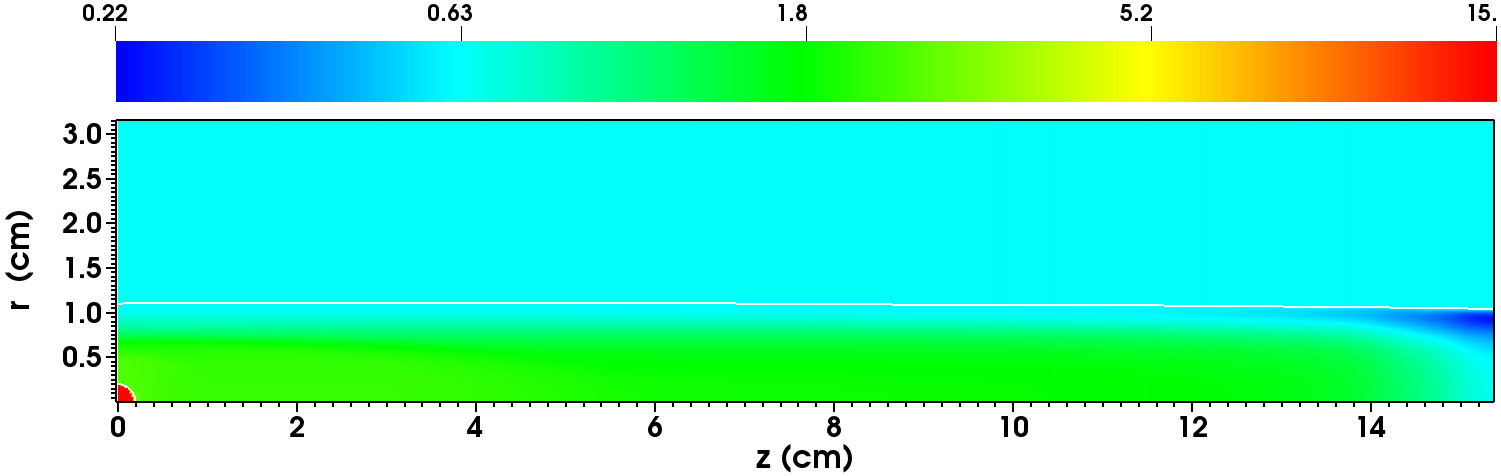}
                  \captionof{figure}{Pressure (b)}
                  \label{fig:9T_pres_2d}
                \end{subfigure}
                \begin{subfigure}{1\linewidth}
                  \includegraphics[width = 0.95\textwidth, height = 0.15\textheight]{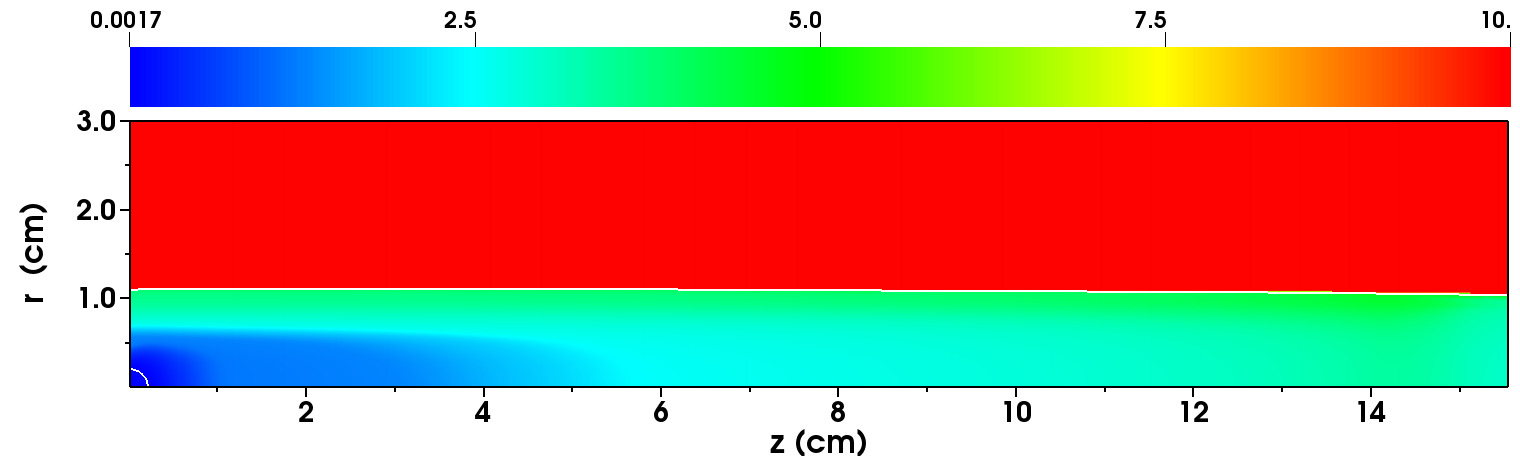}
                  \captionof{figure}{Temperature (eV)}
                  \label{fig:9T_temp_2d}
                \end{subfigure}
                \begin{subfigure}{1\linewidth}
                  \includegraphics[width = 0.95\textwidth, height = 0.15\textheight]{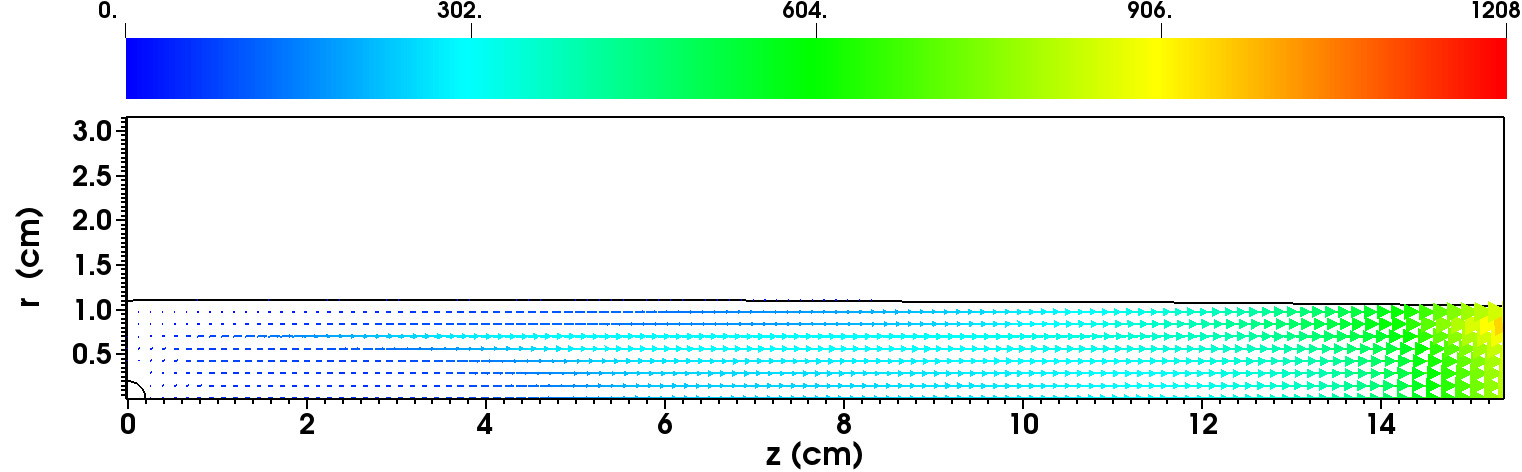}
                  \captionof{figure}{Velocity field (cm/ms)}
                  \label{fig:9T_velo_2d}
                \end{subfigure}
                \caption{States distribution for an ionized neon gas ablatant with
                  parameters $T_{e\infty}=2$ keV, $n_{e\infty}=10^{14}$, $\mathrm{r_p}=2$ mm, $B=9T$ in axisymmetric approximation.}
                \label{fig:9T}
          \end{figure}

            The establishment of the ablation channel for a neon
            pellet during the first $20\mu$s is shown
            in figure~\ref{fig:channel_formation} for $B = 2$ T and $B = 6$ T. The cloud starts
            loosing its spherical shape in the first $5\mu$s
            and begins to reorient itself along the
            magnetic field lines. Near the pellet, the neutral cloud flow
            remains spherical. In table ~\ref{tab:channel_width}, we record the channel width for varying field strengths.
            The channeling of the flow is sensitive to the strength of the magnetic field as can be seen in figure~\ref{fig:rperp_FT}. At low field strengths, the  channel narrows rapidly as the field is increased and undergoes a contraction of 32\% when the magnetic field increases from 2T to 4T. The contraction levels off at higher field strengths, the channel width is decreased by $\sim4.5\%$ when the magnetic field changes from 6T to 9T.

          The 2D distributions of  flow field variables, plotted in
          in figures~\ref{fig:2T},~\ref{fig:4T},\ref{fig:6T},~\ref{fig:9T}, show the structure of the
          ablation flow. The explicit tracking of the cloud boundary
          (represented by the white, almost straight, line in the 2D plots) was
          found to be necessary to unequivocally determine the location of the
          ablated material, and, as result, to evaluate the channel width.  In accordance with the treatment
          of the interface as a propagating contact discontinuity, the pressure
          across the boundary is continuous, while the density, temperature and radial velocity are discontinuous.

          \begin{figure*}
            \begin{subfigure}[t]{0.8\linewidth}
             \centering \includegraphics[width = \linewidth,
              height = 0.45\textheight]{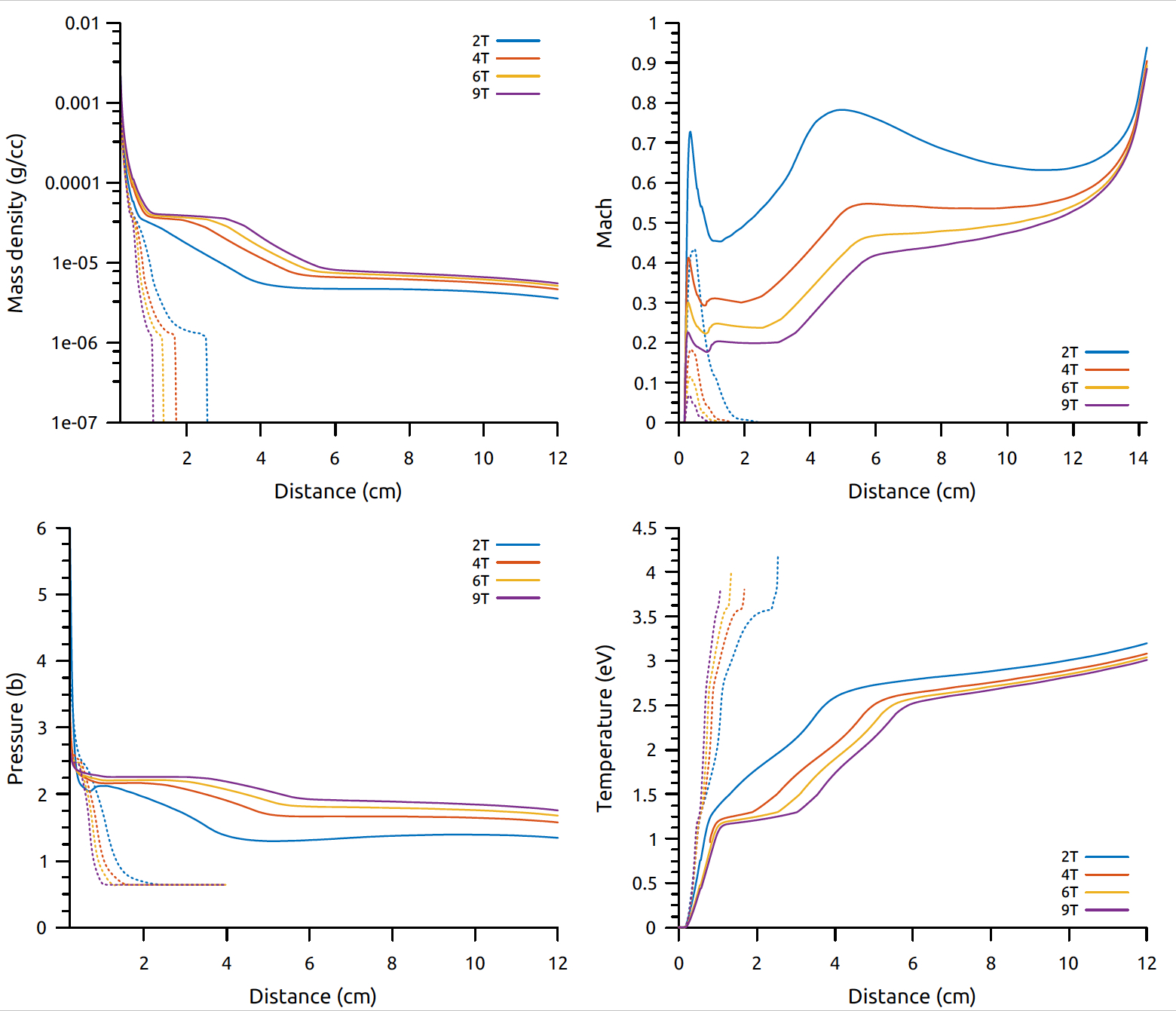}
              \caption{$Ne$.\vspace*{3em}}
              \label{fig:long_rad_all_states_B_mach}
            \end{subfigure}
            
            \begin{subfigure}[t]{0.8\linewidth}
              \centering \includegraphics[width = \linewidth,
              height = 0.45\textheight]{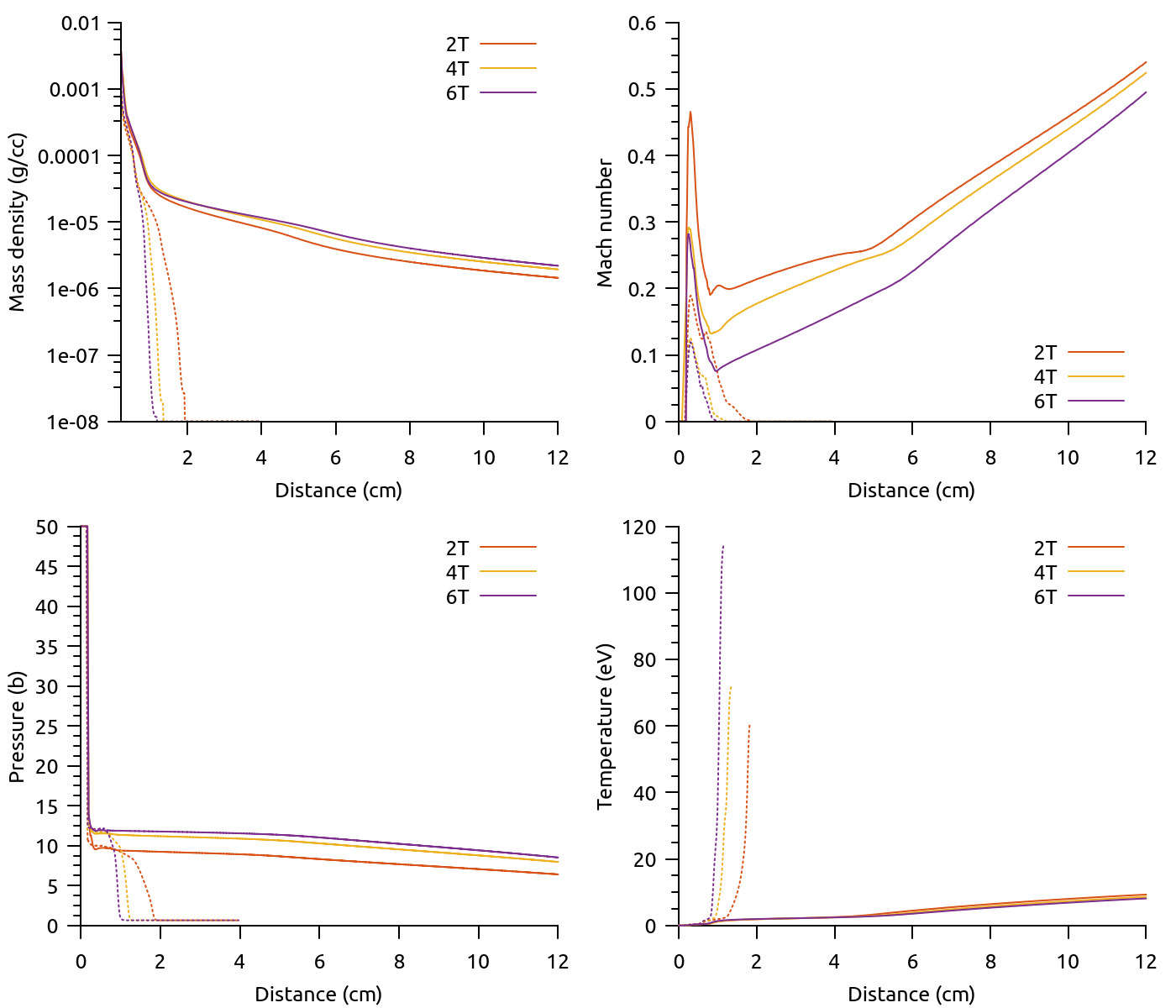}
              \caption{$D_2$.}
              \label{fig:long_rad_D2}
            \end{subfigure}
            \caption{Longitudinal (solid) and radial (dashed) flow variables profiles along the r=0 plane and z=0 plane respectively, for a neon~\ref{fig:long_rad_all_states_B_mach} and molecular deuterium~\ref{fig:long_rad_D2} pellet and varying field values.}
            \end{figure*}

            Radial (dashed) and longitudinal (solid) profiles of the ablation flows are
            presented in figure~\ref{fig:long_rad_all_states_B_mach}. The sudden
            drop in the radial density profiles indicates the edge of the
            channel. Very close to the pellet, as remarked above, the flow
            variables are similarly distributed in both directions. The density and pressure fall rapidly in the radial
            direction  while the temperature  increases. For any value of \textbf{B}, the pressure falls smoothly to
            that of the background plasma $P_{edge} = P_{\infty} = 0.64$ bars with
            a steeper descent gradient at higher field strengths. The
            longitudinial pressure tend to an almost constant value in the
            ablation channel (with a very weak wariation) as the magnetic field
            increases. The $B = 2$ T case shows a negative pressure gradient from
            the pellet surface extending to 4 cm and stabilizing at the (almost)
            constant pressure 1.35 bars. Pressure gradients inside the channel
            tend to vanish as the field strength is increased. When $B = 9$ T, we
            can see that the pressure is constant at 2.25 bars in the region extending from the pellet surface to 4 cm. 
            A weak pressure gradient then
            brings the pressure at 2 bars and then very slowly tapers off. Similar
            effects of the \textbf{B} field are observed on the longitudinal
            density profiles. For $B = 2$ T, the mass density rapidly decreases
            from the pellet surface until 4 cm and remains almost constant
            at $1.45\times10^{-6}$ g/cm$^{-3}$ = $4.32\times10^{16} $ cm$^{-3}$
            until the end of the channel. The $B = 9$ T simulation shows that after a
            strong negative gradient at the surface, the density is almost constant from
            1 cm to 4 cm with value $\rho = 2.4\times10^{-5}$ g/cm$^{-3}$ ($n_t =
            7.16\times10^{17}$ cm$^{-3}$). After a rarefaction region between 4
            and 6 cm the density seems to be stabilizing around $10^{5}$
            g/cm$^{-3}$ = $3\times10^{17}$ cm$^{-3}$. The pressure and density are
            also uniformly higher as the ablation channel contracts more and more
            under the effects of increasing MHD forces. The Mach number and the
            temperature display non linear behaviors, which are caused by
            ionization events starting a few millimeters away from the pellet,
            which reduce the kinetic energy of the flow. The Lorentz force
            squizes the ablation channel forcing a denser flow with higher sound
            speed and lower Mach number, as can be seen in the
            figure~\ref{fig:long_rad_all_states_B_mach}. The flow always remain
            subsonic preventing the formation of shocks, as opposed to the 1D
            spherically symmetric model or the hydrodynamic Saha model. It can
            also be seen that the radial Mach number is consistently at most half
            the value of the longitudinal Mach number. We observe the stagnation point in the transverse direction
            on the z=0 midplane  which coincides
            with the ablation channel radius and where the velocity
            $\mathbf{V}=0$. This is in agreement
            with~\cite{Parks_1991,Samulyak07}.  The temperature increases in both
            directions and just like in the hydrodynamic case, the pellet casts a
            shadow reducing the heat deposition behind it. This difference between
            transverse and longitudinal temperature becomes more pronounced as the
            field strength is increased.

\begin{figure}[!h]
\centering \includegraphics[width=0.95\linewidth,height=0.3\textheight]{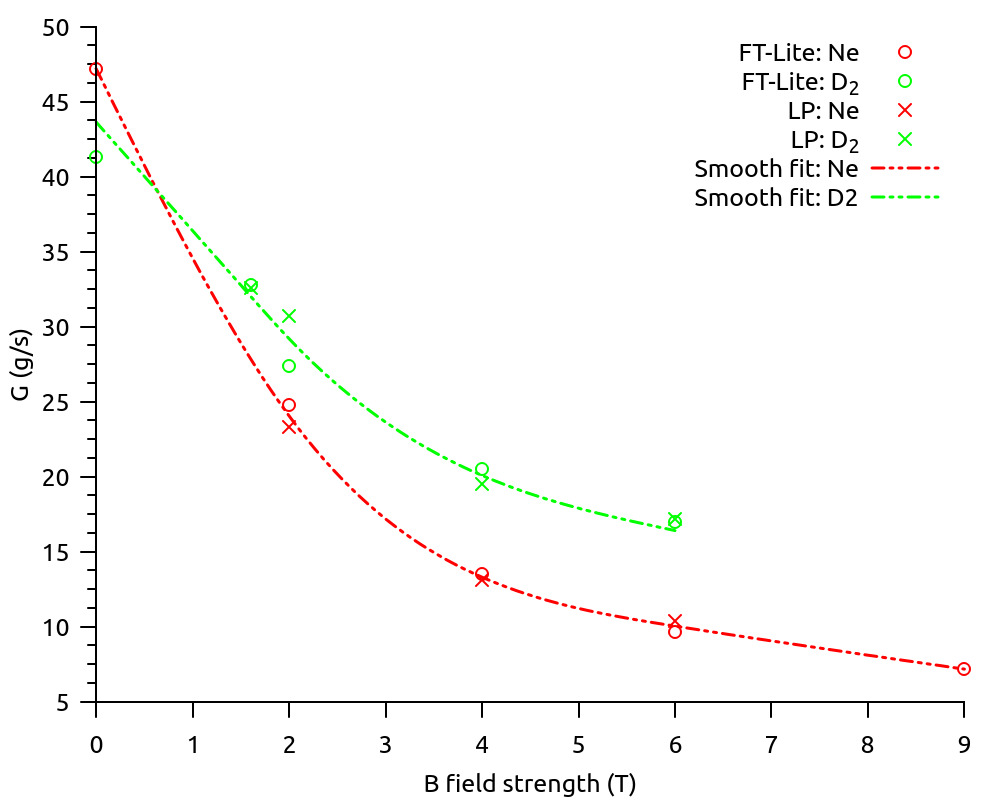}
\caption[Ablation rate as a function of $B$ for $D_2$ and $Ne$ using FronTier-Lite and LP.] {Ablation rate as a function of \textbf{B} for $Ne$ (red) and $D_2$ (green) pellets computed from the FronTier-Lite and LP pellet codes (LP data courtesy of R. Samulyak, S. Yuan and N. Naithlo~\cite{Samulyak_IAEA_2020}).}
\label{fig:G_FT_LP}
\end{figure}

            The radial and longitudinal profiles for ablating $D_2$ pellets are shown in figure~\ref{fig:long_rad_D2} . 
            These plots confirm the
            observations made for a neon pellet concerning the response of the
            ablated flow to the action of the $\mathbf{J}\times\mathbf{B}$
            force. The flow density experiences a strong gradient at the pellet
            surface until $r=1$ cm and then continues to decrease slowly in the
            channel. This is to be contrasted with the density profiles for $Ne$,
            which undergo regions of constant density (from $\sim 1$cm to 4 cm)
            and again from $\sim 5$ cm until the exit. The channel is more narrow
            for $D_2$ and the radial density falls off rapidly. The mean densities
            for both materials have also comparable values. This is not the case
            for the pressure distribution, which has higher values for $D_2$
            pellets. For $B = 2$ T, the mean pressure in the channel is $\sim 1.5$
            bar for $Ne$ and $\sim7.5$ bar for $D_2$, and for $B = 6$ T, $\sim 1.85$ bar
            and $\sim 11$ bar for $Ne$ and $D_2$, respectively. The slow and constant
            decay of the density in the field direction is accompanied by a
            similar decline in the pressure profile. Finally, we note that  the cloud keeps its pressure 
            in the radial direction; it is only near the
            cloud edge that the pressure plunges to the background plasma pressure
            $P_\infty$. This effect is more pronounced as the field increases. The
            $D_2$ pellet cloud is also hotter than the corresponding neon cloud. Main differences between neon and  $D_2$  pellet clouds are 
            explained by
            the absence of volume radiation in deuterium, as well
            as by the multiple ionization potentials of neon. The high temperatures in the channel 
            for $D_2$ pellets also leads to heighten sensitivity to small numerical errors not seen 
            for $Ne$ pellets (figure~\ref{fig:G_FT_LP},~\ref{fig:rperp_FT}).             
            
          Finally, we summarize our studies of the reduction of neon and deuterium pellet ablation rates in magnetic fields in tables \ref{tab:G1},\ref{tab:G2} 
          and in  figure \ref{fig:G_FT_LP}. FronTier results are compared with fully three-dimensional simulations performed with the Lagrangian particle 
          pellet / SPI code \cite{Samulyak_IAEA_2020}. For the code comparison purpose, the grad-B drift model in the Lagrangian particle code
          was turned off and the same fixed shielding length of the ablation cloud was imposed. Both codes are in good agreement for both pellet 
          materials over the magnetic field range from zero to 2 T.  Simulation results of both codes are shown with 
          discrete points,  and the continuous line was obtained by an optimal polynomial fit to the numerical data. Both codes predict strong reduction 
          of the ablation rate in magnetic fields of increasing strength, caused by increased densities in narrower ablation channels, increasing the 
          pellet shielding, and lower ablation flow velocities. The reduction rate is the most significant at low values of the magnetic field.

\begin{table}[h!]
\centering \vspace{2 mm}
\begin{tabular}{>{\centering\arraybackslash}p{2cm}>{\centering\arraybackslash}p{2cm}>{\centering\arraybackslash}p{2cm}}
  \toprule
  $B$ (T)  & $G_{\textrm{FronTier}}$ (g/s) & $G_{\textrm{LP}}$ (g/s) \\\hline
  2        &   24.8       & 23.3 \\
  4        &   13.5       & 13.1 \\
  6        &   9.68       & 10.4 \\
  \hline
\end{tabular}
\caption{Ablation rates (in g/s) comparison for a $Ne$ pellet at different \textbf{B} field values using FT and LP based pellet codes (LP data courtesy of R. Samulyak, S. Yuan and N. Naithlo~\cite{Samulyak_IAEA_2020}).}
\label{tab:G1}
\end{table}

\begin{table}
\begin{tabular}{>{\centering\arraybackslash}p{2cm}>{\centering\arraybackslash}p{2cm}>{\centering\arraybackslash}p{2cm}}
  \toprule
  $B$ (T)  & $G_{\textrm{FronTier}}$ (g/s) & $G_{\textrm{LP}}$ (g/s) \\\hline
  1.6      &   32.8       & 32.6 \\
  2        &   27.4       & 30.7 \\
  4        &   20.5       & 19.5 \\
  6        &   17         & 17.2 \\
  \hline
\end{tabular}
\caption{Ablation rates (in g/s) comparison for a $D_2$ pellet at different \textbf{B} field values using FT and LP based pellet codes (LP data courtesy of R. Samulyak, S. Yuan and N. Naithlo~\cite{Samulyak_IAEA_2020}).}
\label{tab:G2}
\end{table}

\section{Conclusion}
\label{sec:conclusion}

In this work, near-field physics processes relevant to the pellet ablation problem in the context of plasma disruptions mitigation and refueling have been studied numerically using the pellet model based on the FronTier code. The pellet model resolves the pellet surface ablation, the formation of a dense ablated cloud, the anisotropic heating from deposition of hot plasma electrons along magnetic field lines, the ionization of the cloud and redirection of the flow under the effects of MHD forces. Owing to the plasma and pellet parameters considered in this work, we developed a numerical model where MHD effects near the pellet are resolved in the low magnetic Reynold's number approximation ($R_m \ll 1$) and where conditions away from the pellet support the low magnetic $\beta = 2\:\mu_0\:P/B^2$  assumption. The kinetic heat flux model of Parks, updated for neon and deuterium, was implemented as well as a recent conductivity model for the partially ionized ablation cloud of a noble impurity pellet. 

A tabulated LTE equation of state based on solutions of the Saha system for ionization fractions has been used for partially ionized neon and deuterium. To account for deviations from the ideal gas law in high-density, low-temperature ablated material near the pellet surface, a numerical EOS model based on Redlich-Kwong equations has been implemented.  For the pellet sizes and plasma parameters considered here the Redlich-Kwong EOS had negligible effect on the ablation rate but we expect deviation from the ideal case to be important at high temperatures and densities of the background plasma and and small neon pellets where high surface pressure would develop.

Our pellet code is a client of the \textit{FronTier-Lite} application programming interface, which refactors the front tracking libraries of the \textit{FronTier} package. The resulting code is a lightweight computational software for MHD of free surface flows in the low magnetic Reynolds number approximation with interface tracking capability and ablating pellet physics specific routines. In particular, the code is able to resolve the pellet/cloud interface but also the cloud/plasma interface propagating them as contact discontinuities.

In the spherically symmetric approximation for neon pellets, the code has been validated against the improved NGS model  which uses a kinetic solution of the electron distribution function to obtain the heat flux moment for incident Maxwellian electrons and for all light element pellets.in the spherically symmetric approximation in the case of neon. 
Furthermore, results obtained in~\cite{Parks_Ishizaki04,Samulyak07} for deuterium pellets have been confirmed. Additional investigations on the influence of atomic processes have also been conducted and found that the reduction of ablation rates for neon pellets was significant only for larger pellets. Cross-validation of the two dimensional axisymmetric version of the code has been performed against the full 3D Lagrangian particle pellet code and satisfying agreement was obtained over a range of increasing magnetic field strength. Finally, simulation results using the two dimensional axisymmetric approximations have quantified the influence of MHD forces on the cloud evolution and predicted a strong reduction of steady state ablation rates in magnetic fields of increasing strength, suggesting longer pellet lifetime at higher field strength.

Future work will focus on the study of additional physical phenomena (grad B drift, $\mathbf{E}\times\mathbf{B}$ rotation) in full 3D numerical models, the refinement of the EOS models to enable simulations of neon-deuterium mixtures, development and implementation of models for the heating of pellets by runaway electrons, and the resolution of multiple pellet fragments for the disruption mitigation system by shattered pellet injection (SPI).

\vskip5mm
{\bf Acknowledgement.}
This research has been supported by the DOE SciDAC Center for Tokamak Transient Simulations.

\bibliography{BIB} \bibliographystyle{unsrt}
\end{document}